\documentclass[11pt]{article}
\usepackage{a4wide}
\usepackage{amsmath,amssymb,amsbsy}
\usepackage{latexsym,layout,fontenc,xcolor,amsthm,multirow,amsfonts,bm,textcomp}
\usepackage{hyperref}
\usepackage{pstricks}
\usepackage{graphicx}
\usepackage{comment}
\usepackage[utf8x]{inputenc}
\usepackage[english]{babel}
\usepackage[footnotesize]{caption}
\usepackage{indentfirst}
\usepackage{fancyhdr}
\usepackage{vmargin}
\usepackage{yfonts}
\usepackage{lscape}
\usepackage[metapost]{mfpic}
\usepackage{fancybox}
\usepackage[verbose]{wrapfig}
\usepackage{appendix}
\usepackage{subfigure}
\usepackage{enumitem}
\usepackage{array}
\usepackage{colortbl}
\usepackage{calc}
\usepackage{subfloat}
\usepackage{cmll}
\usepackage{multicol}
\usepackage{bookmark}
\usepackage{verbatim}
\usepackage{pst-grad}
\usepackage{pst-plot}
\usepackage{transparent}
\usepackage{color,soul}
\usepackage{cancel}
\usepackage{placeins}

\definecolor{color1}{RGB}{204,0,51}
\definecolor{color2}{RGB}{159,182,205}

\newsavebox{\uuunit}
\sbox{\uuunit}
    {\setlength{\unitlength}{0.825em}
     \begin{picture}(0.6,0.7)
        \thinlines
        \put(0,0){\line(1,0){0.5}}
        \put(0.15,0){\line(0,1){0.7}}
        \put(0.35,0){\line(0,1){0.8}}
       \multiput(0.3,0.8)(-0.04,-0.02){12}{\rule{0.5pt}{0.5pt}}
     \end {picture}}

\def\2{\frac12}
\def\4{\frac14}
\def\ie{{\it i.e. }}

\newcommand{\ket}[1]{\lvert \, #1 \, \rangle}

\def\equationautorefname~#1\null{eq.~(#1)\null
}
\begin{document}

\begin{titlepage}
\begin{center}

 \hfill DFPD/2015/TH/1

\vskip 2cm

{\Large \bf Real Weights, Bound States and Duality Orbits}

\vskip 1cm

{\bf Alessio Marrani\,$^{1,2}$, Fabio Riccioni\,$^3$  and Luca Romano\,$^4$}

\vskip 20pt

{\em $^1$ Centro Studi e Ricerche ``Enrico Fermi'',\\
Via Panisperna 89A, I-00184, Roma, Italy \vskip 5pt }

\vskip 10pt

{\em $^2$ Dipartimento di Fisica e Astronomia ``Galileo Galilei'', \\Universit\`a di Padova,\\ Via Marzolo 8, I-35131 Padova, Italy \vskip 5pt }

{email: {\tt Alessio.Marrani@pd.infn.it}} \\

\vskip 10pt

{\em $^3$ \hskip -.1truecm
 INFN Sezione di Roma,   Dipartimento di Fisica,\\ Universit\`a di Roma ``La Sapienza'',\\ Piazzale Aldo Moro 2, I-00185 Roma, Italy
 \vskip 5pt }

{email: {\tt Fabio.Riccioni@roma1.infn.it}} \\

\vskip 10pt

{\em $^4$ \hskip -.1truecm  Dipartimento di Fisica and INFN Sezione di Roma,\\ Universit\`a di Roma ``La Sapienza'',\\ Piazzale Aldo Moro 2, I-00185 Roma, Italy
 \vskip 5pt }

{email: {\tt  Luca.Romano@roma1.infn.it}} \\

\end{center}

\vskip 0.2cm

\begin{center} {\bf ABSTRACT}\\[3ex]
\end{center}
We show that the duality orbits of extremal black holes in supergravity theories with symmetric scalar
manifolds can be derived by studying the stabilizing subalgebras of suitable
representatives, realized as \textit{bound states} of specific weight vectors of the
corresponding representation of the duality symmetry group.  The weight vectors always correspond to weights that are {\it real}, where the reality properties are derived from the Tits-Satake diagram that identifies the real form of the Lie algebra of the duality symmetry group.
Both $\mathcal{N}=2$ magic Maxwell-Einstein supergravities and the semisimple infinite
sequences of $\mathcal{N}=2$ and $\mathcal{N}=4$ theories in $D=4$ and $5$
are considered, and various results, obtained over the years in the literature
using different methods, are retrieved. In particular, we show that the stratification of the orbits of these theories occurs because of very specific properties of the representations: in the case of the theory based on the real numbers, whose symmetry group is maximally non-compact and therefore all the weights are real, the stratification is due to the presence of weights of different length, while in the other cases it is due to the presence of complex weights.

\vskip 0.7cm


\end{titlepage}

\newpage \pagestyle{plain}

\tableofcontents

\newpage

\numberwithin{equation}{section}

\section{\label{Intro}Introduction}

In recent years, the quest for a consistent theory of quantum gravity led to
an intense study of black hole (BH) solutions within supergravity theories.
A well understood issue is how to physically discriminate among
asymptotically flat branes, depending on their background fluxes. For
example, in $D=4$ space-time dimensions, extremal BHs have electric and
magnetic charges (namely, the fluxes of the 2-form Abelian field strengths
and their duals), which sit in a representation $\mathbf{R}$ of the
electric-magnetic (U-) duality\,\footnote{Throughout the present investigation, we work in the (semi)classical regime
for which the electromagnetic charges take values in the real numbers. Here U-duality is referred to as the ``continuous'' symmetries of
\cite{Cremmer:1979up}. Their discrete versions are the non-perturbative U-duality string theory symmetries studied in \cite{Hull:1994ys}. The orbit
classification of the discrete stringy U-duality groups was started for
the maximally supersymmetric $D=6,5,4$ theories in \cite{Borsten:2009zy,Borsten:2010aa}. Moreover, for $D=4,\mathcal{N}=8$
supergravity it has recently been observed that some of the orbits of $E_{7(7)}(\mathbb{Z})$ should play an important role in counting microstates
of this theory \cite{Bianchi:2009wj,Bianchi:2009mj}. The importance of
discrete invariants and orbits to the dyon spectrum of string theory has
been the subject of much investigation \cite{Dabholkar:2007vk,Sen:2007qy,Banerjee:2007sr,Banerjee:2008pu,Banerjee:2008ri,Sen:2008ta,Sen:2008sp}. For a recent investigation, \textit{cfr.} \cite{Carbone:2014hla}.}
symmetry group $G_{4}$, defining - by virtue of a Theorem due to Dynkin \cite{Dynkin} - the embedding of $G_{4}$ into $Sp(2n,\mathbb{R})$, which
is the largest group acting linearly on the fluxes. The $\mathbf{R}$-representation space of the U-duality group generally exhibits a
stratification into disjoint classes of orbits, which can be identified by
means of suitable sets of constraints on the $G$-invariant (homogeneous of
degree four) polynomial $I_{4}$ \cite{Ferrara:1997uz,D'Auria:1999fa,Ferrara:1997ci}.

BHs with charges fitting into different orbits of $G_{4}$ correspond to
physically distinct solutions. Furthermore, the attractor mechanism \cite{Ferrara:1995ih,Strominger:1996kf,Ferrara:1996dd,Ferrara:1996um,Ferrara:1997tw}
ensures the BH entropy to be independent of the scalars at infinity, and to
be a function of the electric and magnetic charges only. Thus, the
Bekenstein-Hawking \cite{Bekenstein:1973ur,Bekenstein:1973ur-1} entropy can be expressed in
terms of the invariant $I_{4}$ itself. It is here worth recalling the
important distinction between the so-called ``large'' and ``small''
orbits. While the former have $I_{4}\neq 0$ and support an attractor
behavior of the scalar fields in the BH near-horizon geometry, for the
latter the attractor mechanism does not hold, because they correspond to $I_{4}$ $=0$, thus yielding a vanishing Bekenstein-Hawking entropy (at least
at the Einsteinian two-derivative level) \cite{Ferrara:1995ih}.

It is then easy to realize that the classification of U-duality charge
orbits plays a key role in the structure of solutions to gravity theories,
since it captures remarkable properties of the spectrum of possible BHs (and
more generally, of asymptotically flat brane solutions), in turn hinting to
interesting string or M-theoretic insights. The orbits of the $\mathcal{N}=8$
supergravity \cite{Cremmer:1979up} and of the \textit{magic}\ $\mathcal{N}=2$
supergravity based on octonions $\mathbb{O}$ \cite{Gunaydin:1983rk,Gunaydin:1983bi} were obtained in $4$ and $5$ dimensions in
\cite{Ferrara:1997uz}  for both ``large'' and ``small'' BHs exploiting
Jordan algebraic techniques, based on the analysis
of the Freudenthal triple systems defined by the charges. The orbits of the maximal supergravity theories
were independently derived in \cite{Lu:1997bg} performing an
analysis of the weight space of the U-representation $\mathbf{R}$.

The analysis started in \cite{Ferrara:1997uz} was then extended in \cite{Bellucci:2006xz} to the ``large'' orbits of the $\mathcal{N}=2$ Maxwell-Einstein supergravities coupled to vector multiplets,
which also include the three non-exceptional magic theories (based on
Hamilton's quaternions $\mathbb{H}$, on the complex numbers $\mathbb{C}$ and
on the reals $\mathbb{R}$). The
``small'' orbits of the triality-symmetric $STU$ model \cite{Cvetic:1995uj,Duff:1995sm,Cvetic:1995bj,Cvetic:1996zq,Behrndt:1996hu,Bellucci:2007zi,Bellucci:2008sv}
were analyzed in \cite{Borsten:2009yb}. For the infinite sequences of $\mathcal{N}=4$ and $\mathcal{N}=2$ theories with symmetric scalar manifolds,
the U-duality invariant constraints determining the stratification into
distinct orbits as well as the corresponding properties of supersymmetry
breaking, were obtained in \cite{Ferrara:1997ci,Cerchiai:2009pi}, and then
further investigated in \cite{Andrianopoli:2010bj,Ceresole:2010nm,Borsten:2011nq}.
For what concerns the relation between U-invariant BPS conditions and
charge orbits in $D=5$, in which magnetic black strings are the duals of
electric BHs, it was studied along the years in \cite{Ferrara:1997uz,Ferrara:1997ci,Lu:1997bg,Andrianopoli:1997hb,D'Auria:1999fa,
Ferrara:2006xx,Andrianopoli:2007kz,Cerchiai:2010xv}.
Then, in \cite{Borsten:2011ai} the analysis of symmetric supergravity
theories in $D=4$ and $D=5$ was completed. In particular, by exploiting
results and methods from \cite{Ferrara:1997uz,Cerchiai:2009pi,Ceresole:2010nm,Ferrara:2006xx,Ceresole:2007rq,Cerchiai:2010xv,Borsten:2011nq} and \cite{Krutelevich,Shukuzawa:2006ec}, all
``small'' orbits were classified for non-exceptional
magic supergravities, for $\mathcal{N}=2,4$ supergravity
coupled to an arbitrary number of vector multiplets including the special
cases of the $STU$, $ST^{2}$ and $T^{3}$ models, as well as for \textit{minimally coupled} $\mathcal{N}=2$, matter coupled $\mathcal{N}=3$, and
``pure'' $\mathcal{N}=5$ theories. For a review, see also
\cite{Marrani:2010bn}.
Finally,
the analysis of U-invariant BPS conditions and orbits in $D=6$ supergravity
theories, in which electric BHs and magnetic black two-branes are duals to
each other and asymptotically flat dyonic black strings exist, was performed
in \cite{Ferrara:1997ci,Lu:1997bg,Ferrara:2006xx,Andrianopoli:2007kz,Borsten:2010aa}.

The asymptotically flat solutions of supergravity theories discussed so far describe branes with at least three transverse directions, but  in general in string theory there are  also branes that have less than three transverse directions.
 One can consider for instance the D7- and D9-branes of the IIB theory and the D8-brane of the IIA theory. Although a single D7-brane does not have finite energy \cite{Greene:1989ya,Gibbons:1995vg}, one can construct multiple brane configurations which include orientifolds to obtain finite-energy solutions. Similarly, the IIA D8-brane can be viewed as a solution of the massive IIA theory \cite{Romans} whose consistency also requires orientifolds \cite{Polchinski:1995df}. Finally, the space-filling D9-brane of the IIB theory plays a crucial role in the Type-I orientifold construction \cite{Polchinski:1995mt,Angelantonj:2002ct}. All these objects are 1/2-BPS, and they give rise that  a  world-volume effective action which is $\kappa$-symmetric.

In $D$ dimensions, the branes with two, one or zero transverse directions are electrically charged under potentials that are $(D-2)$, $(D-1)$ and $D$-forms respectively. While the  $(D-2)$-forms are dual to the scalars, the other potentials are not propagating and their existence can only be determined by requiring the closure of the supersymmetry algebra. This was done for the ten-dimensional maximal theories in \cite{Bergshoeff:2005ac,Bergshoeff:2005ac-1,Bergshoeff:2005ac-2}, and then all the possible 1/2-BPS branes of the IIB theory were determined in \cite{Bergshoeff:2006gs} requiring the existence of a  $\kappa$-symmetric effective action. The outcome of that analysis is that there are less asymptotically non-flat branes than the corresponding components of the potential. For instance the IIB theory describes an $SL(2,\mathbb{R})$ quadruplet of 10-forms, but only two components can couple to 1/2-BPS 9-branes.

The classification of all the forms of the maximal supergravity theories in any dimension was performed in \cite{Riccioni:2007au,Bergshoeff:2007qi} using the Kac-Moody algebra $E_{11}$ \cite{West:2001as}.
Based on this, a complete classification of 1/2-BPS branes in maximal supergravity theories was obtained in \cite{Bergshoeff:2010xc,Bergshoeff:2011qk,Kleinschmidt:2011vu,Bergshoeff:2012ex}. The brane charges that are selected in this way correspond to particular components of the representations of the fields, and as already mentioned for the IIB case, for branes with two or less transverse directions the number of these components is less than the dimension of the representation. In  \cite{Bergshoeff:2013sxa} it was then understood that the 1/2-BPS branes correspond to the longest weights of the representations of the corresponding potential. In the maximal theory the potentials associated to the asymptotically flat branes have all weights of the same length, while those associated to the asymptotically non-flat ones have weights of different length. A simple example is the $SL(2,\mathbb{R})$  quadruplet of 10-forms of IIB mentioned above: this representation has two long weights and two short ones, and the 1/2-BPS
9-branes correspond to the long weights.

In \cite{Bergshoeff:2012jb,Bergshoeff:2012jb-1} it was shown that the aforementioned classification of branes can also be applied to the half-maximal theories. Although the orthogonal symmetry groups of these theories are not split, and the analysis of \cite{Bergshoeff:2013sxa} cannot be straightforwardly applied, the 1/2-BPS branes are still classified as specific components, identified by a set of light-like conditions, of the representations of the orthogonal symmetry group to which the charges of the branes belong. It was then understood in \cite{Bergshoeff:2014lxa} that  such conditions can be reformulated as the precise group-theory statement that the branes correspond only to the longest weights that are real, where the reality properties are defined by the Tits-Satake diagram that describes the real form of the orthogonal group. This result can then be naturally conjectured to apply to any theory with scalars parametrizing a symmetric manifold, and in particular to all the magic $\mathcal{N}=2$ theories discussed at the beginning of this introduction \cite{Bergshoeff:2014lxa}.

Focusing only on the branes that are asymptotically flat, like BHs in four and five dimensions, the fact that we
have identified from a group-theory perspective what are the weights that correspond to a single 1/2-BPS brane  ({\it i.e.} a rank-1 or 1-charge solution) in the magic  $\mathcal{N}=2$ supergravities means that one can now
 extend the analysis of \cite{Lu:1997bg} to these theories. This is precisely the aim of this paper. We will show that all the orbits of these theories correspond to \textit{bound states} of single 1/2-BPS states associated to the real longest weights.
We will first consider the theory with
\textit{split} U-duality group, which is the magic supergravity based on the
simple rank-3 Jordan algebra on the reals $J_{3}^{\mathbb{R}}$ in $D=5$ and $4$. We will show that the stratification in different orbits of solutions of a given rank is due to the fact that that representations to which the BH charges belong contain weights of different length.
Moreover, as an example of supergravity with a \textit{non-split} U-duality group, we will carefully investigate the orbits of magic
supergravity based on the simple rank-3 Jordan algebra on the complex
numbers $J_{3}^{\mathbb{C}}$ in $D=5$ and $4$ (similar results hold for the
magic theories based on the rank-3 Jordan algebras $J_{3}^{\mathbb{H}}$ and $J_{3}^{\mathbb{O}}$, on the quaternions and octonions, respectively). In this case we will show that the stratification arises because not all the weights in the representation are real.
To summarize, we will be able to compute the stabilizers of various
``large'' and ``small'' orbits from the stability algebra
of \textit{bound states} of weight vectors of the corresponding
representation space.\footnote{In this respect, and as far as the space-like ``large''
4-charge duality orbit in $D=4$ is concerned, it is worth mentioning that
our approach to orbit representatives may be considered the Lie algebraic
analogue of the constituent model of $D=4$ non-BPS extremal BHs
proposed in \cite{Gimon:2009gk}.} These results not only give an alternative method to compute the various orbits of extremal black holes, but more importantly they confirm the validity of the conjecture presented in \cite{Bergshoeff:2014lxa}. It should be here remarked that our
study provides and alternative approach with respect to the analysis based
on nilpotent orbits of symmetry groups characterizing the $D=3$ time-like
reduced gravity theories \cite{Bossard:2009at,Bossard:2009mz,Fre:2011ns}.

It turns out that the generalization of these results to the case of branes with two or less transverse directions is not straightforward. In particular, in three dimensions a codimension-two object ({\it i.e.} a defect brane) is a 0-brane, and in \cite{deBoer:2014iba} a complete classification of the types of such supersymmetric solutions in $D=3$ maximal supergravity  was performed. This  classification did not give rise to a simple criterion for supersymmetry in terms of the
charges of the objects. We will comment on how such a criterion could in principle be derived from our method at least for the case of defect branes in the maximal theories, while the extension to the defect branes of the ${\cal N}=2$ theories could be more complicated due to the structure of the weights of the representations involved.

The plan of the paper is as follows. In \autoref{sectionreviewmaximalcase} we give a brief review of \cite{Lu:1997bg} that will be needed to understand the rest of the paper.
In \autoref{magicn2sugraR} we analyze in detail the orbit stratification of the BH
representations in the $\mathcal{N}=2$ magic theory based on $J_{3}^{\mathbb{R}}$ in five and four dimensions.
Then, in \autoref{secmagicalCHO}, after quickly reviewing how the Tits-Satake diagram of a given real form is defined, we consider the $\mathcal{N}=2$ magic supergravities whose U-duality Lie algebra is not maximally non-compact (\textit{i.e.}, \textit{non-split}), dealing in detail with the simplest case of the theory based on
$J_{3}^{\mathbb{C}}$, again in $D=5$ and $D=4$.
\autoref{sectioninfinieseries} is devoted to the analysis of the supergravity theories based on
the \textit{semisimple} rank-3 Jordan algebras $\mathbb{R}\oplus \mathbf{\Gamma }_{m-1,n-1}$. In particular, we analyze in depth the illustrative
example of the $\mathcal{N}=2$ theory in four dimensions whose U-duality symmetry is $SL(2,\mathbb{R})\times SO(2,4)$. In \autoref{centralcharges} we comment on how our results could be extended to the case of defect branes.
\autoref{conclusions} contains the conclusions. Three appendices conclude the paper.
In \autoref{appendixshortweights} we show why for split real forms the splitting of the orbits is due to the presence of short weights in the representation. \autoref{appendixextraspecial} contains some details about the Cartan involution that are used in the paper. Finally, in \autoref{appendixpictures} we give a simple diagrammatic derivation of the 2-charge orbits for the theory based on $J_{3}^{\mathbb{R}}$ in four dimensions.

\section{\label{sectionreviewmaximalcase}Extremal black holes in maximal supergravity}

The orbits of extremal BH solutions of maximal supergravity theories were determined in \cite{Lu:1997bg} by computing the stabilizing algebra of suitable bound states of weight vectors of the representation of the U-duality symmetry to which the BH charges belong.  In particular, the 1/2-BPS BH solutions have charges that are identified with a single weight vector of the representation,\footnote{As we will emphasize in the rest of the paper, for BHs of the maximal theories any weight corresponds to a 1-charge solution because the symmetry algebra is split and all the weights have the same length \cite{Bergshoeff:2014lxa}.}   and the solutions preserving less supersymmetry correspond to bound states of such 1-charge configurations.
The aim of this section is to give a brief review of  the results of \cite{Lu:1997bg},  that will be then generalized to theories with less supersymmetry in the rest of the paper. In particular, we will only focus on the five-dimensional  and four-dimensional cases.

The orbits of 1/2-BPS BHs can easily be computed by determining the stabilizers of a single weight vector, {\it i.e.} the generators that annihilate it. In particular, one determines the real form of the semisimple part of the stabilizing algebra from the action of the Cartan involution on the generators. It is worth reviewing here that, in general, a real form $\mathfrak{g}$ of a complex Lie algebra $\mathfrak{g}_{\mathbb{C}}$ is characterized by a Cartan involution $\theta$, defined as
an involution such that
\begin{equation}
B_{\theta}(X,Y)= B(X,\theta Y)
\end{equation}
is negative definite, where $B(X,Y)$ is the Killing metric and $X$ and $Y$
are generators of $\mathfrak{g}$. This implies that the compact generators have eigenvalue $+1$ and the non-compact generators have eigenvalue $-1$ under $\theta$.
For instance, if the real form is compact,
then the Killing metric is negative definite and the Cartan involution is
the identity operator. In the case of the maximally non-compact (\ie split) real form, one can
define the action of the Cartan involution to be
\begin{flalign}
 &\theta H_{\alpha}=-H_{\alpha}\qquad \theta E_{\alpha}=-E_{-\alpha} \quad ,
\label{splitthetaactionrootvector}
\end{flalign}
where we denote with $H_\alpha$ the Cartan generators and with $E_\alpha$ the root generators.
This relation implies that the Cartan generators are non-compact, while the root generators combine to form the compact generators $F_\alpha^-$ and the non-compact generators $F_\alpha^+$, both defined as
\begin{flalign}
 &F_{\alpha}^{\pm} \equiv E_{\alpha}\pm E_{-\alpha}\quad .\label{F+F-connectinglongwithshort}
\end{flalign}
In general, from \autoref{splitthetaactionrootvector} one  derives how the Cartan involution acts on any combination of the generators that form a subalgebra, and hence the real form of its semisimple part.

The orbits of less-supersymmetric solutions were  computed in \cite{Lu:1997bg} by determining the stabilizers of
``combinations'' (\textit{bound states}) of weight vectors. For
instance, by considering the stabilizers of a combination of two weight vectors $\ket{W_{1}}$ and $\ket{W_{2}}$ that are not connected by U-duality algebra
transformations, one obtains the orbits of the ``small'' $1/4$-BPS BH solutions, that are \textit{bound states} of two $1/2$-BPS BHs whose
charge corresponds to each of the two weight vectors. The stabilizers are
the generators whose action vanishes on each of the weight vectors (which we
refer to as \textit{common} stabilizers), as well as those that bring them
to two weight vectors that belong to a single weight space (which we refer
to as \textit{conjunction} stabilizers). More specifically, if $E_{\alpha
}\ket{W_{1}}$ and $E_{\beta }\ket{W_{2}}$ define the same weight space with weight $W_{c}$, and $E_{\alpha }\ket{W_{2}}=E_{\beta }\ket{W_{1}}=0$, then the conjunction
stabilizer is the generator $E_{\gamma }$ where $\gamma $ is uniquely
determined to be equal to $\tfrac{1}{2}(\alpha +\beta )$.\footnote{Here one is using the fact that the algebra is simply laced, which is always the case for the maximal theories.}
The construction is then naturally extended to bound states of more than two weight vectors.

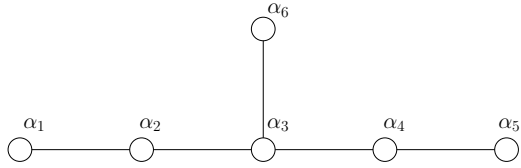
\begin{figure}[h!]
\centering
\scalebox{0.4} {
\begin{pspicture}(0,-0.674844)(16.8,4.7148438)
\psline[linewidth=0.02cm](0.6,-0.25796875)(16.2,-0.25796875)
 \psline[linewidth=0.02cm](8.4,3.9420311)(8.4,-0.25796875)
\pscircle[linewidth=0.02,dimen=outer,fillstyle=solid](0.4,-0.25796875){0.4}
 \pscircle[linewidth=0.02,dimen=outer,fillstyle=solid](8.4,3.7420313){0.4}
\pscircle[linewidth=0.02,dimen=outer,fillstyle=solid](16.4,-0.25796875){0.4}
\pscircle[linewidth=0.02,dimen=outer,fillstyle=solid](8.4,-0.25796875){0.4}
\pscircle[linewidth=0.02,dimen=outer,fillstyle=solid](12.4,-0.25796875){0.4}
\pscircle[linewidth=0.02,dimen=outer,fillstyle=solid](4.4,-0.25796875){0.4}
\rput(0.87640625,0.5651562){\huge $\alpha_1$}
\rput(8.8925,0.5651562){\huge $\alpha_3$}
\rput(12.710468,0.5651562){\huge $\alpha_4$}
\rput(16.48953,0.5651562){\huge $\alpha_5$}
\rput(8.900782,4.365156){\huge $\alpha_6$}
\rput(4.6865625,0.5651562){\huge $\alpha_2$}
\end{pspicture}
}
\caption{The Dynkin diagram of $\mathfrak{e}_{6(6)}$.}
\label{e6dynkindiagram}
\end{figure}

The five-dimensional theory has a global symmetry $E_{6(6)}$, and we draw in \autoref{e6dynkindiagram} the Dynkin diagram of the corresponding Lie algebra $\mathfrak{e}_{6(6)}$.
The BH charges belong to the ${\bf 27}$, whose highest weight is
\begin{equation}
\Lambda_1 = \tfrac{4}{3} \alpha_1 +  \tfrac{5}{3} \alpha_2 + 2 \alpha_3  + \tfrac{4}{3} \alpha_4 +
\tfrac{2}{3} \alpha_5 + \alpha_6 \quad ,
\end{equation}
 corresponding to the Dynkin labels $\boxed{1 \ 0\ 0\ 0\ 0\ 0}$. The generators that annihilate the highest-weight vector $\ket{\Lambda_1}$ form the $\mathfrak{e}_{6(6)}$ maximal triangular subalgebra
$\mathfrak{so}(5,5) \ltimes \mathbb{R}^{16}$,
which is the stabilizing algebra of the ``small'' 1/2-BPS orbit \cite{Ferrara:1997uz,Lu:1997bg}. Clearly, one would get the same algebra considering any other weight in the representation.

Not all the weight vectors of the ${\bf 27}$ can be reached acting with the generators of $\mathfrak{e}_{6(6)}$ on $\ket{\Lambda_1}$. In particular, one weight vector not connected to $\ket{\Lambda_1}$ is $\ket{\Lambda_2}$, with weight
 \begin{equation}
\Lambda_2 = -\tfrac{2}{3} \alpha_1 -  \tfrac{1}{3} \alpha_2  + \tfrac{1}{3} \alpha_4 +
\tfrac{2}{3} \alpha_5  \quad .
\end{equation}
One can then consider the bound state $\ket{\Lambda_1} + \ket{\Lambda_2}$, whose stabilizing algebra is
$\mathfrak{so}(4,5) \ltimes \mathbb{R}^{16}$,
corresponding to the ``small'' 1/4-BPS orbit \cite{Ferrara:1997uz,Lu:1997bg}. There is still one weight vector that is neither connected to $\ket{\Lambda_1}$ nor to $\ket{\Lambda_2}$, which is the lowest-weight vector $\ket{\Lambda_3}$, whose   weight is
 \begin{equation}
\Lambda_3 = -\tfrac{2}{3} \alpha_1 -  \tfrac{4}{3} \alpha_2  -3\alpha_3 - \tfrac{5}{3} \alpha_4 -
\tfrac{4}{3} \alpha_5 - \alpha_6 \quad .
\end{equation}
One can determine the 3-charge orbit from the stabilizers of the bound state $\ket{\Lambda_1} + \ket{\Lambda_2} + \ket{\Lambda_3}$, giving the algebra
$\mathfrak{f}_{4(4)}$,
corresponding to the ``large'' 1/8-BPS orbit. All the weight vectors of the representation are connected to at least one of the three weight vectors just considered, which means that one cannot construct a BH solution of rank higher than 3 in this theory \cite{Lu:1997bg}.

One might ask what happens to the analysis above if one changes the relative sign of one of the weight vectors, which corresponds to changing the sign of one of the charges of the constituents of the bound state. One can show that in general nothing changes for the BH orbits of the maximal theories, because the relative sign of the charges has no effect in determining the real form of the stabilizing algebra.
One single exception to this general rule is the case of the 4-charge orbits in four dimensions.
Indeed, as we will show below, the splitting of the rank-4 solution in two different orbits (the 1/8-BPS and the non-supersymmetric dyonic orbits) is due to the fact that changing the sign of one of the weight vectors in a 4-charge bound state leads to a stabilizing algebra which is a different real form of the same complex algebra. As we will see in the rest of the paper, this feature is completely general in theories with less supersymmetry.

\begin{figure}[h!]
\centering
\scalebox{0.4} {
\begin{pspicture}(0,-0.674844)(20.8,4.7148438)
\psline[linewidth=0.02cm](0.6,-0.25796875)(20.2,-0.25796875)
 \psline[linewidth=0.02cm](8.4,3.9420311)(8.4,-0.25796875)
\pscircle[linewidth=0.02,dimen=outer,fillstyle=solid](0.4,-0.25796875){0.4}
 \pscircle[linewidth=0.02,dimen=outer,fillstyle=solid](8.4,3.7420313){0.4}
\pscircle[linewidth=0.02,dimen=outer,fillstyle=solid](16.4,-0.25796875){0.4}
\pscircle[linewidth=0.02,dimen=outer,fillstyle=solid](20.4,-0.25796875){0.4}
\pscircle[linewidth=0.02,dimen=outer,fillstyle=solid](8.4,-0.25796875){0.4}
\pscircle[linewidth=0.02,dimen=outer,fillstyle=solid](12.4,-0.25796875){0.4}
\pscircle[linewidth=0.02,dimen=outer,fillstyle=solid](4.4,-0.25796875){0.4}
\rput(0.87640625,0.5651562){\huge $\alpha_1$}
\rput(8.8925,0.5651562){\huge $\alpha_3$}
\rput(12.710468,0.5651562){\huge $\alpha_4$}
\rput(16.48953,0.5651562){\huge $\alpha_5$}
\rput(8.900782,4.365156){\huge $\alpha_7$}
\rput(4.6865625,0.5651562){\huge $\alpha_2$}
\rput(20.48953,0.5651562){\huge $\alpha_6$}
\end{pspicture}
}
\caption{The Dynkin diagram of $\mathfrak{e}_{7(7)}$.}
\label{e7dynkindiagram}
\end{figure}
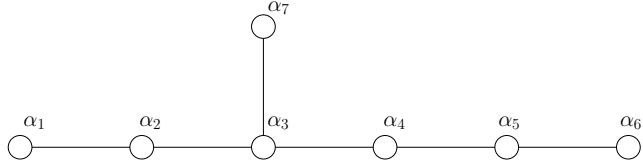

We now move to discussing the orbits of the four-dimensional theory, whose U-duality symmetry is $E_{7(7)}$. We draw in \autoref{e7dynkindiagram} the Dynkin diagram of the Lie algebra $\mathfrak{e}_{7(7)}$.
 The
BH charges belong to the irrep. ${\bf 56}$,
 whose highest weight is
\begin{equation}
\Lambda_1 = \alpha_1 + 2 \alpha_2 + 3 \alpha_3 + \tfrac{5}{2} \alpha_4 + 2 \alpha_5 + \tfrac{3}{2} \alpha_6 + \tfrac{3}{2} \alpha_7  \quad ,
\end{equation}
corresponding to the Dynkin labels $\boxed{0 \ 0\ 0\ 0\ 0\ 1\ 0}$,
using the conventions for the simple roots defined in \autoref{e7dynkindiagram}. The stabilizers of the highest-weight vector $\ket{\Lambda_1}$ form the $\mathfrak{e}_{7(7)}$  maximal triangular subalgebra  $\mathfrak{e}_{6(6)} \ltimes \mathbb{R}^{27}$,
resulting in the 1/2-BPS ``small'' orbit \cite{Ferrara:1997uz,Lu:1997bg}.

Exactly as in five dimensions, one considers bound states of weight vectors that are not connected to each other by transformations in the algebra. In the case of the ${\bf 56}$, the maximum number of such weight vectors is four, and in particular we choose them to be $\ket{\Lambda_1}$ together with $\ket{\Lambda_2}$, $\ket{\Lambda_3}$ and $\ket{\Lambda_4}$, whose weights are
\begin{eqnarray}
& & \Lambda_2 = \alpha_1 + \alpha_2 + \alpha_3 + \tfrac{1}{2} \alpha_4 -\tfrac{1}{2} \alpha_6 + \tfrac{1}{2} \alpha_7 \nonumber \\
& & \Lambda_3 = -\alpha_1 - \alpha_2 - \alpha_3 - \tfrac{1}{2} \alpha_4 -\tfrac{1}{2} \alpha_6 - \tfrac{1}{2} \alpha_7 \nonumber \\
& & \Lambda_4 = -\alpha_1 - 2 \alpha_2 - 3 \alpha_3 - \tfrac{5}{2} \alpha_4 - 2 \alpha_5 - \tfrac{1}{2} \alpha_6 - \tfrac{3}{2} \alpha_7 \quad .
\end{eqnarray}
The 2-charge bound state $\ket{\Lambda_1}+ \ket{\Lambda_2}$ has a stabilizing algebra $(\mathfrak{so}(6,5) \ltimes \mathbb{R}^{32} ) \times \mathbb{R}$, while the 3-charge bound state $\ket{\Lambda_1} + \ket{\Lambda_2} + \ket{\Lambda_3}$ has a stabilizing algebra $\mathfrak{f}_{4(4)} \ltimes \mathbb{R}^{26}$. They correspond to the 1/4-BPS and 1/8-BPS ``small'' orbits \cite{Ferrara:1997uz,Lu:1997bg}.

As already anticipated, the 4-charge orbits are special because as we will show now the result depends on the relative sign on the weight vectors.  In particular, one can consider the bound state
 \begin{equation}
\ket{\Lambda_1} +\ket{\Lambda_2}+\ket{\Lambda_3}+\ket{\Lambda_4}\quad ,\label{1234boundstate56e77susy}
\end{equation}
which will turn out to give the 1/8-BPS ``large'' orbit, and the bound state
\begin{equation}
\ket{\Lambda_1} +\ket{\Lambda_2}+\ket{\Lambda_3}-\ket{\Lambda_4}\quad ,\label{1234boundstate56e77dyonic}
\end{equation}
which differs from the one in \autoref{1234boundstate56e77susy} because we have changed the sign of one of the weight vectors, and which will result in the dyonic orbit. Without going into the details, we can analyze the compactness of the stabilizers using the fact that the Cartan involution acts as in \autoref{splitthetaactionrootvector}. As we will see, it will turn out that some of the stabilizing generators are combinations of the $F$ generators defined in \autoref{F+F-connectinglongwithshort}, and changing the sign in the bound state corresponds to transforming stabilizers that are combinations of $F^-$ generators in stabilizers that are combinations of $F^+$ generators.

We now proceed with a more detailed analysis of the stabilizing algebra.
There are 28 common stabilizers, where four of them are Cartan and thus are non-compact because of \autoref{splitthetaactionrootvector}, while the other 24 are root generators and thus split into 12 compact and 12 non-compact stabilizers. These are the same for the two bound states. Among the conjunction stabilizers, there are 48 root generators that evenly split into 24 compact and 24 non-compact, and again this is not sensitive to the sign of the bound state. There are only two conjunction stabilizers that are compact for the bound state of \autoref{1234boundstate56e77susy} and become non-compact for the one in \autoref{1234boundstate56e77dyonic}. Here we only focus on these generators.
Defining the roots
\begin{eqnarray}
& & \alpha = 2 \alpha_1 + 3 \alpha_2 + 4 \alpha_3 + 3 \alpha_4 + 2 \alpha_5 + \alpha_6 + 2 \alpha_7 \nonumber \\
& & \beta = \alpha_2 + 2 \alpha_3 + 2 \alpha_4 + 2 \alpha_5 + \alpha_6 + \alpha_7 \nonumber \\
& & \gamma= \alpha_6 \quad ,
\end{eqnarray}
these generators can be written as  $F^-_{\beta} +F^-_{\gamma}$ and  $F^-_{\alpha} +F^-_{\gamma}$  for the supersymmetric bound state in \autoref{1234boundstate56e77susy} and $F^+_{\beta} - F^+_{\gamma}$ and  $F^-_{\alpha} +F^-_{\gamma}$  for the dyonic bound state in \autoref{1234boundstate56e77dyonic}.
 The detailed analysis is given in \autoref{thesplittinginthemaximalcase}.

\begin{table}[t!]
\renewcommand{\arraystretch}{1.3}
\par
\begin{center}
\scalebox{1}{
\begin{tabular}{|c||c|c|}
\hline
conj. pairs &$\ket{\Lambda_1}\! +\! \ket{\Lambda_2}\! +\! \ket{\Lambda_3}\! +\! \ket{\Lambda_4}$ &  $\ket{\Lambda_1} \! +\! \ket{ \Lambda_2}\! + \! \ket{\Lambda_3} \! -\! \ket{\Lambda_4}$ \\
\hline
\hline
$(\Lambda_1 , \Lambda_2 )-(\Lambda_3 , \Lambda_4 )$ & $F^-_{\beta} +F^-_{\gamma}$ &  $F^+_{\beta} - F^+_{\gamma}$ \\
 \hline
$(\Lambda_1 , \Lambda_3 )-(\Lambda_2 , \Lambda_4 )$ & $F^-_{\alpha} +F^-_{\gamma}$ &  $F^+_{\alpha} - F^+_{\gamma}$ \\
\hline
$(\Lambda_1 , \Lambda_4 )-(\Lambda_2 , \Lambda_3 )$ & $F^-_{\alpha} -F^-_{\beta}$ &  $F^+_{\alpha} - F^+_{\beta}$ \\
\hline
\end{tabular}
}
\end{center}
\caption{The generators that change compactness in the two 4-charge bound states of the ${\bf 56}$ of $\mathfrak{e}_{7(7)}$. In the first column we list the pairs of weights for which the corresponding generators are conjunction stabilizers. In each of the second and third columns, there are only two independent generators.}
\label{thesplittinginthemaximalcase}
\end{table}

Summarizing, in the  case of the bound state of \autoref{1234boundstate56e77susy} there are 40 non-compact and 38 compact generators, giving the real form $\mathfrak{e}_{6(2)}$, while in the case of the bound state of \autoref{1234boundstate56e77dyonic} two generators become non-compact, giving 42 non-compact and 36 compact generators, leading to the split real form $\mathfrak{e}_{6(6)}$. These correspond to the 1/8-BPS and the dyonic ``large'' orbits.
The latter orbit can also be obtained as the bound state $\ket{\Lambda_1} + \ket{\Lambda_5}$, where
$\Lambda_5 = - \Lambda_1$ is the lowest weight \cite{Lu:1997bg}. One can show that no further stratification occurs, and  in particular the bound state $\ket{\Lambda_1} +\ket{\Lambda_2}-\ket{\Lambda_3}-\ket{\Lambda_4}$ is the same as the one in \autoref{1234boundstate56e77susy}.

\section{\label{magicn2sugraR}Magic $\mathcal{N}=2$ supergravity based on $J_{3}^{\mathbb{R}}$}

The classification of orbits of extremal BHs as bound states of 1/2-BPS
objects performed in \cite{Lu:1997bg} for the case of maximal theories can
be naturally extended to those particular $\mathcal{N}=2$ theories whose
U-duality groups are maximally non-compact (\textit{i.e.} split).
In particular, we consider in this section the four and five-dimensional
theories resulting from the uplift of the $D=3$ theory with global symmetry $F_{4(4)}$. BH orbits in $D=4,5$
were obtained in \cite{Ferrara:1997uz,Bellucci:2006xz,Borsten:2011ai} by
analyzing the symmetries of the attractor equations (for \textquotedblleft
large\textquotedblright\ orbits) as well as the corresponding U-invariant
constraints and Freudenthal triple system. In this section we will derive
the same orbits as resulting from \textit{bound states} of the weight vectors associated to the longest
weights of the global symmetry representation to which the BH
charges belong. We will first consider the $D=5$ case, with symmetry $SL(3,\mathbb{R})$, and then the $D=4$ case, with symmetry $Sp(6,\mathbb{R})$.

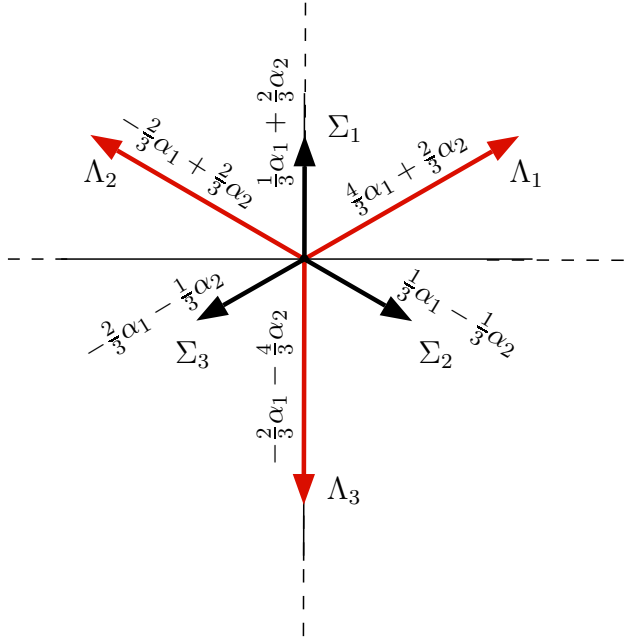
\begin{figure}[t]
\centering
\scalebox{1} 
{\
\begin{pspicture}(0,-6.1296544)(11.604319,2.1296754)
\definecolor{color180}{rgb}{0.8549019607843137,0.054901960784313725,0.0}
\psline[linewidth=0.06cm,linecolor=color180,arrowsize=0.05291667cm 4.0,arrowlength=1.4,arrowinset=0.0]{->}(5.7536855,-0.08980105)(2.9377182,1.5571133)
\psline[linewidth=0.06cm,linecolor=color180,arrowsize=0.05291667cm 4.0,arrowlength=1.4,arrowinset=0.0]{->}(5.7536855,-0.08980105)(8.577701,1.5432748)
\psline[linewidth=0.02cm](8.754422,-0.08931123)(2.5544224,-0.08931123)
\psline[linewidth=0.02cm](5.743871,-4.089789)(5.754422,2.1106887)
\psline[linewidth=0.02cm,linestyle=dashed,dash=0.16cm 0.16cm](5.7679167,3.3101814)(5.762519,1.1101881)
\psline[linewidth=0.02cm,linestyle=dashed,dash=0.16cm 0.16cm](5.7448525,-3.4897902)(5.7409267,-5.0897856)
\psline[linewidth=0.02cm,linestyle=dashed,dash=0.16cm 0.16cm](9.953667,-0.10550432)(8.153672,-0.10108778)
\psline[linewidth=0.02cm,linestyle=dashed,dash=0.16cm 0.16cm](2.6344223,-0.08931123)(1.8544223,-0.08931123)
\psdots[dotsize=0.12,dotangle=179.85942](5.7536855,-0.08980105)
\usefont{T1}{ppl}{m}{n}
\rput{89.93101}(7.127371,-3.6461434){\rput(5.3589535,1.7606888){$\frac{1}{3}\alpha_{1}+\frac{2}{3}\alpha_{2}$}}
\usefont{T1}{ppl}{m}{n}
\rput{-32.132645}(0.008459474,2.4391885){\rput(4.2089534,1.2206888){$-\frac{2}{3}\alpha_{1}+\frac{2}{3}\alpha_{2}$}}
\usefont{T1}{ppl}{m}{n}
\rput{-30.994379}(1.5553126,3.8991597){\rput(7.7789536,-0.83931124){$\frac{1}{3}\alpha_{1}-\frac{1}{3}\alpha_{2}$}}
\usefont{T1}{ppl}{m}{n}
\rput{31.240772}(0.12812975,-2.0884645){\rput(3.7689536,-0.7993112){$-\frac{2}{3}\alpha_{1}-\frac{1}{3}\alpha_{2}$}}
\usefont{T1}{ppl}{m}{n}
\rput{90.00887}(3.624694,-7.134318){\rput(5.3489537,-1.7393112){$-\frac{2}{3}\alpha_{1}-\frac{4}{3}\alpha_{2}$}}
\usefont{T1}{ppl}{m}{n}
\rput{30.055645}(1.5019294,-3.4245937){\rput(7.0989537,1.1006888){$\frac{4}{3}\alpha_{1}+\frac{2}{3}\alpha_{2}$}}
\psline[linewidth=0.06cm,linecolor=color180,arrowsize=0.05291667cm 4.0,arrowlength=1.4,arrowinset=0.0]{->}(5.7536855,-0.08980105)(5.7456865,-3.3497913)
\psline[linewidth=0.06cm,arrowsize=0.05291667cm 4.0,arrowlength=1.4,arrowinset=0.0]{->}(5.7536855,-0.08980105)(7.1716695,-0.91328275)
\psline[linewidth=0.06cm,arrowsize=0.05291667cm 4.0,arrowlength=1.4,arrowinset=0.0]{->}(5.7536855,-0.08980105)(4.331678,-0.90631443)
\psline[linewidth=0.06cm,arrowsize=0.05291667cm 4.0,arrowlength=1.4,arrowinset=0.0]{->}(5.7536855,-0.08980105)(5.7577095,1.550194)
\usefont{T1}{ptm}{m}{n}
\rput(6.28786,1.6456888){\large $\Sigma_{1}$}
\usefont{T1}{ptm}{m}{n}
\rput(7.4878597,-1.3543112){\large $\Sigma_{2}$}
\usefont{T1}{ptm}{m}{n}
\rput(4.28786,-1.3543112){\large $\Sigma_{3}$}
\usefont{T1}{ptm}{m}{n}
\rput(8.677859,1.0456887){\large $\Lambda_{1}$}
\usefont{T1}{ptm}{m}{n}
\rput(3.0778599,1.0456887){\large $\Lambda_{2}$}
\usefont{T1}{ptm}{m}{n}
\rput(6.2778597,-3.1543112){\large $\Lambda_{3}$}
\end{pspicture}
}
\caption{The weights of the $\mathbf{6}$ of $\mathfrak{sl}(3,\mathbb{R})$.
We have painted in red the three long weights. }
\label{weightsof6ofsl3}
\end{figure}

\subsection{\label{5dimtheoryorbits}$D=5$}

Besides the spin-2 graviton, the massless bosonic spectrum of the ungauged magic
$\mathcal{N}=2$, $D=5$ Maxwell-Einstein supergravity based on $J_{3}^{\mathbb{R}}$ (coupled to 5 vector
multiplets) consists of 5 real scalars, parametrizing the symmetric coset $SL(3,\mathbb{R})/SO(3)$, and 5 Abelian vectors which, together with the
vector in the gravity multiplet (\textit{graviphoton}) transform\,\footnote{In fact, among all Maxwell-Einstein supergravities with \textit{homogeneous}
scalar manifold, magic supergravity theories are the only \textit{unified}
theories in $D=5$ \cite{Gunaydin:2003yx}.} in the  $\mathbf{6}$ (rank-$2$ symmetric) of $SL(3,\mathbb{R})$ \cite{Gunaydin:1983rk,Gunaydin:1983bi}.\footnote{In $D=5$, there are two classes of asymptotically flat branes: electric
black holes (0-branes) and magnetic black strings (1-branes), respectively
sitting in the $\mathbf{6}$ and $\mathbf{6}^{\prime }$ of $SL(3,\mathbb{R})$, using opposite conventions on irreps. of $SL(3,\mathbb{R})$ and
$SL(3,\mathbb{C})$ with respect to the ones used in \cite{Cerchiai:2010xv}.}

\subsubsection{1-charge orbit}

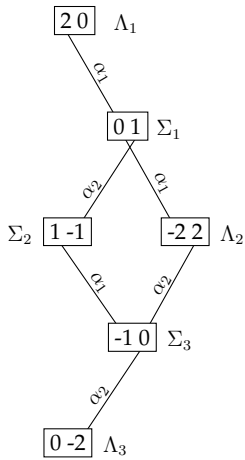
\begin{figure}[t]
\centering
\scalebox{0.7} 
{\
\begin{pspicture}(0,-4.348073)(7.286875,4.348073)
\psline[linewidth=0.02cm](3.8434374,-2.124323)(2.7034376,-3.844323)
\psline[linewidth=0.02cm](4.8634377,-0.084322914)(3.8634374,-1.8843229)
\psline[linewidth=0.02cm](2.1834376,-0.12432291)(3.4234376,-1.924323)
\psline[linewidth=0.02cm](2.3034375,3.895677)(3.3034375,2.155677)
\psline[linewidth=0.02cm](3.4034376,1.8756771)(4.3634377,0.15567708)
\psline[linewidth=0.02cm](3.7434375,1.8756771)(2.6634376,0.19567709)
\usefont{T1}{ppl}{m}{n}
\rput(2.396875,-4.009323){\large \psframebox[linewidth=0.02,fillstyle=solid]{0 -2}}
\usefont{T1}{ppl}{m}{n}
\rput(3.6079688,-2.009323){\large \psframebox[linewidth=0.02,fillstyle=solid]{-1 0}}
\usefont{T1}{ppl}{m}{n}
\rput(4.600625,-0.009322917){\large \psframebox[linewidth=0.02,fillstyle=solid]{-2 2}}
\usefont{T1}{ppl}{m}{n}
\rput(3.5185938,1.9906771){\large \psframebox[linewidth=0.02,fillstyle=solid]{0 1}}
\usefont{T1}{ppl}{m}{n}
\rput(2.5226562,3.990677){\large \psframebox[linewidth=0.02,fillstyle=solid]{2 0}}
\usefont{T1}{ppl}{m}{n}
\rput(2.3889062,-0.009322917){\large \psframebox[linewidth=0.02,fillstyle=solid]{1 -1}}
\usefont{T1}{ppl}{m}{n}
\rput(3.486875,3.9306772){\large $\Lambda_{1}$}
\usefont{T1}{ppl}{m}{n}
\rput(3.286875,-4.069323){\large $\Lambda_{3}$}
\usefont{T1}{ppl}{m}{n}
\rput(5.486875,-0.069322914){\large $\Lambda_{2}$}
\usefont{T1}{ppl}{m}{n}
\rput(4.296875,1.930677){\large $\Sigma_{1}$}
\usefont{T1}{ppl}{m}{n}
\rput(1.496875,-0.069322914){\large $\Sigma_{2}$}
\usefont{T1}{ppl}{m}{n}
\rput(4.496875,-2.0693228){\large $\Sigma_{3}$}
\usefont{T1}{ptm}{m}{n}
\rput{-54.33173}(2.062156,1.9982002){\rput(2.9479687,-0.9943229){$\alpha_{1}$}}
\usefont{T1}{ptm}{m}{n}
\rput{56.023384}(-1.2311541,-3.8658836){\rput(2.9879687,-3.074323){$\alpha_{2}$}}
\usefont{T1}{ptm}{m}{n}
\rput{62.090515}(1.3580568,-4.236356){\rput(4.1679688,-0.9743229){$\alpha_{2}$}}
\usefont{T1}{ptm}{m}{n}
\rput{55.4818}(1.9144267,-2.0203378){\rput(2.8479688,0.8256771){$\alpha_{2}$}}
\usefont{T1}{ptm}{m}{n}
\rput{-61.164623}(1.390566,4.0927677){\rput(4.127969,0.8856771){$\alpha_{1}$}}
\usefont{T1}{ptm}{m}{n}
\rput{-59.138016}(-1.079539,4.0370636){\rput(2.9879687,2.985677){$\alpha_{1}$}}
\end{pspicture}
}
\caption{The weights of the $\mathbf{6}$ of $\mathfrak{sl}(3,\mathbb{R})$.
The weights are represented by boxes, and the entries of each box are the
Dynkin labels of the corresponding weight.}
\label{6sl3}
\end{figure}

The orbits of ``small'' extremal BH solutions
with vanishing quadratic constraint on the charges, \textit{i.e.} the highest-weight orbits, are called rank-$1$ orbits in the language of Jordan triple systems \cite{Ferrar,Krutelevich}.
The $\mathbf{6}$ of $\text{SL}(3,\mathbb{R})$, which is the representation
with two symmetric fundamental indices, contains three long weights and
three short weights, as shown in \autoref{6sl3}, and the highest-weight
orbit can easily be computed by determining the generators that stabilize
each long-weight vector \cite{Bergshoeff:2013sxa}. In components, the black-hole charge is $Q_{MN}=Q_{(MN)}$ ($M,N=1,2,3$) and the long-weight vectors correspond to the components $Q_{11}$, $Q_{22}$ and $Q_{33}$, which are indeed the components for which the
quadratic constraint vanishes, while the other three components are
associated to the short-weight vectors. The Dynkin labels of each weight in \autoref{weightsof6ofsl3} are listed in \autoref{6sl3}. In both figures the long
weights are denoted by $\Lambda _{i}$ while the short weights are denoted by
$\Sigma _{i}$, with $i=1,2,3$. In particular,
\begin{equation}
\Lambda _{1}=\tfrac{4}{3}\alpha _{1}+\tfrac{2}{3}\alpha _{2}
\label{highestweightofthe6ofsl3}
\end{equation}
is the highest weight, where we denote with $\alpha _{1}$ and $\alpha _{2}$
the simple roots of the Lie algebra $\mathfrak{sl}(3,\mathbb{R})$. The
diagram in \autoref{6sl3} is drawn using the conventions explained in \cite{Bergshoeff:2013sxa}, and in particular going down in the right direction
means subtracting $\alpha _{1}$, while going down in the left direction
means subtracting $\alpha _{2}$.

By looking at the diagram in \autoref{6sl3}, it is easy to determine the
generators $E_{\alpha }$ that stabilize each single weight vector $\ket{W}$ with
weight $W$, by observing the roots $\alpha $ that do not give another weight
in the diagram when summed to $W$, together with the Cartan generators that
annihilate the weight vector. In general, we are interested in all the
weight vectors that are not connected by transformations of the Lie algebra, and in
the particular case of the $\mathbf{6}$ of $\text{SL}(3,\mathbb{R})$ a
choice of such vectors is $\ket{\Lambda _{i}}$ $(i=1,2,3)$,  whose corresponding stabilizing
generators are listed in \autoref{stabilizers6sl3}.\footnote{We also list in the table the stabilizing generators for the short-weight vector $\ket{\Sigma _{1}}$, which is not connected to $\ket{\Lambda_3}$.} The semisimple part of
the stabilizing algebra is identified by the subset of the stabilizers
closed under the action of the Cartan involution, which acts as in \autoref{splitthetaactionrootvector} because the algebra is split. The final result
is that the stabilizing algebra for each long weight $\Lambda _{i}$ is $\mathfrak{sl}(2,\mathbb{R})\ltimes \mathbb{R}^{2}$, which is precisely the
rank-$1$ orbit in the language of \cite{Ferrar,Krutelevich}. The stabilizing
algebra of the short weights is instead $\mathfrak{so}(1,1)\ltimes \mathbb{R}^{2}$.

\subsubsection{2-charge orbits}

As reviewed in the previous section,
in \cite{Lu:1997bg} the orbits of BH solutions of maximal supergravity with
rank higher than 1 were computed by determining the stabilizers of
bound states of weight vectors.   In $\mathcal{N}=8$ maximal supergravity the representations of the BH
charges always have a single dominant weight, which means that all weights
have the same length. This is clearly not always the case for $\mathcal{N}=2$
theories. As we will see below, this implies that the analysis of orbits of
bound states of 1/2-BPS BHs has to be refined.

\begin{table}[h]
\renewcommand{\arraystretch}{1.2}
\par
\begin{center}
\begin{tabular}{|c|c|c|c|c|}
\hline
height & $\ket{\Lambda_{1}}$ & $\ket{\Lambda_{2}}$ & $\ket{\Lambda_{3}}$ & $\ket{\Sigma_1}$ \\
\hline\hline
2 & $E_{\alpha_{1}+\alpha_{2}}$ & $E_{\alpha_{1}+\alpha_{2}}$ &  & $E_{\alpha_{1}+\alpha_{2}}$ \\
1 & $E_{\alpha_{1}}\quad E_{\alpha_{2}}$ & $E_{\alpha_{2}}$ & $E_{\alpha_{1}}
$ & $E_{\alpha_{2}}$ \\
0 & $H_{\alpha_{2}}$ & $H_{\alpha_{1}}+H_{\alpha_{2}}$ & $H_{\alpha_{1}}$ & $H_{\alpha_{1}}$ \\
$-1$ & $E_{-\alpha_{2}}$ & $E_{-\alpha_{1}}$ & $E_{-\alpha_{1}}\quad
E_{-\alpha_{2}}$ &  \\
$-2$ &  & $E_{-\alpha_{1}-\alpha_{2}}$ & $E_{-\alpha_{1}-\alpha_{2}}$ &  \\
\hline
\end{tabular}
\end{center}
\caption{Stabilizers for the weight vectors of the $\mathbf{6}$ of $\mathfrak{sl}(3,\mathbb{R})$. The first three (long) weights have a stabilizing algebra $\mathfrak{sl}(2,\mathbb{R})\ltimes \mathbb{R}^{2}$, while for the last
(short) weight $\Sigma _{1}$ one gets $\mathfrak{so}(1,1)\ltimes \mathbb{R}^{2}$. We list in the first column the sum of the coefficients of the simple
roots that occur in a given root, which we dub its \textit{height } \cite{Bergshoeff:2014lxa}.}
\label{stabilizers6sl3}
\end{table}

Knowing that the 1-charge 1/2-BPS BHs correspond to the longest
weights \cite{Bergshoeff:2013sxa}, we want to derive the 2-charge (\textit{i.e.} rank-2 \cite{Ferrar,Krutelevich}) orbits as those that stabilize the combination of two
such weights, in the very same way as in the maximal
theory. Schematically, given the 1-charge BHs associated to the longest
weights $\Lambda _{1}$ and $\Lambda _{2}$, we write their bound state as $\ket{\Lambda _{1}}+\ket{\Lambda _{2}}$. As in maximal supergravity, this state is
annihilated by the common stabilizers and by the conjunction stabilizers.
From \autoref{stabilizers6sl3}, one obtains that the common stabilizers are $E_{\alpha _{1}+\alpha _{2}}$ and $E_{\alpha _{2}}$, while from \autoref{6sl3}
it is clear that there is only one weight vector that can be reached from $\ket{\Lambda_{1}}$ by acting with $E_{-\alpha _{1}}$ as well as from
$\ket{\Lambda _{2}}$ by
acting with $E_{\alpha _{1}}$, namely the short-weight vector $\ket{\Sigma _{1}}$.
Correspondingly, the conjunction stabilizer is $E_{\alpha _{1}}-E_{-\alpha
_{1}}$. It is worth observing that, with respect to the same analysis for
the maximal supergravity \cite{Lu:1997bg}, in this case the
two roots are one the opposite of the other, which means that the
conjunction stabilizer is not associated to a combination of roots.
Moreover, if one formally takes the linear combination $\ket{\Lambda _{1}}-\ket{\Lambda
_{2}}$, the common stabilizers are the same as before, while the conjunction
stabilizer becomes $E_{\alpha _{1}}+E_{-\alpha _{1}}$. We recognize in these two combinations the generators $F_\alpha^-$ and $F_\alpha^+$ defined in \autoref{F+F-connectinglongwithshort}. Remembering the generator $F_\alpha^-$ is compact while the generator $F_\alpha^+$ is non-compact,
 we get that the
stabilizing algebra is $\mathfrak{so}(2)\ltimes \mathbb{R}^{2}$ for the $\ket{\Lambda _{1}}+\ket{\Lambda _{2}}$ orbit and $\mathfrak{so}(1,1)\ltimes \mathbb{R}^{2}$ for the $\ket{\Lambda _{1}}-\ket{\Lambda _{2}}$ orbit. This is summarized in \autoref{6sl3l1+l2stabalgb}.

\begin{table}[h!]
\renewcommand{\arraystretch}{1.2}
\par
\begin{center}
\begin{tabular}{|c|c|c|}
\hline
Common & $\ket{\Lambda_{1}}+\ket{\Lambda_{2}}$ Conjunction & $\ket{\Lambda_{1}}-\ket{\Lambda_{2}}$
Conjunction \\ \hline\hline
$E_{\alpha_{1}+\alpha_{2}}$ & $F_{\alpha_{1}}^{-}$ & $F_{\alpha_{1}}^{+}$ \\
$E_{\alpha_{2}}$ &  &  \\ \hline
\end{tabular}
\end{center}
\caption{Generators of the $\ket{\Lambda_{1}}+\ket{\Lambda_{2}}$ and $\ket{\Lambda_{1}}-
\ket{\Lambda_{2}}$ stabilizing algebras, that are $\mathfrak{so}(2)\ltimes \mathbb{R}^{2}$ and $\mathfrak{so}(1,1)\ltimes \mathbb{R}^{2}$ respectively.}
\label{6sl3l1+l2stabalgb}
\end{table}

It should be stressed that the existence of two 2-charge orbits is
here exactly due to the fact that there are short and long weights  in the representation. Indeed,
whenever this occurs, one gets two long weights that are connected to a
short weight by $E_{\alpha }$ and $E_{-\alpha }$ respectively, and therefore
one can get both $F_{\alpha }^{-}$ and $F_{\alpha }^{+}$ as suitable
conjunction stabilizers.\footnote{The fact that in the maximal theories all the weights have the same length explains in this perspective why in the maximal five-dimensional supergravity the splitting of the orbits does not occur.}
This is explained in detail in \autoref{appendixshortweights}.
As we will see, this will also be the cause of the splitting of the 3-charge configurations in two different orbits in this theory, as well as of the existence of more
than one U-orbit (with rank $>1$) in the corresponding four-dimensional
theory.

Another crucial point to observe, which will be used throughout the paper,
is the fact that the $\ket{\Lambda _{1}}-\ket{\Lambda _{2}}$ orbit coincides with that
of $\ket{\Sigma _{1}}$, as can be seen from \autoref{stabilizers6sl3}. The
comparison with the literature reveals that the orbit with a more compact
stabilizer, \textit{i.e.} $\ket{\Lambda _{1}}+\ket{\Lambda _{2}}$, is 1/2-BPS, whereas
the $\ket{\Lambda _{1}}-\ket{\Lambda _{2}}$ orbit is non-supersymmetric. This result
also will turn out to be completely general: the non-supersymmetric orbit
can always be obtained as a bound state of weights where  one of them is short.

\subsubsection{3-charge orbits}

We now move to the ``large'' 3-charge (or rank-$3$ \cite{Ferrar,Krutelevich})
orbits. Repeating the construction above, we want to derive these
orbits as \textit{bound states} of three long-weight vectors. Up to an (irrelevant) overall sign, there are two
possibilities for the choice of representatives, \textit{i.e.} $\ket{\Lambda
_{1}}+\ket{\Lambda _{2}}+\ket{\Lambda _{3}}$ and $\ket{\Lambda _{1}}+\ket{\Lambda _{2}}-\ket{\Lambda _{3}}$. We see from \autoref{stabilizers6sl3} that there are no common stabilizers, while from \autoref{6sl3} one derives that the conjunction
stabilizers of each pair of long-weight vectors annihilates the third,
which implies that they are stabilizers of the bound state of three long-weight vectors.
Finally, there are no weight vectors that are connected by the algebra to all of
the three long-weight vectors; this translates to the statement that \textit{there
are no 3-conjunction stabilizers}.\footnote{In general, an $n$-conjunction stabilizer occurs when $n$ weights are
connected to a single weight by transformations of the algebra.} The
overall result is summarized in \autoref{threechargeorbitsl3}, which shows
that while all generators in the first case are compact, in the second case
the conjunction stabilizers involving the weight $\Lambda _{3}$ become
non-compact. As a result, we get the stabilizing algebra $\mathfrak{su}(2)$
in the first case and $\mathfrak{sl}(2,\mathbb{R})$ in the second, in
precise agreement with the literature \cite{Ferrara:2006xx}. In particular,
the $\ket{\Lambda _{1}}+\ket{\Lambda _{2}}+\ket{\Lambda _{3}}$ orbit is 1/2-BPS while the $\ket{\Lambda _{1}}+\ket{\Lambda _{2}}-\ket{\Lambda _{3}}$ orbit is non-supersymmetric.

\begin{table}[h]
\renewcommand{\arraystretch}{1.5}
\par
\begin{center}
\begin{tabular}{|c|c|c|}
\hline
2-conj. & {$\ket{\Lambda_{1}}+\ket{\Lambda_{2}}+\ket{\Lambda_{3}}$ Stabilizers} & {$\ket{\Lambda_{1}}+\ket{\Lambda_{2}}-\ket{\Lambda_{3}}$ Stabilizers} \\ \hline\hline
{$\Lambda_{1},\Lambda_{2}$} & $F^{-}_{\alpha_{1}}$ & $F^{-}_{\alpha_{1}}$ \\
\hline
{$\Lambda_{1},\Lambda_{3}$} & $F^{-}_{\alpha_{1}+\alpha_{2}}$ & $
F^{+}_{\alpha_{1}+\alpha_{2}}$ \\ \hline
{$\Lambda_{2},\Lambda_{3}$} & $F^{-}_{\alpha_{2}}$ & $F^{+}_{\alpha_{2}}$ \\
\hline
\end{tabular}
\end{center}
\par
\caption{Stabilizers of $\ket{\Lambda _{1}}+\ket{\Lambda _{2}}+\ket{\Lambda _{3}}$ and $
\ket{\Lambda _{1}}+\ket{\Lambda _{2}}-\ket{\Lambda _{3}}$ bound states, resulting in $
\mathfrak{su}(2)$ and $\mathfrak{sl}(2,\mathbb{R})$ respectively. In the
first column we list the weights of the states for which the corresponding
operator is a conjunction stabilizer. }
\label{threechargeorbitsl3}
\end{table}

Finally, we can determine the stabilizers of the bound state of a long-weight vector and a short-weight vector. Considering \textit{e.g.} $\ket{\Lambda _{3}}+\ket{\Sigma
_{1}}$, it can be checked using \autoref{stabilizers6sl3} and \autoref{6sl3}
that the stabilizers are those listed in \autoref{longweightshortweightsl3},
leading to the algebra $\mathfrak{so}(1,2)$, which is isomorphic to $
\mathfrak{sl}(2,\mathbb{R})$. We therefore get the same orbit as for the
bound state $\ket{\Lambda _{1}}+\ket{\Lambda _{2}}-\ket{\Lambda _{3}}$, in agreement with the
aforementioned general rule that a short-weight vector is equivalent to the
difference of two long-weight vectors. This is also confirmed by the fact that the
bound state $\ket{\Lambda _{3}}-\ket{\Sigma _{1}}$ leads again to the same stabilizer,
as can be checked using the same procedure.

\begin{table}[h!]
\renewcommand{\arraystretch}{1.2}
\par
\begin{center}
\begin{tabular}{|c|c|}
\hline
\multicolumn{2}{|c|}{$\ket{\Lambda_{3}}+\ket{\Sigma_{1}}$ Stabilizers} \\ \hline
Common & Conjunction \\ \hline\hline
$H_{\alpha_{1}}$ & $E_{\alpha_{1}+\alpha_{2}}-E_{-\alpha_{2}}$ \\
& $E_{\alpha_{2}}-E_{-\alpha_{1}-\alpha_{2}}$ \\ \hline
\end{tabular}
\end{center}
\caption{Generators of the $\mathfrak{so}(1,2)$ stabilizing algebra of $
\ket{\Lambda_{3}}+\ket{\Sigma_{1}}$. }
\label{longweightshortweightsl3}
\end{table}

This completes the analysis of the orbits for the $\mathbf{6}$ of $\mathfrak{sl}(3,\mathbb{R})$. We summarize the results in \autoref{summarysl3},
matching those reported in Table II of \cite{Borsten:2011ai} (\textit{cfr.} also Refs. therein).

In the next subsection we will show that all the orbits of the  $\mathbf{14^{\prime }}$ of $\mathfrak{sp}(6,\mathbb{R})$, pertaining to
extremal BHs of the $\mathcal{N}=2$, $D=4$ Maxwell-Einstein theory based on $J_{3}^{\mathbb{R}}$, can be computed as bound states of longest-weight vectors,
exactly as in $D=5$, and in particular we will again see that the existence
of more than one orbit (with rank $>1$) is related to the presence of short
weights in the representation, and the orbits that can be obtained as bound
states involving short-weight vectors are always non-supersymmetric ones.
\begin{table}[h]
\renewcommand{\arraystretch}{1.7}
\par
\begin{center}
\begin{tabular}{|c|c|c|}
\hline
\multicolumn{2}{|c|}{State} & Stabilizer \\ \hline\hline
\multirow{-1}{*}{\rotatebox{90}{\begin{scriptsize}1-state\end{scriptsize}}}
& $\ket{\Lambda_{1}}$ & $\mathfrak{sl}(2,\mathbb{R})\ltimes \mathbb{R}^{2}$ \\
\hline
& $\ket{\Lambda_{1}}+\ket{\Lambda_{2}}$ & $\mathfrak{so}(2)\ltimes \mathbb{R}^{2}$ \\
& $\ket{\Lambda_{1}}-\ket{\Lambda_{2}}$ & $\mathfrak{so}(1,1)\ltimes \mathbb{R}^{2}$ \\
\multirow{-3}{*}{\rotatebox{90}{\begin{scriptsize}2-state\end{scriptsize}}}
& $\ket{\Sigma_{1}}$ & $\mathfrak{so}(1,1)\ltimes \mathbb{R}^{2}$ \\ \hline
& $\ket{\Lambda_{1}}+\ket{\Lambda_{2}}+\ket{\Lambda_{3}}$ & $\mathfrak{su}(2)\sim \mathfrak{so}
(3)$ \\
& $\ket{\Lambda_{1}}+\ket{\Lambda_{2}}-\ket{\Lambda_{3}}$ & $\mathfrak{sl}(2,\mathbb{R})\sim
\mathfrak{so}(1,2)$ \\
& $\ket{\Lambda_{3}}+\ket{\Sigma_{1}}$ & $\mathfrak{so}(1,2)$ \\
\multirow{-4}{*}{\rotatebox{90}{\begin{scriptsize}3-state\end{scriptsize}}}
& $\ket{\Lambda_{3}}-\ket{\Sigma_{1}}$ & $\mathfrak{so}(1,2)$ \\ \hline
\end{tabular}
\end{center}
\caption{Stabilizers in the $\mathbf{6}$ of $\mathfrak{sl}(3,\mathbb{R})$. }
\label{summarysl3}
\end{table}

\subsection{\label{4dimtheoryorbits}$D=4$}

The $\mathcal{N}=2,D=4$ Maxwell-Einstein supergravity theory based on $
J_{3}^{\mathbb{R}}$ (coupled to 6 vector multiplets) has a global U-duality symmetry $Sp(6,\mathbb{R})$ \cite{Gunaydin:1983rk,Gunaydin:1983bi}.
In order to define our conventions for the simple roots of the $\mathfrak{sp}(6,\mathbb{R})$ algebra, we draw its Dynkin diagram in
\autoref{sp6dynkin}.\footnote{In \autoref{appendixextraspecial} we give a detailed derivation of the structure constants of this algebra that are used in this section.}
\begin{figure}[h]
\centering
\scalebox{0.4} 
{\
\begin{pspicture}(0,-0.9225)(12.849063,0.8925)
\definecolor{color114b}{rgb}{0.996078431372549,0.996078431372549,0.996078431372549}
\psline[linewidth=0.02cm](1.941875,0.2825)(2.141875,0.292813)
\psline[linewidth=0.02cm](6.341875,0.092813)(10.141875,0.092813)
\psline[linewidth=0.02cm](6.141875,0.492813)(9.941875,0.492813)
\psline[linewidth=0.02cm](2.341875,0.292813)(6.141875,0.292813)
\pscircle[linewidth=0.02,dimen=outer,fillstyle=solid,fillcolor=color114b](6.141875,0.2825){0.4}
\pscircle[linewidth=0.02,dimen=outer,fillstyle=solid](2.141875,0.2825){0.4}
\pscircle[linewidth=0.02,dimen=outer,fillstyle=solid](10.141875,0.2825){0.4}
\psline[linewidth=0.02cm](7.941875,0.2825)(8.541875,0.8825)
\psline[linewidth=0.02cm](7.941875,0.2825)(8.541875,-0.3175)
\usefont{T1}{ppl}{m}{n}
\rput(2.2464063,-0.5075){\huge $\alpha_{1}$}
\usefont{T1}{ppl}{m}{n}
\rput(6.4464064,-0.5075){\huge $\alpha_{2}$}
\usefont{T1}{ppl}{m}{n}
\rput(10.446406,-0.5075){\huge $\alpha_{3}$}
\end{pspicture}
}
\caption{The Dynkin diagram of $\mathfrak{sp}(6,\mathbb{R})$.}
\label{sp6dynkin}
\end{figure}
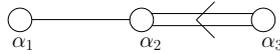
As in the $D=5$ case discussed above, the symmetry group is split, which
implies that the Cartan involution acts on the generators as in \autoref{splitthetaactionrootvector}.
The electric and magnetic charges of the extremal black holes
in the theory transform in the (rank-$3$ anti-symmetric skew-traceless)
irrep. $\mathbf{14^{\prime }}$, that is the representation whose
highest weight is
\begin{equation}
\Lambda _{1}=\alpha _{1}+2\alpha _{2}+\tfrac{3}{2}\alpha _{3}\quad ,
\label{highestweightofthe14ofsp6}
\end{equation}
with Dynkin labels $\boxed{ 0 \ 0\ 1}$.
\begin{figure}[tbp]
\centering
\scalebox{0.5} 
{\
\begin{pspicture}(0,-12.726406)(11.486875,8.426406)
\psline[linewidth=0.02cm](3.9634376,-1.5601562)(6.3634377,-3.9601562)
\usefont{T1}{ppl}{m}{n}
\rput{-45.49773}(3.5120575,3.0038974){\rput(5.3079686,-2.6701562){$\alpha_{1}$}}
\psline[linewidth=0.02cm](2.1634376,1.0398438)(4.5634375,-1.3601563)
\usefont{T1}{ppl}{m}{n}
\rput{-45.49773}(1.1193662,2.4976587){\rput(3.5079687,-0.07015625){$\alpha_{1}$}}
\psline[linewidth=0.02cm](5.7634373,1.0398438)(8.163438,-1.3601563)
\usefont{T1}{ppl}{m}{n}
\rput{-45.49773}(2.195991,5.0652604){\rput(7.107969,-0.07015625){$\alpha_{1}$}}
\psline[linewidth=0.02cm](3.5634375,3.8398438)(5.9634376,1.4398438)
\usefont{T1}{ppl}{m}{n}
\rput{-45.49773}(-0.45896974,4.3335457){\rput(4.9079685,2.7298439){$\alpha_{1}$}}
\psline[linewidth=0.02cm](3.5634375,6.639844)(5.9634376,4.239844)
\usefont{T1}{ppl}{m}{n}
\rput{-45.49773}(-2.4559932,5.1709204){\rput(4.9079685,5.529844){$\alpha_{1}$}}
\psline[linewidth=0.02cm](4.5634375,-1.5601562)(4.5634375,-3.9601562)
\usefont{T1}{ppl}{m}{n}
\rput{91.81141}(1.5925226,-7.434829){\rput(4.3679686,-2.9301562){$\alpha_{2}$}}
\psline[linewidth=0.02cm](6.3634377,3.8398438)(6.3634377,1.4398438)
\usefont{T1}{ppl}{m}{n}
\rput{91.81141}(8.846722,-3.6632366){\rput(6.1679688,2.4698439){$\alpha_{2}$}}
\psline[linewidth=0.02cm](4.1634374,6.639844)(4.1634374,4.239844)
\usefont{T1}{ppl}{m}{n}
\rput{91.81141}(9.375781,1.4241719){\rput(3.9679687,5.2698436){$\alpha_{2}$}}
\psline[linewidth=0.02cm](4.1634374,9.439844)(4.1634374,7.0398436)
\usefont{T1}{ppl}{m}{n}
\rput{91.81141}(12.174382,4.31268){\rput(3.9679687,8.069843){$\alpha_{2}$}}
\psline[linewidth=0.02cm](7.1634374,-9.760157)(4.7634373,-12.160156)
\usefont{T1}{ppl}{m}{n}
\rput{43.80002}(-5.785549,-7.0998755){\rput(5.9079685,-10.730156){$\alpha_{3}$}}
\psline[linewidth=0.02cm](8.963437,-1.5601562)(6.5634375,-3.9601562)
\usefont{T1}{ppl}{m}{n}
\rput{43.80002}(0.39085853,-6.0641656){\rput(7.7079687,-2.5301561){$\alpha_{3}$}}
\psline[linewidth=0.02cm](6.9634376,1.0398438)(4.5634375,-1.3601563)
\usefont{T1}{ppl}{m}{n}
\rput{43.80002}(1.6339514,-3.956455){\rput(5.7079687,0.06984375){$\alpha_{3}$}}
\psline[linewidth=0.02cm](4.7634373,3.8398438)(2.3634374,1.4398438)
\usefont{T1}{ppl}{m}{n}
\rput{43.80002}(2.959825,-1.6546673){\rput(3.5079687,2.8698437){$\alpha_{3}$}}
\psline[linewidth=0.02cm](6.9634376,12.239843)(4.5634375,9.839844)
\usefont{T1}{ppl}{m}{n}
\rput{43.80002}(9.385958,-0.84016705){\rput(5.7079687,11.269844){$\alpha_{3}$}}
\psline[linewidth=0.02cm](3.9634376,-4.360156)(6.3634377,-6.760156)
\usefont{T1}{ppl}{m}{n}
\rput{-45.49773}(5.5090814,2.1665223){\rput(5.3079686,-5.470156){$\alpha_{1}$}}
\psline[linewidth=0.02cm](6.5634375,-6.9601564)(6.5634375,-9.360156)
\usefont{T1}{ppl}{m}{n}
\rput{91.81141}(-1.7415595,-15.004523){\rput(6.3679686,-8.330156){$\alpha_{2}$}}
\psline[linewidth=0.02cm](6.5634375,-4.360156)(6.5634375,-6.760156)
\usefont{T1}{ppl}{m}{n}
\rput{91.81141}(0.85714155,-12.322337){\rput(6.3679686,-5.7301564){$\alpha_{2}$}}
\usefont{T1}{ppl}{m}{n}
\rput(7.686875,12.374844){\large $\Lambda_{1}$}
\usefont{T1}{ppl}{m}{n}
\rput(7.886875,-9.625156){\large $\Lambda_{7}$}
\usefont{T1}{ppl}{m}{n}
\rput(7.886875,-4.2251563){\large $\Lambda_{6}$}
\usefont{T1}{ppl}{m}{n}
\rput(1.686875,1.1748438){\large $\Lambda_{4}$}
\usefont{T1}{ppl}{m}{n}
\rput(5.096875,6.7748437){\large $\Sigma_{1}$}
\usefont{T1}{ppl}{m}{n}
\rput(4.168594,9.574843){\large \psframebox[linewidth=0.02,fillstyle=solid,framesep=0.1]{0 2 -1}}
\usefont{T1}{ppl}{m}{n}
\rput(6.4885936,12.374844){\large \psframebox[linewidth=0.02,fillstyle=solid,framesep=0.1]{0 0 1}}
\usefont{T1}{ppl}{m}{n}
\rput(4.0845313,6.7748437){\large \psframebox[linewidth=0.02,fillstyle=solid,framesep=0.1]{1 0 0}}
\usefont{T1}{ppl}{m}{n}
\rput(4.1670313,3.9748437){\large \psframebox[linewidth=0.02,fillstyle=solid,framesep=0.1]{2 -2 1}}
\usefont{T1}{ppl}{m}{n}
\rput(6.377969,3.9748437){\large \psframebox[linewidth=0.02,fillstyle=solid,framesep=0.1]{-1 1 0}}
\usefont{T1}{ppl}{m}{n}
\rput(6.3685937,1.1748438){\large \psframebox[linewidth=0.02,fillstyle=solid,framesep=0.1]{0 -1 1}}
\usefont{T1}{ppl}{m}{n}
\rput(2.7670312,1.1748438){\large \psframebox[linewidth=0.02,fillstyle=solid,framesep=0.1]{2 0 -1}}
\usefont{T1}{ppl}{m}{n}
\rput(4.568594,-1.4251562){\large \psframebox[linewidth=0.02,fillstyle=solid,framesep=0.1]{0 1 -1}}
\usefont{T1}{ppl}{m}{n}
\rput(8.372344,-1.4251562){\large \psframebox[linewidth=0.02,fillstyle=solid,framesep=0.1]{-2 0 1}}
\usefont{T1}{ppl}{m}{n}
\rput(4.5645313,-4.2251563){\large \psframebox[linewidth=0.02,fillstyle=solid,framesep=0.1]{1 -1 0}}
\usefont{T1}{ppl}{m}{n}
\rput(6.452344,-4.2251563){\large \psframebox[linewidth=0.02,fillstyle=solid,framesep=0.1]{-2 2 -1}}
\usefont{T1}{ppl}{m}{n}
\rput(6.5779686,-6.825156){\large \psframebox[linewidth=0.02,fillstyle=solid,framesep=0.1]{-1 0 0}}
\usefont{T1}{ppl}{m}{n}
\rput(6.568594,-9.625156){\large \psframebox[linewidth=0.02,fillstyle=solid,framesep=0.1]{0 -2 1}}
\usefont{T1}{ppl}{m}{n}
\rput(4.3685937,-12.425157){\large \psframebox[linewidth=0.02,fillstyle=solid,framesep=0.1]{0 0 -1}}
\usefont{T1}{ppl}{m}{n}
\rput(7.696875,-6.825156){\large $\Sigma_{6}$}
\usefont{T1}{ppl}{m}{n}
\rput(3.496875,-4.2251563){\large $\Sigma_{5}$}
\usefont{T1}{ppl}{m}{n}
\rput(3.496875,-1.4251562){\large $\Sigma_{4}$}
\usefont{T1}{ppl}{m}{n}
\rput(7.496875,3.9748437){\large $\Sigma_{2}$}
\usefont{T1}{ppl}{m}{n}
\rput(7.496875,1.1748438){\large $\Sigma_{3}$}
\usefont{T1}{ppl}{m}{n}
\rput(5.286875,9.574843){\large $\Lambda_{2}$}
\usefont{T1}{ppl}{m}{n}
\rput(3.086875,3.9748437){\large $\Lambda_{3}$}
\usefont{T1}{ppl}{m}{n}
\rput(9.686875,-1.4251562){\large $\Lambda_{5}$}
\usefont{T1}{ppl}{m}{n}
\rput(5.486875,-12.425157){\large $\Lambda_{8}$}
\end{pspicture}
}
\caption{The weights of the $\mathbf{14^{\prime }}$ of $\mathfrak{sp}(6,\mathbb{R})$. }
\label{14ofsp6}
\end{figure}
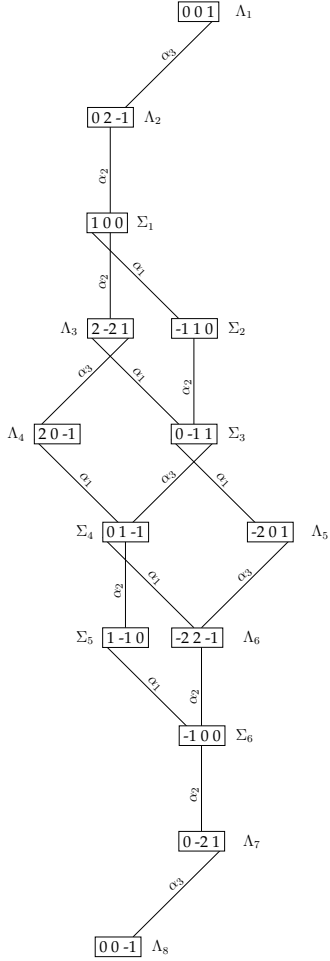
In \autoref{14ofsp6} we draw the weights of the $\mathbf{14^{\prime }}$, as
well as the simple roots that have to be subtracted to any given weight in
order to get the weights below them in the figure. As above, the long
weights are denoted by $\Lambda $ and the short ones by $\Sigma $. As it can
be seen from the figure, eight weights are long and six are short.

\subsubsection{1-charge orbit}

The 1-charge (\textit{i.e. }rank-1 \cite{Ferrar,Krutelevich}) BH orbits
can be computed from the stabilizing algebra of each long-weight vector. Here we
want to show that all the orbits of rank higher than one can be computed as
bound states of long-weight vectors, precisely as occurring in $D=5$. In
particular, we consider bound states of weight vectors that are not connected to
each other by an infinitesimal transformation, \textit{i.e.} by a
transformation in the algebra $\mathfrak{sp}(6,\mathbb{R})$. Starting from
the highest-weight vector $\ket{\Lambda _{1}}$, without any loss of generality, we choose
the other long-weight vectors to be $\ket{\Lambda _{4}}$, $\ket{\Lambda _{6}}$ and $\ket{\Lambda _{7}}
$. We list in \autoref{stab14sp6uno}  the
generators $E_{\alpha }$ that stabilize these weight vectors, as well as the
Cartan stabilizers $H$. We also list the
stabilizers for the short-weight vector $\ket{\Sigma _{6}}$, not connected to the weight vectors
$\ket{\Lambda _{1}}$ and $\ket{\Lambda _{4}}$, because we want to show that, as it holds
in $D=5$, the existence of two 2-charge and two 3-charge orbits is
ultimately due to the presence of short weights, and that non-supersymmetric
orbits can always be obtained as a bound state involving  a
short-weight vector. Finally, in \autoref{stab14sp6uno} we also list the stabilizers
of the lowest-weight vector $\ket{\Lambda _{8}}$, that will be needed for the rank-4
dyonic orbit. As can be deduced from \autoref{stab14sp6uno}, for each long
weight one obtains the stabilizing algebra $\mathfrak{sl}(3,\mathbb{R})\ltimes \mathbb{R}^{6}$, in agreement with the literature \cite{Borsten:2011ai}.

On the other hand, the stabilizing algebra of each short weight is $\mathfrak{so}(2,2)\ltimes (\mathbb{R}\times \mathbb{R}^{4})$. Since in this
analysis a short-weight vector can be traded for a difference of two long-weight vectors,
we will see below that $\mathfrak{so}(2,2)\ltimes (\mathbb{R}\times \mathbb{R}^{4})$ is indeed the stabilizer of a 2-charge (\textit{i.e.} rank-2) orbit
of $\mathfrak{sp}(6,\mathbb{R})$.

\begin{table}[h!]
\renewcommand{\arraystretch}{1.3}
\par
\begin{center}
\scalebox{0.7}{
\begin{tabular}{|c|c|c|c|c|c|c|c|}
\hline
height&$\ket{\Lambda_{1}}$&$\ket{\Lambda_{4}}$&$\ket{\Lambda_{6}}$&$\ket{\Lambda_{7}}$&$\ket{\Sigma_{6}}$&$\ket{\Lambda_{8}}$\\
\hline\hline
5&$E_{2\alpha_{1}+2\alpha_{2}+\alpha_{3}}$&$E_{2\alpha_{1}+2\alpha_{2}+\alpha_{3}}$&&&&\\

\hline

4&
\begin{tabular}{c}
 $E_{\alpha_{1}+2\alpha_{2}+\alpha_{3}}$
\end{tabular}&
\begin{tabular}{c}
 $E_{\alpha_{1}+2\alpha_{2}+\alpha_{3}}$
\end{tabular}&
\begin{tabular}{c}
 $E_{\alpha_{1}+2\alpha_{2}+\alpha_{3}}$
\end{tabular}
&&&\\

\hline

3&
\begin{tabular}{c}
 $E_{\alpha_{1}+\alpha_{2}+\alpha_{3}}$\\
 $E_{2\alpha_{2}+\alpha_{3}}$
\end{tabular}&
\begin{tabular}{c}
 $E_{\alpha_{1}+\alpha_{2}+\alpha_{3}}$
\end{tabular}&
\begin{tabular}{c}
 $E_{2\alpha_{2}+\alpha_{3}}$
\end{tabular}&
\begin{tabular}{c}
 $E_{\alpha_{1}+\alpha_{2}+\alpha_{3}}$
\end{tabular}&
\begin{tabular}{c}
 $E_{2\alpha_{2}+\alpha_{3}}$
\end{tabular}&\\

\hline

2&
\begin{tabular}{c}
 $E_{\alpha_{1}+\alpha_{2}}$\\
 $E_{\alpha_{2}+\alpha_{3}}$
\end{tabular}
&\begin{tabular}{c}
 $E_{\alpha_{1}+\alpha_{2}}$
\end{tabular}
&
\begin{tabular}{c}
 $E_{\alpha_{1}+\alpha_{2}}$\\
 $E_{\alpha_{2}+\alpha_{3}}$
\end{tabular}
&\begin{tabular}{c}
 $E_{\alpha_{2}+\alpha_{3}}$
\end{tabular}&&
\begin{tabular}{c}
 $E_{\alpha_{1}+\alpha_{2}}$
\end{tabular}
\\

\hline

1&\begin{tabular}{c}
 $E_{\alpha_{1}}$\\
 $E_{\alpha_{2}}$\\
 $E_{\alpha_{3}}$
\end{tabular}&
\begin{tabular}{c}
 $E_{\alpha_{1}}$\\
 $E_{\alpha_{2}}$
\end{tabular}&
\begin{tabular}{c}
 $E_{\alpha_{2}}$
\end{tabular}&
\begin{tabular}{c}
 $E_{\alpha_{1}}$\\
 $E_{\alpha_{3}}$
\end{tabular}&
\begin{tabular}{c}
 $E_{\alpha_{3}}$
\end{tabular}&
\begin{tabular}{c}
 $E_{\alpha_{1}}$\\
 $E_{\alpha_{2}}$
\end{tabular}\\

\hline

0&
\begin{tabular}{c}
 $H_{\alpha_{1}}$\\
 $H_{\alpha_{2}}$
\end{tabular}&
\begin{tabular}{c}
 $H_{\alpha_{2}} $\\
 $H_{\alpha_{1}}+2H_{\alpha_{3}}$
\end{tabular}&
\begin{tabular}{c}
 $H_{\alpha_{1}}+H_{\alpha_{2}}$\\
 $H_{\alpha_{2}}+2H_{\alpha_{3}}$
\end{tabular}&
\begin{tabular}{c}
 $H_{\alpha_{1}}$\\
 $H_{\alpha_{2}}+2H_{\alpha_{3}}$
\end{tabular}&
\begin{tabular}{c}
 $H_{\alpha_{2}}$\\
 $H_{\alpha_{3}}$
\end{tabular}&
\begin{tabular}{c}
 $H_{\alpha_{1}}$\\
 $H_{\alpha_{2}}$
\end{tabular}\\

\hline

$-1$&
\begin{tabular}{c}
    $E_{-\alpha_{1}}$\\
    $E_{-\alpha_{2}}$
\end{tabular}&
\begin{tabular}{c}
 $E_{-\alpha_{2}}$\\
 $E_{-\alpha_{3}}$
\end{tabular}&
\begin{tabular}{c}
 $E_{-\alpha_{1}}$\\
 $E_{-\alpha_{3}}$
\end{tabular}&
\begin{tabular}{c}
 $E_{-\alpha_{1}}$\\
 $E_{-\alpha_{2}}$
\end{tabular}&
\begin{tabular}{c}
 $E_{-\alpha_{1}}$\\
 $E_{-\alpha_{3}}$
\end{tabular}&
\begin{tabular}{c}
 $E_{-\alpha_{1}}$\\
 $E_{-\alpha_{2}}$\\
 $E_{-\alpha_{3}}$
\end{tabular}
\\

\hline

$-2$&
\begin{tabular}{c}
 $E_{-\alpha_{1}-\alpha_{2}}$
\end{tabular}&
\begin{tabular}{c}
 $E_{-\alpha_{2}-\alpha_{3}}$
\end{tabular}&
\begin{tabular}{c}
 $E_{-\alpha_{1}-\alpha_{2}}$\\
 $E_{-\alpha_{2}-\alpha_{3}}$
\end{tabular}&
\begin{tabular}{c}
 $E_{-\alpha_{1}-\alpha_{2}}$\\
 $E_{-\alpha_{2}-\alpha_{3}}$
\end{tabular}&
\begin{tabular}{c}
 $E_{-\alpha_{1}-\alpha_{2}}$
\end{tabular}&
\begin{tabular}{c}
 $E_{-\alpha_{1}-\alpha_{2}}$\\
 $E_{-\alpha_{2}-\alpha_{3}}$
\end{tabular}\\

\hline

$-3$&&
\begin{tabular}{c}
 $E_{-\alpha_{1}-\alpha_{2}-\alpha_{3}}$\\
 $E_{-2\alpha_{2}-\alpha_{3}}$
\end{tabular}&
\begin{tabular}{c}
 $E_{-\alpha_{1}-\alpha_{2}-\alpha_{3}}$
\end{tabular}&
\begin{tabular}{c}
 $E_{-\alpha_{1}-\alpha_{2}-\alpha_{3}}$\\
 $E_{-2\alpha_{2}-\alpha_{3}}$
\end{tabular}&
\begin{tabular}{c}
 $E_{-\alpha_{1}-\alpha_{2}-\alpha_{3}}$\\
 $E_{-2\alpha_{2}-\alpha_{3}}$
\end{tabular}&
\begin{tabular}{c}
 $E_{-\alpha_{1}-\alpha_{2}-\alpha_{3}}$\\
 $E_{-2\alpha_{2}-\alpha_{3}}$
\end{tabular}\\

\hline

$-4$&&
\begin{tabular}{c}
 $E_{-\alpha_{1}-2\alpha_{2}-\alpha_{3}}$
\end{tabular}&
\begin{tabular}{c}
 $E_{-\alpha_{1}-2\alpha_{2}-\alpha_{3}}$
\end{tabular}&
\begin{tabular}{c}
 $E_{-\alpha_{1}-2\alpha_{2}-\alpha_{3}}$
\end{tabular}&
\begin{tabular}{c}
 $E_{-\alpha_{1}-2\alpha_{2}-\alpha_{3}}$
\end{tabular}&
\begin{tabular}{c}
 $E_{-\alpha_{1}-2\alpha_{2}-\alpha_{3}}$
\end{tabular}\\

\hline

$-5$&&&\begin{tabular}{c}
$E_{-2\alpha_{1}-2\alpha_{2}-\alpha_{3}}$
\end{tabular}&
\begin{tabular}{c}
$E_{-2\alpha_{1}-2\alpha_{2}-\alpha_{3}}$
\end{tabular}
&\begin{tabular}{c}
$E_{-2\alpha_{1}-2\alpha_{2}-\alpha_{3}}$
\end{tabular}&
\begin{tabular}{c}
$E_{-2\alpha_{1}-2\alpha_{2}-\alpha_{3}}$
\end{tabular}
\\

\hline
\end{tabular}
}
\end{center}
\caption{Stabilizers for the weight vectors of the $\mathbf{14^\prime}$ of $\mathfrak{sp}(6,\mathbb{R})$ that are used in the analysis of the paper. }
\label{stab14sp6uno}
\end{table}

\subsubsection{2-charge orbits}
We now move to consider the multi-charge states as being associated to
combinations of long-weight vectors. In particular, the 2-charge orbits
result from the bound states\,\footnote{In \autoref{appendixpictures} we give a simple argument that shows that one obtains the same orbits considering any other pair of long-weight vectors.}
\begin{flalign}
 &\ket{\Lambda_{1}}+\ket{\Lambda_{4}} \quad  \qquad \ket{\Lambda_{1}}-\ket{\Lambda_{4}} \quad  .\label{twochargeboundstate14prime}
\end{flalign}
The stabilizers of each orbit can be read from \autoref{14stab2states}. In general, one
can deduce the real form of the semisimple part of the stabilizing algebra
by looking at how the Cartan involution acts on the corresponding generators
following \autoref{splitthetaactionrootvector}.

In the $\ket{\Lambda_{1}}+\ket{\Lambda_{4}}$ case, one can recombine all the conjunction
stabilizers and the common stabilizers $H_{\alpha_2}$, $E_{\alpha_2}$ and $E_{-\alpha_2}$ in the form
\begin{flalign}
 &H_{\beta_{1}}=\frac{1}{2}\bigl( H_{\alpha_{2}}+i F^{-}_{\alpha_{2}+\alpha_{3}}\bigr) \nonumber \\
 &E_{\beta_{1}}=\frac{1}{2}\biggl[E_{\alpha_{2}}-i\bigl(E_{2\alpha_{2}+\alpha_{3}}-E_{-\alpha_{3}}\bigr)\biggr]\nonumber \\
 &E_{-\beta_{1}}=\frac{1}{2}\biggl[E_{-\alpha_{2}}-i\bigl(E_{\alpha_{3}}-E_{-2\alpha_{2}-\alpha_{3}}\bigr)\biggr]\nonumber \\
 &H_{\beta_{2}}=\frac{1}{2} \bigl( H_{\alpha_{2}}-i  F^{-}_{\alpha_{2}+\alpha_{3}}\bigr)\nonumber \\
 &E_{\beta_{2}}=\frac{1}{2}\biggl[E_{\alpha_{2}}+i\bigl(E_{2\alpha_{2}+\alpha_{3}}-E_{\alpha_{3}}\bigr)\biggr] \nonumber \\
 &E_{-\beta_{2}}=\frac{1}{2}\biggl[E_{-\alpha_{2}}+i\bigl(E_{\alpha_{3}}-E_{-2\alpha_{2}-\alpha_{3}}\bigr)\biggr] \ \ .\label{so13generatorssp6}
 \end{flalign}
The generators $H_{\beta_1}$, $E_{\beta_1}$ and $E_{-\beta_1}$ and the
generators $H_{\beta_2}$, $E_{\beta_2}$ and $E_{-\beta_2}$ form two separate
$\mathfrak{su}(2)$ algebras, and from \autoref{splitthetaactionrootvector} it
follows that
\begin{flalign}
 &\theta H_{\beta_1} =- H_{\beta_2} \qquad \theta E_{\beta_1} = - E_{-\beta_2 } \ \ .
  \end{flalign}
As a consequence, the generators in \autoref{so13generatorssp6} form the
algebra $\mathfrak{sl}(2, \mathbb{C})$ which is isomorphic to $\mathfrak{so}(1,3)$. The remaining stabilizers in the first column of \autoref{14stab2states} transform in the $\mathbf{4 \oplus 1}$ of $\mathfrak{so}(1,3)
$, where the singlet is the generator $E_{2\alpha_1 + 2 \alpha_2 + \alpha_3}$.
Thus, the stabilizing algebra in the $\ket{\Lambda_{1}}+\ket{\Lambda_{4}}$
case is $\mathfrak{so}(1,3) \ltimes ( \mathbb{R}^{1}\times \mathbb{R}^{4})$.

Similarly, in the case of the $\ket{\Lambda_{1}}-\ket{\Lambda_{4}}$ bound state one
deduces from \autoref{14stab2states} that the semisimple part of the
stabilizing algebra is generated by
\begin{flalign}
 &H_{\gamma_{1}}=\frac{1}{2}\bigl(H_{\alpha_{2}}+F_{\alpha_{2}+\alpha_{3}}^{+} \bigr) \nonumber \\
 &E_{\gamma_{1}}=\frac{1}{\sqrt{2}}\biggl[E_{\alpha_{2}}-\bigl(E_{2\alpha_{2}+\alpha_{3}}+E_{-\alpha_{3}}\bigr)\biggr]\nonumber \\
 &E_{-\gamma_{1}}=\frac{1}{\sqrt{2}}\biggl[E_{-\alpha_{2}}-\bigl(E_{\alpha_{3}}+E_{-2\alpha_{2}-\alpha_{3}}\bigr)\biggr]\nonumber \\
 &H_{\gamma_{2}}=\frac{1}{2}\bigl(H_{\alpha_{2}}-F_{\alpha_{2}+\alpha_{3}}^{+} \bigr)\nonumber \\
 &E_{\gamma_{2}}=\frac{1}{\sqrt{2}}\biggl[E_{\alpha_{2}}+\bigl(E_{2\alpha_{2}+\alpha_{3}}+E_{-\alpha_{3}}\bigr)\biggr]\nonumber \\
 &E_{-\gamma_{2}}=\frac{1}{\sqrt{2}}\biggl[E_{-\alpha_{2}}+\bigl(E_{\alpha_{3}}+E_{-2\alpha_{2}-\alpha_{3}}\bigr)\biggr] \quad .
 \end{flalign}
 \begin{table}[t!]
\renewcommand{\arraystretch}{1.2}
\begin{center}
\begin{tabular}{|c|c|c|}

\hline
Common& $\ket{\Lambda_{1}}+\ket{\Lambda_{4}}$ Conjunction & $\ket{\Lambda_{1}}-\ket{\Lambda_{4}}$ Conjunction\\
\hline \hline
$E_{2\alpha_{1}+2\alpha_{2}+\alpha_{3}}$&$F_{\alpha_{2}+\alpha_{3}}^{-}$ &$F_{\alpha_{2}+\alpha_{3}}^{+}$ \\
$E_{\alpha_{1}+2\alpha_{2}+\alpha_{3}}$&$E_{2\alpha_{2}+\alpha_{3}}-E_{-\alpha_{3}}$ & $E_{2\alpha_{2}+\alpha_{3}}+E_{-\alpha_{3}}$\\
$E_{\alpha_{1}+\alpha_{2}+\alpha_{3}}$&$E_{\alpha_{3}}-E_{-2\alpha_{2}-\alpha_{3}}$  &
$E_{\alpha_{3}}+E_{-2\alpha_{2}-\alpha_{3}}$\\
$E_{\alpha_{1}+\alpha_{2}}$& &\\
$E_{\alpha_{1}}\quad E_{\alpha_{2}}$& &\\
$H_{\alpha_{2}}$& &\\
$E_{-\alpha_{2}}$& & \\
\hline
\end{tabular}
\caption{The stabilizers of the bound states of two long-weight vectors, namely $\ket{\Lambda_{1}}+\ket{\Lambda_{4}}$ and $\ket{\Lambda_{1}}-\ket{\Lambda_{4}}$. In the first case
one gets the algebra $\mathfrak{so}(1,3) \ltimes (\mathbb{R}^{1}\times
\mathbb{R}^{4})$, while the second case gives $\mathfrak{so}(2,2) \ltimes (\mathbb{R}^{1}\times \mathbb{R}^{4})$ (this latter case matches the stabilizer of each short weight).\label{14stab2states}
}
\end{center}
\end{table}
Again, the first three generators commute with the last three, but in this
case using \autoref{splitthetaactionrootvector} one gets
\begin{flalign}
 &\theta H_{\gamma_i} = - H_{\gamma_i} \qquad \theta E_{\gamma_i} = - E_{-\gamma_i } \qquad i=1,2 \ ,
  \end{flalign}
which implies that the semisimple part of the stabilizing algebra is $\mathfrak{sl}(2,\mathbb{R})\oplus \mathfrak{sl}(2,\mathbb{R})$, that is
isomorphic to $\mathfrak{so}(2,2)$. The full stabilizer of the $\ket{\Lambda_1} -
\ket{\Lambda_4}$ orbit is $\mathfrak{so}(2,2) \ltimes (\mathbb{R}^{1}\times
\mathbb{R}^{4} )$.

\begin{table}[t!]
\renewcommand{\arraystretch}{1.3}
\par
\begin{center}
\begin{tabular}{|c||c|c|c|}
\hline
Common & 2-conj. & $\ket{\Lambda_{1}}+\ket{\Lambda_{4}}+\ket{\Lambda_{6}}$ & $
\ket{\Lambda_{1}}+\ket{\Lambda_{4}}-\ket{\Lambda_{6}}$ \\ \hline\hline
$E_{\alpha_{1}+2\alpha_{2}+\alpha_{3}}$ &  & $F^{-}_{\alpha_{2}+\alpha_{3}}$
& $F^{-}_{\alpha_{2}+\alpha_{3}}$ \\
$E_{\alpha_{1}+\alpha_{2}}$ & \multirow{-2}{*}{\rotatebox{90}{$\Lambda_{1},
\Lambda_{4}$}} & $E_{2\alpha_{2}+\alpha_{3}}-E_{-\alpha_{3}}$ & $E_{2\alpha_{2}+\alpha_{3}}-E_{-\alpha_{3}}$ \\ \cline{2-4}
$E_{\alpha_{2}}$ &  & $F^{-}_{\alpha_{1}+\alpha_{2}+\alpha_{3}}$ & $F^{+}_{\alpha_{1}+\alpha_{2}+\alpha_{3}}$ \\
& \multirow{-2}{*}{\rotatebox{90}{$\Lambda_{1},\Lambda_{6}$}} & $E_{2\alpha_{1}+2\alpha_{2}+\alpha_{3}}+E_{-\alpha_{3}}$ & $E_{2\alpha_{1}+2\alpha_{2}+\alpha_{3}}-E_{-\alpha_{3}}$ \\ \cline{2-4}
&  & $F_{\alpha_{1}}^{-}$ & $F_{\alpha_{1}}^{+}$ \\
& \multirow{-2}{*}{\rotatebox{90}{$\Lambda_{4},\Lambda_{6}$}} & $E_{2
\alpha_1+ 2\alpha_{2}+\alpha_{3}}+E_{2\alpha_2+ \alpha_{3}}$ & $E_{2
\alpha_1 + 2\alpha_{2}+\alpha_{3}}- E_{2 \alpha_2 + \alpha_{3}}$ \\ \hline
\end{tabular}
\end{center}
\caption{The stabilizers of the bound states of three long-weight vectors, namely $\ket{\Lambda_{1}}+\ket{\Lambda_{4}}+\ket{\Lambda_{6}}$ and $\ket{\Lambda_{1}}+\ket{\Lambda_{4}}-\ket{\Lambda_{6}}$, giving the algebra $\mathfrak{su}(2)\ltimes\mathbb{R}^{5}$ and $\mathfrak{su}(1,1)\ltimes\mathbb{R}^{5}$
respectively. The 2-conjunction stabilizers connect the two weight vectors whose weights are listed
in the second column, and annihilate the third weight vector. }
\label{stab14614sp6}
\end{table}

As holding in the $D=5$ treatment performed in the previous subsection, the $D=4$ analysis
of the 2-charge states of long weights reveals that the existence of more than one rank-2 orbit
can be traced back to the presence of the conjunction
stabilizers $F^\pm_{\alpha_2 +\alpha_3}$ in \autoref{14stab2states}. As resulting from \autoref{14ofsp6}, these operators transform the states with weight $\Lambda_1$ and the states with weight $\Lambda_4$ to a state with weight $\Sigma_1$, which is short. Hence, as is the five-dimensional case, the existence of two 2-charge orbits is due to the presence of short weights in the
relevant U-representation. By comparing with the literature \cite{Borsten:2011ai}, we observe that the $\ket{\Lambda_1} + \ket{\Lambda_4}$ orbit, whose stabilizer is more compact, is supersymmetric, while the $\ket{\Lambda_1} - \ket{\Lambda_4}$ orbit, whose
stabilizer has a maximally non-compact semisimple part, is
non-supersymmetric. As is the five-dimensional case, the
latter orbit can also be obtained as the one stabilizing a single short
weight, and indeed from \autoref{stab14sp6uno} one gets the stabilizing
algebra $\mathfrak{so}(2,2)\ltimes ( \mathbb{R} \times \mathbb{R}^{4})$ for
the short weight $\Sigma_6$ (and actually for each short weight).

\subsubsection{3-charge orbits}

The 3-charge orbits can be obtained as the stabilizers of the bound
states of the three long-weight vectors
\begin{flalign}
 &\ket{\Lambda_{1}}+\ket{\Lambda_{4}} + \ket{\Lambda_6}\quad  \qquad \ket{\Lambda_{1}}+\ket{\Lambda_{4}} - \ket{\Lambda_6} \ \ . \label{threechargeboundstates}
\end{flalign}
We first consider the $\ket{\Lambda_{1}}+\ket{\Lambda_{4}} +\ket{ \Lambda_6}$ orbit. The
common stabilizers are the subset of the stabilizers in the first column of
\autoref{14stab2states} which stabilize $\ket{\Lambda_6}$ as well, and by observing
\autoref{14ofsp6} one deduces that these are the generators $E_{\alpha_1
+ 2 \alpha_2 + \alpha_3}$, $E_{\alpha_1 + \alpha_2}$ and $E_{\alpha_2}$.
Furthermore, out of the 2-conjunction stabilizers of $\ket{\Lambda_1} +\ket{ \Lambda_4}$, only the
first two in the second column of \autoref{14stab2states}, namely $F^-_{\alpha_2 + \alpha_3}$ and $E_{2\alpha_2 + \alpha_3} - E_{-\alpha_3}$,
stabilize $\ket{\Lambda_6}$. By repeating the same analysis for the generators
that are 2-conjunction stabilizers for $\ket{\Lambda_1} +\ket{\Lambda_6}$ and are
stabilizers of $\ket{\Lambda_4}$, one gets the generators $F^-_{\alpha_1 +
\alpha_2 + \alpha_3}$ and $E_{2 \alpha_1 + 2 \alpha_2 + \alpha_3 } + E_{-
\alpha_3}$.\footnote{The conventions for the signs  are all consistent with the structure constants defined in \autoref{appendixextraspecial}.  Choosing whether the $F$ generators are compact or not, which corresponds to our choices of the relative signs in the bound states, imposes the other stabilizers to be exactly those in \autoref{stab14614sp6}. In particular, defining the weight vector $\ket{\Lambda_2}$ by the relation $E_{-\alpha_3}\ket{\Lambda_1} = \ket{\Lambda_2}$ implies that $E_{2\alpha_1 + 2\alpha_3 +\alpha_3} \ket{\Lambda_6} = - \ket{\Lambda_2}$, giving the stabilizer $E_{2 \alpha_1 + 2 \alpha_2 + \alpha_3 } + E_{-
\alpha_3}$ for the bound state $\ket{\Lambda_{1}}+\ket{\Lambda_{4}} +\ket{ \Lambda_6}$.} Finally, the 2-conjunction stabilizers of $\ket{\Lambda_4} + \ket{\Lambda_6}$
that also stabilize $\ket{\Lambda_1}$ are $F^-_{\alpha_1}$ and $E_{2\alpha_1 + 2
\alpha_2 + \alpha_3} + E_{2\alpha_2 + \alpha_3}$. The latter generator is
not linearly independent, and therefore we do not consider it in the
stabilizing algebra. Actually, there also exists a 3-conjunction stabilizer,
because the states with weights $\Lambda_{1}$, $\Lambda_{4}$ and $\Lambda_6$
can be transformed to the state of weight $\Lambda_2$ by acting with the
generators $E_{-\alpha_3}$, $E_{2\alpha_2 + \alpha_3}$ and $E_{2\alpha_1 + 2
\alpha_2 + \alpha_3}$ respectively, but this stabilizer is not independent
because each pair of generators is a 2-conjunction stabilizer for a pair of
states that also stabilizes the third state.  To summarize, in the
first and third columns of \autoref{stab14614sp6} we list the independent stabilizers of
the $\ket{\Lambda_{1}}+\ket{\Lambda_{4}} + \ket{\Lambda_6}$ bound state.

Repeating the same
analysis for the $\ket{\Lambda_{1}}+\ket{\Lambda_{4}} - \ket{\Lambda_6}$ case, one has to
modify the 2-conjunction stabilizers that involve the weight $\Lambda_6$ in
a way similar to what happens in \autoref{14stab2states} for the bound states of two long-weight vectors, \ie $F^-_{\alpha_1}$, $F^-_{\alpha_1 + \alpha_2 +\alpha_3}$ and $E_{2\alpha_1 +
2\alpha_2 + \alpha_3} + E_{-\alpha_3}$ become $F^+_{\alpha_1}$, $F^+_{\alpha_1 + \alpha_2 +\alpha_3}$ and $E_{2\alpha_1 + 2\alpha_2 +
\alpha_3} - E_{-\alpha_3}$, respectively. The stabilizers of the $\ket{\Lambda_{1}}+\ket{\Lambda_{4}}
- \ket{\Lambda_6}$ bound state are listed in the first and fourth column of \autoref{stab14614sp6}.

\begin{table}[t!]
\renewcommand{\arraystretch}{1.2}
\begin{center}
\begin{tabular}{|c|c|c|}
\hline
Common&$\ket{\Lambda_{1}}+\ket{\Sigma_{6}}$ Conjunction&$\ket{\Lambda_{1}}-\ket{\Sigma_{6}}$ Conjunction\\
\hline \hline
$E_{2\alpha_{2}+\alpha_{3}}$&$E_{\alpha_{2}+\alpha_{3}}-2E_{-2\alpha_{1}-2\alpha_{2}-\alpha_{3}}$ & $E_{\alpha_{2}+\alpha_{3}}+2E_{-2\alpha_{1}-2\alpha_{2}-\alpha_{3}}$\\
$E_{\alpha_{3}}$&$E_{\alpha_{1}+2\alpha_{2}+\alpha_{3}}-E_{-\alpha_{1}-\alpha_{2}-\alpha_{3}}$ & $E_{\alpha_{1}+2\alpha_{2}+\alpha_{3}}+E_{-\alpha_{1}-\alpha_{2}-\alpha_{3}}$\\
$H_{\alpha_{2}}$&$E_{\alpha_{1}+\alpha_{2}+\alpha_{3}}-E_{-\alpha_{1}-2\alpha_{2}-\alpha_{3}}$ & $E_{\alpha_{1}+\alpha_{2}+\alpha_{3}}+E_{-\alpha_{1}-2\alpha_{2}-\alpha_{3}}$\\
$E_{-\alpha_{1}}$&&\\
$E_{-\alpha_{1}-\alpha_{2}}$& &  \\
\hline
\end{tabular}

\caption{The stabilizers of the $\ket{\Lambda_{1}}+\ket{\Sigma_{6}}$  and $\ket{\Lambda_{1}}-\ket{\Sigma_{6}}$ orbits. \label{14sp6lambda1sigma6}}
\end{center}
\end{table}

The semisimple part of the stabilizing algebra of the 3-charge states
is given by the generators $F^\pm$ in \autoref{stab14614sp6}. In the $\ket{\Lambda_{1}}+\ket{\Lambda_{4}}+\ket{\Lambda_{6}}$ case, this yields the compact algebra $\mathfrak{su}(2)$, and thus the full stabilizing algebra is $\mathfrak{su}(2)\ltimes\mathbb{R}^{5}$. On the other hand, in the $\ket{\Lambda_{1}}+\ket{\Lambda_{4}}-\ket{\Lambda_{6}}$
case, there are two non-compact and one compact generators in the
semisimple part of the stabilizing algebra, which therefore leads to the algebra $\mathfrak{sl}(2,\mathbb{R})$, and consequently the full stabilizing algebra is $\mathfrak{sl}(2,\mathbb{R})\ltimes\mathbb{R}^{5}$. As in the 2-charge orbits, the
generators that change the compactness of the stabilizing algebra have the form given by \autoref{F+F-connectinglongwithshort},
connecting each long-weight vector to a short-weight vector (see \autoref{appendixshortweights} for details), and
therefore the existence of more than one orbit is again due to the presence of short
weights in the representation. By comparison with the literature \cite{Borsten:2011ai}, the 1/2-BPS orbit is related to $\ket{\Lambda_{1}}+ \ket{\Lambda_{4}}+\ket{\Lambda_{6}}$, and in this 3-charge case has a compact
stabilizer, while the non-supersymmetric orbit is related to $\ket{\Lambda_{1}}+\ket{\Lambda_{4}}+\ket{\Lambda_{6}}$, whose (semisimple part of the) stabilizer is maximally non-compact.

As in all previous cases, one can obtain the non-supersymmetric orbit as a
bound state involving a short-weight vector. In particular, one can consider in
this case the bound states $\ket{\Lambda_1} \pm \ket{\Sigma_6}$, whose stabilizers are
listed in \autoref{14sp6lambda1sigma6}. One can notice that the stabilizing
algebra is in both cases $\mathfrak{sl}(2,\mathbb{R})\ltimes\mathbb{R}^{5}$, where $\mathfrak{sl}(2,\mathbb{R})$ is generated by $H_{\alpha_2}$, $E_{\alpha_1 + 2\alpha_2
+ \alpha_3} \pm E_{-\alpha_1 - \alpha_2 -\alpha_3}$ and $E_{-\alpha_1 -
2\alpha_2 - \alpha_3} \pm E_{\alpha_1 + \alpha_2 +\alpha_3}$. By looking at
\autoref{splitthetaactionrootvector} one notices that the first generator is mapped into
minus itself under the Cartan involution, while the latter two go each to
minus the other, and hence there are one compact and two non-compact
generators in total, leading to the real form $\mathfrak{sl}(2,\mathbb{R})$. The
remaining generators in \autoref{14sp6lambda1sigma6} transform in the $\mathbf{5}$ of $\mathfrak{sl}(2,\mathbb{R})$.

\subsubsection{4-charge orbits}

\begin{table}[t!]
\renewcommand{\arraystretch}{1.2}
\par
\begin{center}
\scalebox{0.95}{
\begin{tabular}{|c|c|c|c|}
\hline
2-conj. & $\ket{\Lambda_1}\! + \!\ket{\Lambda_4} \!+ \!\ket{\Lambda_6}\! +\! \ket{\Lambda_7}$ & $\ket{\Lambda_1} \!+\!
\ket{\Lambda_4 }\!+\!\ket{\Lambda_6 }\!-\! \ket{\Lambda_7}$ & $\ket{\Lambda_1} \!+\!\ket{ \Lambda_4 }\!-\! \ket{\Lambda_6} \!-\!
\ket{\Lambda_7}$ \\ \hline\hline
$\Lambda_{1},\Lambda_{4}$ & $F^{-}_{\alpha_{2}+\alpha_{3}}$ & $
F^{-}_{\alpha_{2}+\alpha_{3}}$ & $F^{-}_{\alpha_{2}+\alpha_{3}}$ \\
& $F^-_{2\alpha_{2}+\alpha_{3}}+F^-_{\alpha_{3}}$ & $F^+_{2\alpha_{2}+
\alpha_{3}}-F^+_{\alpha_{3}}$ & $F^-_{2\alpha_{2}+\alpha_{3}}+F^-_{
\alpha_{3}}$ \\ \hline
$\Lambda_{1},\Lambda_{6}$ & $F^{-}_{\alpha_{1}+\alpha_{2}+\alpha_{3}}$ & $
F^{-}_{\alpha_{1}+\alpha_{2}+\alpha_{3}}$ & $F^{+}_{\alpha_{1}+\alpha_{2}+
\alpha_{3}}$ \\
& $F^-_{2\alpha_{1}+2\alpha_{2}+\alpha_{3}} - F^-_{\alpha_{3}}$ & $
F^+_{2\alpha_{1}+2\alpha_{2}+\alpha_{3}} + F^+_{\alpha_{3}}$ & $
F^-_{2\alpha_{1}+2\alpha_{2}+\alpha_{3}} + F^-_{\alpha_{3}}$ \\ \hline
$\Lambda_{1},\Lambda_{7}$ & $F^{-}_{\alpha_{1}+2\alpha_{2}+\alpha_{3}}$ & $
F^{+}_{\alpha_{1}+2\alpha_{2}+\alpha_{3}}$ & $F^{+}_{\alpha_{1}+2\alpha_{2}+
\alpha_{3}}$ \\
& $F^-_{2\alpha_{1}+ 2\alpha_{2}+\alpha_{3}}+ F^-_{2\alpha_{2}+\alpha_{3}}$
& $F^+_{2\alpha_{1}+ 2\alpha_{2}+\alpha_{3}}+ F^+_{2\alpha_{2}+\alpha_{3}}$
& $F^-_{2\alpha_{1}+ 2\alpha_{2}+\alpha_{3}}- F^-_{2\alpha_{2}+\alpha_{3}}$
\\ \hline
$\Lambda_{4},\Lambda_{6}$ & $F^{-}_{\alpha_{1}}$ & $F^{-}_{\alpha_{1}}$ & $
F^{+}_{\alpha_{1}}$ \\
& $F^-_{2\alpha_{1}+2\alpha_{2}+\alpha_{3}}+F^-_{2\alpha_{2}+\alpha_{3}}$ & $
F^+_{2\alpha_{1}+2\alpha_{2}+\alpha_{3}}+F^+_{2\alpha_{2}+\alpha_{3}}$ & $
F^-_{2\alpha_{1}+2\alpha_{2}+\alpha_{3}}-F^-_{2\alpha_{2}+\alpha_{3}}$ \\
\hline
$\Lambda_{4},\Lambda_{7}$ & $F^{-}_{\alpha_{1}+\alpha_{2}}$ & $
F^{+}_{\alpha_{1}+\alpha_{2}}$ & $F^{+}_{\alpha_{1}+\alpha_{2}}$ \\
& $F^-_{2\alpha_{1}+2\alpha_{2}+\alpha_{3}}-F^-_{\alpha_{3}}$ & $
F^+_{2\alpha_{1}+2\alpha_{2}+\alpha_{3}}+F^+_{\alpha_{3}}$ & $
F^-_{2\alpha_{1}+2\alpha_{2}+\alpha_{3}}+F^-_{\alpha_{3}}$ \\ \hline
$\Lambda_{6},\Lambda_{7}$ & $F^{-}_{\alpha_{2}}$ & $F^{+}_{\alpha_{2}}$ & $
F^{-}_{\alpha_{2}}$ \\
& $F^-_{2\alpha_{2}+\alpha_{3}}+F^-_{\alpha_{3}}$ & $F^+_{2\alpha_{2}+
\alpha_{3}}-F^+_{\alpha_{3}}$ & $F^-_{2\alpha_{2}+\alpha_{3}}+F^-_{
\alpha_{3}}$ \\ \hline
\end{tabular}
}
\end{center}
\caption{The stabilizers of the 4-charge orbits in the $\mathbf{14^{\prime }}$ of $\mathfrak{sp}(6,\mathbb{R})$. In the first column we
list the pair of states for which the operator in the first line of each row
is a 2-conjunction stabilizer. In any column, there are only two independent
generators among those in the second line of each row; in total, the number of independent generators for each 4-charge orbit is 8. }
\label{fourchargeorbitssp14}
\end{table}

Finally, we consider 4-charge (\textit{i.e. }rank-4 \cite{Ferrar,Krutelevich}) ``large'' orbits.
Following our method, we want to obtain these orbits as bound states of the long-weight vectors $\ket{\Lambda_1}$, $\ket{\Lambda_4}$, $\ket{\Lambda_6}$ and $\ket{\Lambda_7}$. Up to an (irrelevant)
overall sign, there are three possibilities, namely
\begin{equation}
 \ket{\Lambda_{1}}\!+\!\ket{\Lambda_{4}} \!+\! \ket{\Lambda_6} \!+\!\ket{\Lambda_7} \quad \ket{\Lambda_{1}}\!+\!\ket{\Lambda_{4}} \!+\! \ket{\Lambda_6} \!-\!\ket{\Lambda_7} \quad \ket{\Lambda_{1}}\!+\!\ket{\Lambda_{4}} \!-\!\ket{ \Lambda_6} \!-\!\ket{\Lambda_7} \ . \label{fourchargeboundstates}
\end{equation}
From \autoref{stab14sp6uno} it can be noticed that there are no common
stabilizers. We consider the 2-conjunction stabilizers for each pair of
weight vectors. If one considers for example the pair $\ket{\Lambda_1} +\ket{\Lambda_4}$, one
can notice that among the 2-conjunction stabilizers listed in \autoref{14stab2states}, only $F^-_{\alpha_2 +\alpha_3}$ vanishes on each of the
other two weight vectors. Moreover, the operator $E_{2\alpha_2 +\alpha_3}- E_{-\alpha_3}$ only
vanishes on $\ket{\Lambda_6}$, while it maps $\ket{\Lambda_7}$ to
the bound state $\ket{\Lambda_5} + \ket{\Lambda_8}$.\footnote{As for the 3-charge bound states, the conventions for the signs are all consistent with the structure constants defined in \autoref{appendixextraspecial}.} Similarly, the operator $E_{\alpha_3} -
E_{-2\alpha_2 -\alpha_3}$ vanishes on $\ket{\Lambda_7}$ and maps $\ket{\Lambda_6}$ to $-\ket{\Lambda_5} - \ket{\Lambda_8}$. In the $\ket{\Lambda_{1}}+\ket{\Lambda_{4}} + \ket{\Lambda_6} +
\ket{\Lambda_7} $ orbit, one then obtains a stabilizer by summing the two
operators above, yielding
\begin{flalign}
 & F^-_{2\alpha_2 + \alpha_3} + F^-_{\alpha_3} \quad .
\end{flalign}
Analogously, in the $\ket{\Lambda_{1}}+\ket{\Lambda_{4}} +\ket{ \Lambda_6 }-\ket{ \Lambda_7} $ orbit
it is the difference of the two operators that gives a stabilizer, which is
\begin{flalign}
 & F^+_{2\alpha_2 + \alpha_3} - F^+_{\alpha_3} \quad .
\end{flalign}
Finally, in the $\ket{\Lambda_{1}}+\ket{\Lambda_{4}} -\ket{ \Lambda_6} - \ket{\Lambda_7} $ it is
again the sum that gives a stabilizer. This analysis can be applied to any
pair of weight vectors, obtaining for each pair a 2-conjunction stabilizer that is
also a stabilizer for each of the other two weight vectors, and a 2-conjunction
stabilizer that is also a 2-conjunction stabilizer for the other two weight vectors.
The complete outcome of this analysis is summarized in \autoref{fourchargeorbitssp14}.

\begin{table}[t!]
\renewcommand{\arraystretch}{1.2}
\begin{center}
\begin{tabular}{|c|c|c|}
\hline
\multicolumn{3}{|c|}{$\ket{\Lambda_{1}}+\ket{\Lambda_{4}}+\ket{\Sigma_{6}}$  Stabilizers}\\
\hline
\hline
Common&2-Conjunction&3-Conjunction\\
\hline
$H_{\alpha_{2}}$&$E_{2\alpha_{2}+\alpha_{3}}-E_{-\alpha_{3}}$&$F_{\alpha_{2}+\alpha_{3}}^{+}-2F_{2\alpha_{1}+2\alpha_{2}+\alpha_{3}}^{+}$\\
&$E_{\alpha_{3}}-E_{-2\alpha_{2}-\alpha_{3}}$&\\
&$E_{\alpha_{1}+2\alpha_{2}+\alpha_{3}}-E_{-\alpha_{1}-\alpha_{2}-\alpha_{3}}$&\\
&$E_{\alpha_{1}+\alpha_{2}+\alpha_{3}}-E_{-\alpha_{1}-2\alpha_{2}-\alpha_{3}}$&\\
&$E_{\alpha_{1}+\alpha_{2}}-E_{-\alpha_{1}}$&\\
&$E_{\alpha_{1}}-E_{-\alpha_{1}-\alpha_{2}}$&\\
\hline
\end{tabular}

\caption{Stabilizers of $\ket{\Lambda_{1}}+\ket{\Lambda_{4}}+
\ket{\Sigma_{6}}$ that generate the algebra $\mathfrak{sl}(3,\mathbb{R})$, matching the stabilizer of the non-BPS space-like 4-charge orbit whose representative is related to the bound state $\ket{\Lambda_{1}}+\ket{\Lambda_{4}} + \ket{\Lambda_6 }- \ket{\Lambda_7 }$.\label{14istabs}}
\end{center}
\end{table}

\begin{table}[b!]
\renewcommand{\arraystretch}{1.2}
\begin{center}
\begin{tabular}{|c|c|c|}
\hline
\multicolumn{3}{|c|}{$\ket{\Lambda_{1}}-\ket{\Lambda_{4}}+\ket{\Sigma_{6}}$  Stabilizers}\\
\hline
\hline
Common&2-Conjunction&3-Conjunction\\
\hline
$H_{\alpha_{2}}$&$E_{2\alpha_{2}+\alpha_{3}}+E_{-\alpha_{3}}$&$F_{\alpha_{2}+\alpha_{3}}^{-}+2F_{2\alpha_{1}+2\alpha_{2}+\alpha_{3}}^{-}$\\
&$E_{\alpha_{3}}+E_{-2\alpha_{2}-\alpha_{3}}$&\\
&$E_{\alpha_{1}+2\alpha_{2}+\alpha_{3}}-E_{-\alpha_{1}-\alpha_{2}-\alpha_{3}}$&\\
&$E_{\alpha_{1}+\alpha_{2}+\alpha_{3}}-E_{-\alpha_{1}-2\alpha_{2}-\alpha_{3}}$&\\
&$E_{\alpha_{1}+\alpha_{2}}+E_{-\alpha_{1}}$&\\
&$E_{\alpha_{1}}+E_{-\alpha_{1}-\alpha_{2}}$&\\
\hline
\end{tabular}

\caption{The $\mathfrak{su}(1,2)$ generators of the
stabilizing algebra of $\ket{\Lambda_{1}}-\ket{\Lambda_{4}}+\ket{\Sigma_{6}}$, matching the stabilizer of the non-BPS time-like 4-charge orbit whose representative is related to the bound state $\ket{\Lambda_{1}}+\ket{\Lambda_{4} }- \ket{\Lambda_6 }-\ket{ \Lambda_7} $.\label{1l-4l+6s}}
\end{center}
\end{table}

\begin{table}[t!]
\renewcommand{\arraystretch}{1.2}
\par
\begin{center}
\begin{tabular}{|c|}
\hline
{$\ket{\Lambda_{1}}+\ket{\Lambda_{8}}$ Stabilizers (Common)} \\ \hline\hline
$E_{\alpha_{1}+\alpha_{2}}$ \\
$E_{\alpha_{1}}\qquad E_{\alpha_{2}}$ \\
$H_{\alpha_{1}}\quad H_{\alpha_{2}}$ \\
$E_{-\alpha_{1}}\qquad E_{-\alpha_{2}}$ \\
$E_{-\alpha_{1}-\alpha_{2}}$ \\ \hline
\end{tabular}
\end{center}
\caption{ Stabilizers of $\ket{\Lambda_{1}}+\ket{\Lambda_{8}}$, generating  the algebra $\mathfrak{sl}(3,\mathbb{R})$.}
\label{dyonic14}
\end{table}

\begin{table}[b!]
\renewcommand{\arraystretch}{1.5}
\begin{center}
\begin{tabular}{|c|c|c|}
\hline
\multicolumn{2}{|c|}{{State}}&{Stabilizer}\\ \hline\hline
\multirow{-1}{*}{\rotatebox{90}{\tiny 1-state}}&$\ket{\Lambda_{1}}$&$\mathfrak{sl}(3,\mathbb{R})\ltimes \mathbb{R}^{6}$\\ \hline
&$\ket{\Lambda_{1}}+\ket{\Lambda_{4}}$& $\mathfrak{so}(1,3)\ltimes (\mathbb{R}\times\mathbb{R}^{4})$\\
&$\ket{\Lambda_{1}}-\ket{\Lambda_{4}}$&$\mathfrak{so}(2,2)\ltimes (\mathbb{R}\times\mathbb{R}^{4})$\\
\multirow{-3}{*}{\rotatebox{90}{\tiny 2-state}}&$\ket{\Sigma_{6}}$&$\mathfrak{so}(2,2)\ltimes (\mathbb{R}\times\mathbb{R}^{4})$\\ \hline
&$\ket{\Lambda_{1}}+\ket{\Lambda_{4}}+\ket{\Lambda_{6}}$&$\mathfrak{su}(2)\ltimes \mathbb{R}^{5}$\\
&$\ket{\Lambda_{1}}+\ket{\Lambda_{4}}-\ket{\Lambda_{6}}$ &$\mathfrak{sl}(2,\mathbb{R})\ltimes \mathbb{R}^{5}$\\
&$\ket{\Lambda_{1}}+\ket{\Sigma_{6}}$&$\mathfrak{sl}(2,\mathbb{R})\ltimes \mathbb{R}^{5}$\\
\multirow{-4}{*}{\rotatebox{90}{\tiny 3-state}}&$\Lambda_{1}-\Sigma_{6}$&$\mathfrak{sl}(2,\mathbb{R})\ltimes \mathbb{R}^{5}$\\ \hline
&$\ket{\Lambda_{1}}+\ket{\Lambda_{4}}+\ket{\Lambda_{6}}+\ket{\Lambda_{7}}$&$\mathfrak{su}(3)$\\
&$\ket{\Lambda_{1}}+\ket{\Lambda_{4}}+\ket{\Lambda_{6}}-\ket{\Lambda_{7}}$&$\mathfrak{sl}(3,\mathbb{R})$\\
&$\ket{\Lambda_{1}}+\ket{\Lambda_{4}}-\ket{\Lambda_{6}}-\ket{\Lambda_{7}}$&$\mathfrak{su}(1,2)$\\
&$\ket{\Lambda_{1}}+\ket{\Lambda_{4}}+\ket{\Sigma_{6}}$&$\mathfrak{sl}(3,\mathbb{R})$\\
&$\ket{\Lambda_{1}}-\ket{\Lambda_{4}}+\ket{\Sigma_{6}}$&$\mathfrak{su}(1,2)$\\
\multirow{-6}{*}{\rotatebox{90}{\tiny 4-state}}&$\ket{\Lambda_{1}}+\ket{\Lambda_{8}}$&$\mathfrak{sl}(3,\mathbb{R})$\\

\hline

\hline
\end{tabular}

\caption{The stabilizing algebras of the various orbits of the ${\bf 14^\prime}$ of $Sp(6,\mathbb{R})$. \label{summary14sp6}}
\end{center}
\end{table}

By looking at the table, one notices that the number of independent
generators for each orbit is 8. In the $\ket{\Lambda_{1}}+\ket{\Lambda_{4}} +\ket{ \Lambda_6}
+\ket{ \Lambda_7} $ orbit, all the generators are compact, while the $\ket{\Lambda_{1}}+\ket{\Lambda_{4}} +\ket{ \Lambda_6} -\ket{ \Lambda_7 }$ orbit has three compact
and five non-compact generators and the $\ket{\Lambda_{1}}+\ket{\Lambda_{4}} - \ket{\Lambda_6}
-\ket{ \Lambda_7} $ orbit has four compact and four non-compact generators. We
recognize these as the three different real forms $\mathfrak{su}(3)$, $\mathfrak{sl}(3,\mathbb{R})$ and $\mathfrak{su}(2,1)$ of the complex algebra
$\mathfrak{sl}(3,\mathbb{C})$, and it can be explicitly checked that these
are indeed the algebras generated by the operators in \autoref{fourchargeorbitssp14}. By comparing with the literature \cite{Bellucci:2006xz}, one recognizes $\ket{\Lambda_{1}}+\ket{\Lambda_{4}} +\ket{ \Lambda_6} + \ket{\Lambda_7} $ to be related to the ``large'' time-like 1/2-BPS
orbit, $\ket{\Lambda_{1}}+\ket{\Lambda_{4}} + \ket{\Lambda_6 }- \ket{\Lambda_7} $ to be related to the ``large'' dyonic
non-supersymmetric (non-BPS) space-like orbit, and $\ket{\Lambda_{1}}+\ket{\Lambda_{4}} -\ket{ \Lambda_6} -\ket{ \Lambda_7 }$ to be related to
the ``large'' non-supersymmetric (non-BPS) time-like orbit. Note that the stabilizer of the non-BPS space-like 4-charge orbit generally coincides with the U-duality Lie algebra of the corresponding theory in $D=5$.

We recall that the generators in the
top line of each row in \autoref{fourchargeorbitssp14} are those connecting the long-weight vectors to a short-weight vector, while the generators in the bottom line of the row connect long-weight vectors with long-weight vectors. The $\ket{\Lambda_{1}}+\ket{\Lambda_{4}} + \ket{\Lambda_6} -
\ket{\Lambda_7} $ orbit differs from the 1/2-BPS orbit because generators of both
types swap compactness. On the other hand, in the $\ket{\Lambda_{1}}+\ket{\Lambda_{4}} -
\ket{\Lambda_6 }-\ket{ \Lambda_7} $ orbit only (four of) the generators on the top line
of each row become non-compact, while the generators in the bottom line
remain compact, and hence in this case the generators that swap compactness compared to the supersymmetric case are only conjunction stabilizers to
short-weight vectors. This explains why this splitting of the 4-charge
configuration in three orbits does not occur in maximal $D=4$ supergravity, in which
only the 1/8-BPS and the dyonic orbits are present. Indeed, in the maximally supersymmetric
case the representation $\mathbf{56}$ of the U-duality symmetry $\mathfrak{e}_{7(7)}$ has only long weights.

Precisely as in the 2-charge and 3-charge cases, one can show that the
non-BPS orbits can be obtained as bound states involving short-weight vectors. Without dealing with the details, we list the stabilizers of the $\ket{\Lambda_{1}}+\ket{\Lambda_{4}}+\ket{\Sigma_{6}}$ and $\ket{\Lambda_{1}}-\ket{\Lambda_{4}}+\ket{\Sigma_{6}}$
bound states in \autoref{14istabs} and \autoref{1l-4l+6s}, from which it can
be deduced that the stabilizing algebras are $\mathfrak{sl}(3,\mathbb{R})$
and $\mathfrak{su}(1,2)$ respectively, in agreement with the general rule
that a short-weight vector corresponds to the difference of two long-weight vectors.
Finally, as a peculiar property of the $D=4$ case, the dyonic orbit can also be obtained as the bound state $\ket{\Lambda_1}
+ \ket{\Lambda_8}$, namely of the highest and lowest-weight vectors of the $\mathbf{14^\prime}$ of the U-duality symmetry $\mathfrak{sp}(6,\mathbb{R})$. In this case there are no conjunction stabilizers, and the
common stabilizers, that generate $\mathfrak{sl}(3,\mathbb{R})$, are listed
in \autoref{dyonic14}.

To conclude, we summarize in \autoref{summary14sp6} the stabilizing algebras
for the various bound states of the $\mathbf{14^\prime}$, as they have been
realized and derived above. The results match the ones reported in Table of VI of \cite{Borsten:2011ai} (\textit{cfr.} also Refs. therein).

In the next section we will repeat the same
analysis for the $\mathcal{N} =2$ magic supergravity theories in five and
four dimensions that are based on the division algebra $\mathbb{C}$ (the
same analysis can easily be applied to the theories based on $\mathbb{H}$
and $\mathbb{O}$; all these theories share the property that the real form of their U-duality symmetry algebra is non-maximally non-compact, \ie non-split).

\section{\label{secmagicalCHO} Magic $\mathcal{N}=2$ supergravities based on $J_{3}^{\mathbb{C}}$, $J_{3}^{\mathbb{H}}$ and $J_{3}^{\mathbb{O}}$}

In this section we consider the orbits of the magic Maxwell-Einstein supergravity theories
based on rank-3 simple Jordan algebras constructed over the division algebras $\mathbb{C}$, $\mathbb{H}$ and $\mathbb{O}$ \cite{Gunaydin:1983rk,Gunaydin:1983bi},
that were originally derived in \cite{Ferrara:1997uz,Ferrara:2006xx,Bellucci:2006xz,Borsten:2011ai}. In the case of 1-charge BHs, the
orbits are those of the highest-weight vector of the representation, and in \cite{Bergshoeff:2014lxa} it
was shown that the BH charges correspond to the real-weight vectors, where
the reality properties of roots and weights are deduced from the
Tits-Satake diagram associated to the global U-duality symmetry of the theory itself. Here, we
will show that all extremal BH orbits of these theories are those of bound states of such real-weight vectors. We will consider in detail explicitly
only the theory based on $J_{3}^{\mathbb{C}}$, corresponding to global
symmetries $SU(3,3)$ and $SL(3,\mathbb{C})$ in four and five
dimensions respectively, but the result naturally applies also to the cases based on $J_{3}^{\mathbb{H}}$ and $J_{3}^{\mathbb{O}}$ (whose global symmetries in four and five dimensions are $SO^{\ast }(12)$ and $SU^{\ast }(6)$, and $E_{7(-25)}$ and $E_{6(-26)}$, respectively).

In order to proceed, we first briefly review how the reality properties of
the weights of a representation result from the Tits-Satake diagram that
characterizes a given real form of a complex Lie algebra $\mathfrak{g}_{\mathbb{C}}$ (for a detailed analysis, see {\it e.g.} \cite{grouptheory,grouptheory-1,grouptheory-2}). As already mentioned in \autoref{sectionreviewmaximalcase}, the Cartan involution $\theta$ characterizes the real form $\mathfrak{g}$ in such a way that the compact generators have eigenvalue $+1$ and the non-compact generators have eigenvalue $-1$ under $\theta$.
In the case of the maximally non-compact (\ie split) real form, one can
define the generators $E_\alpha- E_{-\alpha}$ to be compact and the Cartan
generators $H_\alpha$ and the generators $E_\alpha + E_{-\alpha}$ to be
non-compact, so that the Cartan involution acts as in \autoref{splitthetaactionrootvector},
which was used in the previous two sections. In general, classifying the real
forms of a given complex Lie algebra corresponds to classifying all
possible Cartan involutions. The action of $\theta$ on the Cartan generators
induces a dual action on the roots as
\begin{equation}
\theta (\alpha(H))=\alpha(\theta (H)) \quad ,  \label{thetaalphaHalphathetaH}
\end{equation}
and the basic idea that underlies the Tits-Satake diagrams is the fact that
one can insert in the Dynkin diagram the information of how the Cartan
involution acts on the roots themselves.

One can always define a basis of generators of $\mathfrak{g}$ that are
eigenvectors of the Cartan involution $\theta$. We call $\mathfrak{k}$ the
set of compact generators, that are those with eigenvalue $+1$, and $\mathfrak{p}$ the set of non-compact generators, which are those with
eigenvalue $-1$. In the adjoint representation one has
\begin{equation}
\mathrm{ad } (\theta X ) = - (\mathrm{ad} X )^\dagger \quad ,
\end{equation}
so that for a compact generator, that is $X \in \mathfrak{k}$,  $\mathrm{ad}X$ is anti-Hermitian and thus has imaginary eigenvalues, while
for a non-compact generator, that is $X \in \mathfrak{p}$,  $\mathrm{ad}X$ is Hermitian and thus has real eigenvalues. In particular, the
non-zero eigenvalues of the Cartan matrices in the adjoint $\mathrm{ad}H$
are the roots $\alpha (H)$, implying that one can classify the  roots in the
following way:

\begin{itemize}
\item a  root is a \textbf{\textit{real root}} if it takes real values, that
is if it vanishes for $H \in \mathfrak{k}$;

\item a  root is an \textbf{\textit{imaginary root}} if it takes imaginary
values, that is if it vanishes for $H \in \mathfrak{p}$;

\item a root is a \textbf{\textit{complex root}} if it takes complex values
and hence if it does not vanish for $H\in \mathfrak{k}$ or $H\in \mathfrak{p}$.
\end{itemize}

From \autoref{thetaalphaHalphathetaH} it then follows that for a real  root
one has $\theta \alpha = -\alpha$ and for an imaginary  root one has $\theta
\alpha = \alpha$. From the relation $\theta H_\alpha = H_{\theta \alpha}$
it also follows that
\begin{equation}
 ( \theta \alpha_{k},\alpha_{i} )= (\alpha_{k},\theta
\alpha_{i} )  \label{thetaalphaalpha}
\end{equation}
for any pair of  roots, where in general we denote with $( \alpha_{k},\alpha_{i} )$ the scalar product between the roots that is induced by the Killing metric.

It is known that choosing a Cartan subalgebra that is maximally non-compact  (\textit{i.e.} choosing a basis such that the largest possible number of Cartan generators are non-compact),
there are no non-compact generators associated to the imaginary roots. This
means that if $\theta \alpha =\alpha$, then this implies that $\theta
E_{\alpha} = E_{\alpha}$. Assuming that we have made this choice, we define the Tits-Satake
diagram of $\mathfrak{g}$ using the following procedure.  We split the simple roots in
those that are fixed under $\theta$, that we denote with $\alpha^{\rm Im}_n$ (where the suffix ${\rm Im}$ denotes the fact that such roots are imaginary), and the
rest, that we denote with $\alpha_i$. The action of $\theta$ on the simple
roots $\alpha_i$ is then:
\begin{equation}
\theta\alpha_{i}=-\alpha_{\pi(i)}+\sum_{n}a_{in}\alpha^{\rm Im}_{n}\,,
\label{complexrootTitsSatake1}
\end{equation}
where $\pi$ is an involutive ($\pi^{2}=1$) permutation of the indices.
The coefficients $a_{in}$ are determined by imposing
\begin{equation}
( \alpha_{i}+\alpha_{\pi(i)},\alpha^{\rm Im}_{m})=\sum_{n}
a_{in}( \alpha^{\rm Im}_{n},\alpha^{\rm Im}_{m} ) \ ,  \label{thiseqdeterminesain}
\end{equation}
which follows from \autoref{thetaalphaalpha} and the fact that the simple
roots $\alpha^{\rm Im}_n$ are invariant under $\theta$. The Tits-Satake diagram is
then drawn from the corresponding Dynkin diagram with the following
additional rules:\newline
\begin{wrapfigure}[7]{r,h!}{0.4\textwidth}
 \scalebox{0.5}{
\begin{pspicture}(0,-1.89)(10.41,2.85)
\definecolor{color79b}{RGB}{0,0,0}
\psline[linewidth=0.02cm](3.8,1.65)(6.2,1.65)
\pscircle[linewidth=0.02,dimen=outer,fillstyle=solid](5.0,1.65){0.4}
\psline[linewidth=0.02cm](3.8,4)(6.2,4)
\pscircle[linewidth=0.02,dimen=outer,fillstyle=solid,fillcolor=color79b](5.0,4){0.4}
\psline[linewidth=0.02cm,linestyle=dashed,dash=0.16cm 0.16cm](6.0,1.65)(7.0,1.65)
\psline[linewidth=0.02cm,linestyle=dashed,dash=0.16cm 0.16cm](3.0,1.65)(4.0,1.65)
\psline[linewidth=0.02cm,linestyle=dashed,dash=0.16cm 0.16cm](3.0,4)(4.0,4)
\psline[linewidth=0.02cm,linestyle=dashed,dash=0.16cm 0.16cm](6.0,4)(7.0,4)
\psline[linewidth=0.02cm](0.8,-2.55)(3.2,-2.55)
\pscircle[linewidth=0.02,dimen=outer,fillstyle=solid](2.0,-2.55){0.4}
\psline[linewidth=0.02cm,linestyle=dashed,dash=0.16cm 0.16cm](3.0,-2.55)(4.0,-2.55)
\psline[linewidth=0.02cm,linestyle=dashed,dash=0.16cm 0.16cm](0.0,-2.55)(1.0,-2.55)
\psline[linewidth=0.02cm](7.2,-2.55)(9.6,-2.55)
\pscircle[linewidth=0.02,dimen=outer,fillstyle=solid](8.4,-2.55){0.4}
\psline[linewidth=0.02cm,linestyle=dashed,dash=0.16cm 0.16cm](9.4,-2.55)(10.4,-2.55)
\psline[linewidth=0.02cm,linestyle=dashed,dash=0.16cm 0.16cm](6.4,-2.55)(7.4,-2.55)
\psbezier[linewidth=0.06,arrowsize=0.05291667cm 2.0,arrowlength=1.4,arrowinset=0.4]{<->}(2.1036007,-1.9986596)(2.1036007,-0.94999975)(4.113303,-0.45268014)(5.1178346,-0.42567003)(6.1223655,-0.39865994)(8.503601,-0.62723136)(8.533966,-1.99866)
\rput(5.0848436,0.795){\Large \textbf{\textit{complex}}  1-cycle ($\rm{mod} \ \alpha^{\rm Im} \ {\rm roots}$)}
\rput(4.924844,3.145){\Large \textbf{\textit{imaginary}} (compact generator)}
\rput(5.1215625,-3.605){\Large \textbf{\textit{complex}} 2-cycle ($\rm{mod} \ \alpha^{\rm Im} \ {\rm roots}$)}
\end{pspicture}
}
\end{wrapfigure}
\vspace{-1cm}

\begin{enumerate}
\item to each simple root $\alpha^{\rm Im}$ (imaginary simple root) one associates a black
painted node;

\item to each simple root $\alpha_i$ such that $\pi (i) =i$ one associates
an unpainted node;

\item for each two complex simple roots $\alpha_i$ and $\alpha_j$ such that $\pi (i) = j$ one draws an arrow joining the corresponding unpainted nodes.
\end{enumerate}

The behavior of all other roots under $\theta$ clearly follows from the
behavior of the simple roots. This means that from the Tits-Satake diagram
one knows how the Cartan involution acts on all roots, and consequently
how it acts on all the weights of any representation. In the case in which $\pi (i) =i$ and the node associated to the simple root $\alpha_i$ is not
connected to any painted node in the Tits-Satake diagram, then clearly from
\autoref{thiseqdeterminesain} it follows that $a_{in} =0$ and thus $\theta
\alpha_i =-\alpha_i$, which means that the root is real. In particular, in
the case of the split real form one has $\theta \alpha =- \alpha$ on all
roots, which implies that they are all real.

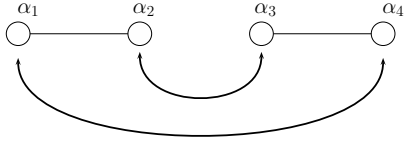
\begin{figure}[h!]
\centering
\scalebox{0.4} 
{\
\begin{pspicture}(0,-2.6959374)(16.309063,2.7059374)
\definecolor{color114b}{rgb}{0.996078431372549,0.996078431372549,0.996078431372549}
\psline[linewidth=0.02cm](9.541875,1.5340625)(13.741875,1.5340625)
\psline[linewidth=0.02cm](1.741875,1.5340625)(5.741875,1.5340625)
\pscircle[linewidth=0.02,dimen=outer,fillstyle=solid](1.741875,1.5340625){0.4}
\pscircle[linewidth=0.02,dimen=outer,fillstyle=solid,fillcolor=color114b](5.741875,1.5340625){0.4}
\pscircle[linewidth=0.02,dimen=outer,fillstyle=solid](9.741875,1.5340625){0.4}
\pscircle[linewidth=0.02,dimen=outer,fillstyle=solid](13.741875,1.5340625){0.4}
\usefont{T1}{ppl}{m}{n}
\rput(2.0764062,2.3440626){\huge $\alpha_{1}$}
\psbezier[linewidth=0.06,arrowsize=0.05291667cm 2.0,arrowlength=1.4,arrowinset=0.4]{<->}(1.741875,0.7340625)(1.741875,-2.6659374)(13.741875,-2.6659374)(13.741875,0.7340625)
\psbezier[linewidth=0.06,arrowsize=0.05291667cm 2.0,arrowlength=1.4,arrowinset=0.4]{<->}(5.741875,0.9340625)(5.741875,-1.0659375)(9.741875,-1.0659375)(9.741875,0.9340625)
\usefont{T1}{ppl}{m}{n}
\rput(5.876406,2.3440626){\huge $\alpha_{2}$}
\usefont{T1}{ppl}{m}{n}
\rput(9.876407,2.3440626){\huge $\alpha_{3}$}
\usefont{T1}{ppl}{m}{n}
\rput(14.0764065,2.3440626){\huge $\alpha_{4}$}
\end{pspicture}
}
\caption{The Tits-Satake diagram of the algebra $\mathfrak{sl}(3,\mathbb{C})$.}
\label{sl3cdynkindiagram}
\end{figure}

One can apply this construction to the U-duality symmetry Lie algebras of the $\mathcal{N}=2$
theories based on $J_{3}^{\mathbb{C}}$, $J_{3}^{\mathbb{H}}$ and $J_{3}^{\mathbb{O}}$. In
particular, the U-duality algebra for the $\mathcal{N}=2$ theory based on $J_{3}^{\mathbb{C}}$ in five dimensions is $\mathfrak{sl}(3,\mathbb{C})$, whose
Tits-Satake diagram is drawn in \autoref{sl3cdynkindiagram}.
From the diagram, one reads that the action of the Cartan involution on the
simple roots is
\begin{equation}
\theta \alpha_1 = - \alpha_4 \qquad \theta \alpha_2 =-\alpha_3 \quad ,
\label{cartaninvsimplerootssl3c}
\end{equation}
from which one derives the action
\begin{equation}
\theta E_{\alpha_1} = - E_{-\alpha_4} \quad \theta E_{\alpha_2}
=-E_{-\alpha_3} \quad \theta H_{\alpha_1} = - H_{\alpha_4} \quad \theta
H_{\alpha_2} = - H_{\alpha_3} \quad ,  \label{cartaninvgeneratorssl3c}
\end{equation}
on the corresponding simple-root generators. The BH (electric) charges of the $D=5$ supergravity theory
transform in the $(\mathbf{\bar{3},3})$ representation, whose highest weight
is
\begin{equation}
\Lambda_1 =\tfrac{1}{3}\alpha_1 +\tfrac{2}{3}\alpha_2 + \tfrac{2}{3}
\alpha_3 + \tfrac{1}{3} \alpha_4 \quad ,  \label{highestweight3bar3}
\end{equation}
which is real because $\theta \Lambda_1 = -\Lambda_1$ as it can be easily
seen using \autoref{cartaninvsimplerootssl3c}. The representation contains
in total three real and six complex weights, and we will show that the
BH U-duality orbits can be derived as the orbits of bound states
of the real weights. As we will see, the conjunction stabilizers take two
real weights to one pair of complex weights connected by $\theta$.
If one relates real weights to long weights and pairs of complex weights to
short weights, this mimics precisely what happens to the bound states of
long weights in the $\mathbf{6}$ of $\mathfrak{sl}(3,\mathbb{R})$, as we discussed in
\autoref{magicn2sugraR}.

The analogy between the real and complex weights of the $(\mathbf{\bar{3},3})$ of $\mathfrak{sl}(3,\mathbb{C})$ and the long and short weights of the
$\mathbf{6}$ of $\mathfrak{sl}(3,\mathbb{R})$
is actually not surprising. Indeed, given a real form $\mathfrak{g}$, one can in
general define the \textit{restricted-root} subalgebra as the maximally
non-compact subalgebra that has as simple roots the \textit{restricted} roots
\begin{equation}
\alpha_R = \tfrac{1}{2} (\alpha - \theta \alpha ) \quad ,
\end{equation}
where the $\alpha$'s are simple roots of $\mathfrak{g}$.
From \autoref{cartaninvsimplerootssl3c}, it can be easily deduced that the restricted-root subalgebra of $\mathfrak{sl}(3,\mathbb{C})$ is
$\mathfrak{sl}(3,\mathbb{R})$, with simple roots
\begin{equation}
(\alpha_R )_1 = \tfrac{1}{2} ( \alpha_2 + \alpha_3 ) \qquad (\alpha_R )_2 =
\tfrac{1}{2} ( \alpha_1 + \alpha_4 ) \quad .
\end{equation}
Moreover, $\Lambda_1$ in \autoref{highestweight3bar3} can be recast as the weight $\tfrac{4}{3} (\alpha_R )_1 + \tfrac{2}{3} (\alpha_R )_2$, which is the
highest weight of the $\mathbf{6}$ of $\mathfrak{sl}(3,\mathbb{R})$ as can
be seen from \autoref{highestweightofthe6ofsl3}. More generally, the
three real weights can be recast as the three long weights, while the complex weights
are projected onto the short weights of the $\mathbf{6}$.

As observed in \cite{Bergshoeff:2014lxa}, this result is completely general. In each space-time dimension, the
restricted-root subalgebra\,\footnote{The projection on the restricted-root subalgebra is also named Tits-Satake
projection (\textit{cfr. e.g.} \cite{Fre:2006eu}, and Refs. therein).} of all the non-compact real forms of the U-duality Lie algebras of the $\mathcal{N}=2$ Maxwell-Einstein theories based on $J_{3}^{\mathbb{C}}$, $J_{3}^{\mathbb{H}}$ and $J_{3}^{\mathbb{O}}$
is the same, and it is the algebra of the theory based on $J_{3}^{\mathbb{R}}$ in the
same dimension. Moreover, the highest weight of the representation of any $p$-brane charge of any theory becomes the highest weight of the representation
of the $p$-brane charge of the theory based on $J_{3}^{\mathbb{R}}$, written in
terms of the restricted roots. In each representation, the real long
weights become the long weights of the representation of the theory on
$J_{3}^{\mathbb{R}}$, while all other weights are mapped onto short weights.
As we will see in detail in the rest of this section, this naturally implies
that the stratification of the orbits of the theories based on $J_{3}^{\mathbb{C}}$, $J_{3}^{\mathbb{H}}$ and $J_{3}^{\mathbb{O}}$ is the same as that of the theory based on $J_{3}^{\mathbb{R}}$.

In the rest of this section we will show how all the orbits of the $\mathcal{N}=2$ Maxwell-Einstein supergravity based on $J_{3}^{\mathbb{C}}$ in $D=5$ and $4$ can be computed as orbits of bound states of real-weight vectors. In \autoref{5dimtheoryonCorbits} we consider the electric BH orbits of the $D=5$ theory with symmetry $SL(3,\mathbb{C})$ as orbits of
bound states of real-weight vectors of the irrep. $(\overline{\mathbf{3}},\mathbf{3})$. Then, in \autoref{4dimtheoryonCorbits} we perform the same detailed analysis for the theory in $D=4$,
and compute the BH orbits as bound states of real-weight vectors of the irrep. $\mathbf{20}$ of the U-duality group $SU(3,3)$. The same approach can be exploited in order to analyze the $D=4$ and $D=5$ Maxwell-Einstein supergravities based on $J_{3}^{\mathbb{H}}$ and $J_{3}^{\mathbb{O}}$.

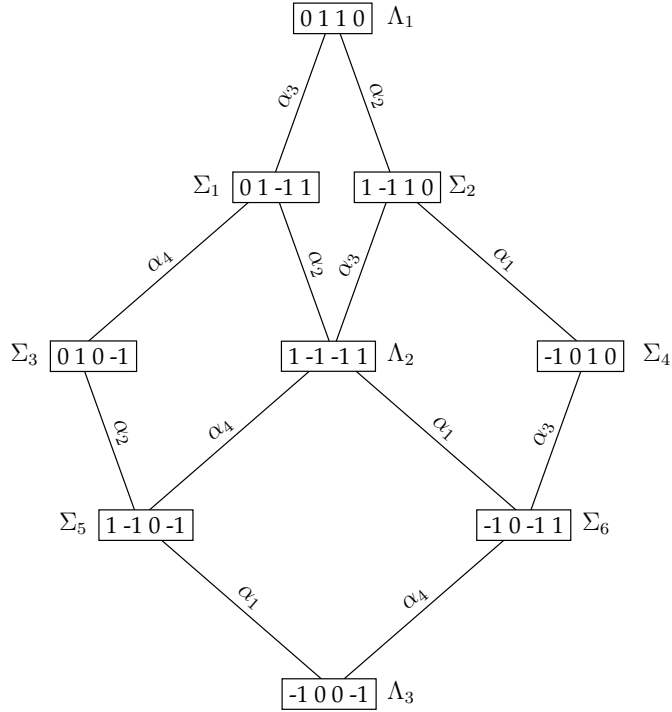
\begin{figure}[h!]
\centering
\scalebox{0.8} 
{\
\begin{pspicture}(0,-5.901927)(13.506875,5.901927)
\usefont{T1}{ppl}{m}{n}
\rput{-40.588333}(3.890698,2.4096527){\rput(5.166875,-4.0360937){\large $\alpha_{1}$}}
\psline[linewidth=0.02cm,fillcolor=black,dotsize=0.07055555cm 2.0]{*-*}(3.3634374,-2.7310936)(6.5634375,-5.5310936)
\psline[linewidth=0.02cm,fillcolor=black,dotsize=0.07055555cm 2.0]{*-*}(2.3634374,0.06890625)(3.3634374,-2.7310936)
\usefont{T1}{ppl}{m}{n}
\rput{-68.655}(3.205273,2.0223513){\rput(3.046875,-1.3160938){\large $\alpha_{2}$}}
\psline[linewidth=0.02cm,fillcolor=black,dotsize=0.07055555cm 2.0]{*-*}(9.763437,-2.7310936)(6.5634375,-5.5310936)
\usefont{T1}{ppl}{m}{n}
\rput{41.345284}(-0.6810696,-6.266558){\rput(7.926875,-4.0160937){\large $\alpha_{4}$}}
\psline[linewidth=0.02cm,fillcolor=black,dotsize=0.07055555cm 2.0]{*-*}(10.763437,0.06890625)(9.763437,-2.7310936)
\usefont{T1}{ppl}{m}{n}
\rput{68.5604}(5.197546,-10.176539){\rput(10.026875,-1.2560937){\large $\alpha_{3}$}}
\usefont{T1}{ppl}{m}{n}
\rput{-40.588333}(2.838871,5.165305){\rput(8.366875,-1.2360938){\large $\alpha_{1}$}}
\psline[linewidth=0.02cm,fillcolor=black,dotsize=0.07055555cm 2.0]{*-*}(6.5634375,0.06890625)(9.763437,-2.7310936)
\psline[linewidth=0.02cm,fillcolor=black,dotsize=0.07055555cm 2.0]{*-*}(5.5634375,2.8689063)(6.5634375,0.06890625)
\usefont{T1}{ppl}{m}{n}
\rput{-68.655}(2.6325922,6.783697){\rput(6.246875,1.4839063){\large $\alpha_{2}$}}
\psline[linewidth=0.02cm,fillcolor=black,dotsize=0.07055555cm 2.0]{*-*}(5.5634375,2.8689063)(2.3634374,0.06890625)
\usefont{T1}{ppl}{m}{n}
\rput{41.345284}(1.9713812,-2.0962152){\rput(3.726875,1.5839063){\large $\alpha_{4}$}}
\psline[linewidth=0.02cm,fillcolor=black,dotsize=0.07055555cm 2.0]{*-*}(7.5634375,2.8689063)(6.5634375,0.06890625)
\usefont{T1}{ppl}{m}{n}
\rput{68.5604}(5.7734604,-5.421425){\rput(6.826875,1.5439062){\large $\alpha_{3}$}}
\usefont{T1}{ppl}{m}{n}
\rput{-40.588333}(1.2577323,6.489594){\rput(9.366875,1.5639062){\large $\alpha_{1}$}}
\psline[linewidth=0.02cm,fillcolor=black,dotsize=0.07055555cm 2.0]{*-*}(7.5634375,2.8689063)(10.763437,0.06890625)
\psline[linewidth=0.02cm,fillcolor=black,dotsize=0.07055555cm 2.0]{*-*}(6.5634375,5.668906)(7.5634375,2.8689063)
\usefont{T1}{ppl}{m}{n}
\rput{-68.655}(0.6606734,9.495951){\rput(7.246875,4.2839065){\large $\alpha_{2}$}}
\psline[linewidth=0.02cm,fillcolor=black,dotsize=0.07055555cm 2.0]{*-*}(6.5634375,0.06890625)(3.3634374,-2.7310936)
\usefont{T1}{ppl}{m}{n}
\rput{41.345284}(0.37097234,-3.4547317){\rput(4.726875,-1.2160938){\large $\alpha_{4}$}}
\psline[linewidth=0.02cm,fillcolor=black,dotsize=0.07055555cm 2.0]{*-*}(6.5634375,5.668906)(5.5634375,2.8689063)
\usefont{T1}{ppl}{m}{n}
\rput{68.5604}(7.74523,-2.7140784){\rput(5.826875,4.3439064){\large $\alpha_{3}$}}
\usefont{T1}{ppl}{m}{n}
\rput(9.702656,-2.8210938){\psframebox[linewidth=0.02,fillstyle=solid]{-1 0 -1 1}}
\usefont{T1}{ppl}{m}{n}
\rput(3.4915626,-2.8210938){\psframebox[linewidth=0.02,fillstyle=solid]{1 -1 0 -1}}
\usefont{T1}{ppl}{m}{n}
\rput(10.637188,-0.02109375){\psframebox[linewidth=0.02,fillstyle=solid]{-1 0 1 0}}
\usefont{T1}{ppl}{m}{n}
\rput(6.4915624,-0.02109375){\psframebox[linewidth=0.02,fillstyle=solid]{1 -1 -1 1}}
\usefont{T1}{ppl}{m}{n}
\rput(2.6295311,-0.02109375){\psframebox[linewidth=0.02,fillstyle=solid]{0 1 0 -1}}
\usefont{T1}{ppl}{m}{n}
\rput(7.626094,2.7789063){\psframebox[linewidth=0.02,fillstyle=solid]{1 -1 1 0}}
\usefont{T1}{ppl}{m}{n}
\rput(5.6295314,2.7789063){\psframebox[linewidth=0.02,fillstyle=solid]{0 1 -1 1}}
\usefont{T1}{ppl}{m}{n}
\rput(6.5640626,5.578906){\psframebox[linewidth=0.02,fillstyle=solid]{0 1 1 0}}
\usefont{T1}{ppl}{m}{n}
\rput(6.5026565,-5.6210938){\psframebox[linewidth=0.02,fillstyle=solid]{-1 0 0 -1}}
\usefont{T1}{ptm}{m}{n}
\rput(7.686875,5.603906){\large $\Lambda_{1}$}
\usefont{T1}{ptm}{m}{n}
\rput(4.496875,2.8039062){\large $\Sigma_{1}$}
\usefont{T1}{ptm}{m}{n}
\rput(8.696875,2.8039062){\large $\Sigma_{2}$}
\usefont{T1}{ptm}{m}{n}
\rput(1.496875,0.00390625){\large $\Sigma_{3}$}
\usefont{T1}{ptm}{m}{n}
\rput(2.296875,-2.7960937){\large $\Sigma_{5}$}
\usefont{T1}{ptm}{m}{n}
\rput(11.896875,0.00390625){\large $\Sigma_{4}$}
\usefont{T1}{ptm}{m}{n}
\rput(7.686875,-5.5960937){\large $\Lambda_{3}$}
\usefont{T1}{ptm}{m}{n}
\rput(7.686875,0.00390625){\large $\Lambda_{2}$}
\usefont{T1}{ptm}{m}{n}
\rput(10.896875,-2.7960937){\large $\Sigma_{6}$}
\end{pspicture}
}
\caption{The weights of the $(\mathbf{\bar{3},3})$ of $\mathfrak{sl}(3,\mathbb{C})$.}
\label{3bar3sl3cdynkintree}
\end{figure}

\subsection{\label{5dimtheoryonCorbits}$D=5$}

The (electric) BH charges of the $\mathcal{N}=2$, $D=5$ Maxwell-Einstein supergravity based
on $J_{3}^{\mathbb{H}}$ (coupled to 8 vector multiplets) transform in the  $(\overline{\mathbf{3}},\mathbf{3})$ of
the U-duality group $SL(3,\mathbb{C})$. We draw in \autoref{3bar3sl3cdynkintree} the Dynkin labels of the weights of such
representation, where the order of the simple roots is as in \autoref{sl3cdynkindiagram}. All the weights have the same length. One can see by
acting with the Cartan involution as in \autoref{cartaninvsimplerootssl3c}
that there are three real weights, that are called $\Lambda_1$, $\Lambda_2$
and $\Lambda_3$ in the figure, while the other six weights, that we call $\Sigma$'s, are complex.

\subsubsection{1-charge orbit}

\begin{table}[t!]
\renewcommand{\arraystretch}{1.2}
\par
\begin{center}
\resizebox{\textwidth}{!}{
\begin{tabular}{|c|c|c|c|c|c|}
\hline
height&$\ket{\Lambda_{1}}$&$\ket{\Lambda_{2}}$&$\ket{\Lambda_{3}}$&$\ket{\Sigma_{1}}$&$\ket{\Sigma_{2}}$\\
\hline\hline
2&\begin{tabular}{c}
$E_{\alpha_{1}+\alpha_{2}}$\\
$E_{\alpha_{3}+\alpha_{4}} $
\end{tabular}&
\begin{tabular}{c}
$E_{\alpha_{1}+\alpha_{2}}$\\
$E_{\alpha_{3}+\alpha_{4}} $
\end{tabular}
&&\begin{tabular}{c}
$E_{\alpha_{1}+\alpha_{2}}$\\
$E_{\alpha_{3}+\alpha_{4}} $
\end{tabular} &
\begin{tabular}{c}
$E_{\alpha_{1}+\alpha_{2}}$\\
$E_{\alpha_{3}+\alpha_{4}} $
\end{tabular}\\

\hline

1&\begin{tabular}{c}
$E_{\alpha_{1}}$ \\
$E_{\alpha_{2}}$\\
$E_{\alpha_{3}}$\\
$E_{\alpha_{4}}$
\end{tabular}
&\begin{tabular}{c}
$E_{\alpha_{1}}$ \\
$E_{\alpha_{4}}$
\end{tabular}&
\begin{tabular}{c}
$E_{\alpha_{2}}$\\
$E_{\alpha_{3}}$
\end{tabular}&
\begin{tabular}{c}
$E_{\alpha_{1}}$ \\
$E_{\alpha_{2}}$\\
$E_{\alpha_{4}}$
\end{tabular}&
\begin{tabular}{c}
$E_{\alpha_{1}}$ \\
$E_{\alpha_{3}}$\\
$E_{\alpha_{4}}$
\end{tabular}\\

\hline

0&\begin{tabular}{c}
$H_{\alpha_{1}}$ \\
$H_{\alpha_{4}}$\\
$H_{\alpha_{2}}-H_{\alpha_{3}}$\\
\end{tabular}&
\begin{tabular}{c}
$H_{\alpha_{1}}+H_{\alpha_{2}}$ \\
$H_{\alpha_{3}}+H_{\alpha_{4}}$\\
$H_{\alpha_{1}}-H_{\alpha_{4}}$\\
\end{tabular}
&\begin{tabular}{c}
$H_{\alpha_{2}}$ \\
$H_{\alpha_{3}}$\\
$H_{\alpha_{1}}-H_{\alpha_{4}}$\\
\end{tabular}&\begin{tabular}{c}
$H_{\alpha_{1}}$ \\
$H_{\alpha_{2}}+H_{\alpha_{3}}$\\
$H_{\alpha_{3}}+H_{\alpha_{4}}$
\end{tabular}
&\begin{tabular}{c}
$H_{\alpha_{4}}$ \\
$H_{\alpha_{1}}+H_{\alpha_{2}}$\\
$H_{\alpha_{2}}+H_{\alpha_{3}}$
\end{tabular}\\

\hline

$-1$&\begin{tabular}{c}
$E_{-\alpha_{1}}$ \\
$E_{-\alpha_{4}}$
\end{tabular}
&\begin{tabular}{c}
$E_{-\alpha_{2}}$\\
$E_{-\alpha_{3}}$
\end{tabular}
&\begin{tabular}{c}
$E_{-\alpha_{1}}$ \\
$E_{-\alpha_{2}}$\\
$E_{-\alpha_{3}}$\\
$E_{-\alpha_{4}}$
\end{tabular}&
\begin{tabular}{c}
$E_{-\alpha_{1}}$ \\
$E_{-\alpha_{3}}$
\end{tabular}&
\begin{tabular}{c}
$E_{-\alpha_{2}}$\\
$E_{-\alpha_{4}}$
\end{tabular}
\\

\hline

$-2$&&\begin{tabular}{c}
$E_{-\alpha_{1}-\alpha_{2}}$\\
$E_{-\alpha_{3}-\alpha_{4}} $
\end{tabular}
&\begin{tabular}{c}
$E_{-\alpha_{1}-\alpha_{2}}$\\
$E_{-\alpha_{3}-\alpha_{4}} $
\end{tabular}&
\begin{tabular}{c}
$E_{-\alpha_{3}-\alpha_{4}}$
\end{tabular}&
\begin{tabular}{c}
$E_{-\alpha_{1}-\alpha_{2}}$
\end{tabular}\\
\hline
\end{tabular}
}
\end{center}
\caption{Stabilizers for the weight vectors of the $\mathbf{(\bar{3},3)}$ of $\mathfrak{sl}(3,\mathbb{C})$ that are used in the paper. }
\label{l1l2l3s1s2stab3bar3sl3c}
\end{table}

\begin{table}[b!]
\renewcommand{\arraystretch}{1.2}
\par
\begin{center}
\begin{tabular}{|c|c|c|}
\hline
Common & $\ket{\Lambda_{1}}+\ket{\Lambda_{2}}$ Conjunction & $\ket{\Lambda_{1}}-\ket{\Lambda_{2}}$
Conjunction \\ \hline\hline
$E_{\alpha_{1}+\alpha_{2}}$ & $E_{\alpha_{2}}-E_{-\alpha_{3}}$ & $E_{\alpha_{2}}+E_{-\alpha_{3}}$ \\
$E_{\alpha_{3}+\alpha_{4}}$ & $E_{\alpha_{3}}-E_{-\alpha_{2}}$ & $E_{\alpha_{3}}+E_{-\alpha_{2}}$ \\
$E_{\alpha_{1}}$ &  &  \\
$E_{\alpha_{4}}$ &  &  \\
$H_{\alpha_{1}}-H_{\alpha_{4}}$ &  &  \\
$H_{\alpha_{2}}-H_{\alpha_{3}}$ &  &  \\ \hline
\end{tabular}
\end{center}
\caption{The generators of the stabilizing algebras $( \mathfrak{su}(2)\oplus \mathfrak{so}(2))\ltimes \mathbb{R}^{(2,2)}$ and $(\mathfrak{sl}(2,\mathbb{R})\oplus \mathfrak{so}(2))\ltimes \mathbb{R}^{(2,2)}$ of the bound
states $\ket{\Lambda_{1}}+\ket{\Lambda_{2}}$ and $\ket{\Lambda_{1}}-\ket{\Lambda_{2}}$.}
\label{l1+l2stabsl3C}
\end{table}

We want to show that the BH orbits correspond to orbits of bound
states involving only real-weight vectors. In \autoref{l1l2l3s1s2stab3bar3sl3c} we list the
stabilizing generators for the three real-weight vectors $\ket{\Lambda_{1}},\ket{\Lambda_{2}},
\ket{\Lambda_{3}}$ and two of the complex-weight vectors, say $\ket{\Sigma_{1}}$ and $\ket{\Sigma_{2}}$ (without any loss of generality, we only consider these two complex weights because all
others are connected to these two by transformations in the algebra). From
the table one deduces that the stabilizing algebra of each real-weight vector is $[\mathfrak{sl}(2,\mathbb{C})\oplus \mathfrak{so}(2)]\ltimes \mathbb{R}^{(2,2)}
$, yielding the highest-weight orbit corresponding to a 1-charge (\ie rank-1 \cite{Ferrar,Krutelevich}) BH duality orbit,
as expected.

\begin{table}[t!]
\renewcommand{\arraystretch}{1.2}
\begin{center}
\begin{tabular}{|c|c|}
\hline
\multicolumn{2}{|c|}{$\ket{\Sigma_{1}}+\ket{\Sigma_{2}}$ Stabilizers}\\
\hline
\hline
Common&Conjunction\\
\hline
$E_{\alpha_{1}+\alpha_{2}}$&$E_{\alpha_2}-E_{\alpha_3}$\\
$E_{\alpha_{3}+\alpha_{4}}$&$E_{-\alpha_2}-E_{-\alpha_3}$\\
$E_{\alpha_{1}}$&\\
$E_{\alpha_{4}}$&\\
$H_{\alpha_{2}}+H_{\alpha_{3}}$&\\
$H_{\alpha_{1}}-H_{\alpha_{3}}-H_{\alpha_{4}}$&\\
\hline
\end{tabular}

\caption{The generators of the $(\mathfrak{sl}(2,\mathbb{R})\oplus \mathfrak{so}(2))\ltimes \mathbb{R}^{(2,2)}$ stabilizing algebra of the bound state $\ket{\Sigma_1}+ \ket{\Sigma_2}$, real combination of two short weights of the $\mathbf{(\bar{3},3)}$ of $\mathfrak{sl}(3,\mathbb{C})$. \label{sigma1sigma2ofsl3C} }
\end{center}
\end{table}

\begin{table}[b!]
\renewcommand{\arraystretch}{1.5}
\begin{center}
\begin{tabular}{|c||c|c|c|}
\hline
Common & 2-conj. & $\ket{\Lambda_{1}}+\ket{\Lambda_{2}}+\ket{\Lambda_{3}}$  &$\ket{\Lambda_{1}}+\ket{\Lambda_{2}}-\ket{\Lambda_{3}}$ \\
\hline
\hline
$H_{\alpha_{1}}-H_{\alpha_{4}}$&&$E_{\alpha_{2}}-E_{-\alpha_{3}}$ &$E_{\alpha_{2}}-E_{-\alpha_{3}}$ \\
$H_{\alpha_{2}}-H_{\alpha_{3}}$&\multirow{-2}{*}{\rotatebox{90}{$\Lambda_{1},\Lambda_{2}$}}&$E_{\alpha_{3}}-E_{-\alpha_{2}}$ &$E_{\alpha_{3}}-E_{-\alpha_{2}}$\\ \cline{2-4}
&&$E_{\alpha_{1}}-E_{-\alpha_{4}}$ &$E_{\alpha_{1}}+E_{-\alpha_{4}}$\\
&\multirow{-2}{*}{\rotatebox{90}{$\Lambda_{2},\Lambda_{3}$}}&$E_{\alpha_{4}}-E_{-\alpha_{1}}$ &$E_{\alpha_{4}}+E_{-\alpha_{1}}$\\ \cline{2-4}
&&$E_{\alpha_{1}+\alpha_{2}}-E_{-\alpha_{3}-\alpha_{4}}$  &$E_{\alpha_{1}+\alpha_{2}}+E_{-\alpha_{3}-\alpha_{4}}$\\
&\multirow{-2}{*}{\rotatebox{90}{$\Lambda_{1},\Lambda_{3}$}}&$E_{\alpha_{3}+\alpha_{4}}-E_{-\alpha_{1}-\alpha_{2}}$ &$E_{\alpha_{3}+\alpha_{4}}+E_{-\alpha_{1}-\alpha_{2}}$\\

\hline
\end{tabular}

\caption{The stabilizers of the $\ket{\Lambda_{1}}+\ket{\Lambda_{2}}+\ket{\Lambda_{3}}$ and $\ket{\Lambda_{1}}+\ket{\Lambda_{2}}-\ket{\Lambda_{3}}$ bound states. The 2-conjunction stabilizers connect the two weight vectors whose weight is listed in the second column, and annihilate the third weight vector. The stabilizing algebra is $\mathfrak{su}(3)$ for the  $\ket{\Lambda_{1}}+\ket{\Lambda_{2}}+\ket{\Lambda_{3}}$ bound state, and $\mathfrak{su}(1,2)$ for the  $\ket{\Lambda_{1}}+\ket{\Lambda_{2}}-\ket{\Lambda_{3}}$ bound state. \label{l1+l2+l3stabsl3C}}
\end{center}
\end{table}

\subsubsection{2-charge orbits}

Following our prescription, we now consider the 2-charge (\ie rank-2 \cite{Ferrar,Krutelevich}) orbits as  bound
states of two real-weight vectors. We consider in particular the bound states $\ket{\Lambda_{1}} +\ket{\Lambda_{2}}$ and $\ket{\Lambda_{1}} -\ket{\Lambda_{2}}$, and we list in \autoref{l1+l2stabsl3C} their stabilizers. From the table, one deduces that
the semisimple part of the stabilizing algebra is generated by $H_{\alpha_1}- H_{\alpha_4}$, $H_{\alpha_2}- H_{\alpha_3}$, $E_{\alpha_2} -
E_{-\alpha_3}$ and $E_{\alpha_3} - E_{-\alpha_2}$ in the $\ket{\Lambda_{1}}
+\ket{\Lambda_{2}}$ case, and by $H_{\alpha_1}- H_{\alpha_4}$, $H_{\alpha_2}-
H_{\alpha_3}$, $E_{\alpha_2} + E_{-\alpha_3}$ and $E_{\alpha_3} +
E_{-\alpha_2}$ in the $\ket{\Lambda_{1}} -\ket{\Lambda_{2}}$ case. From \autoref{cartaninvgeneratorssl3c}, one then obtains that in the first case all generators are compact, while in the second case the generators $
E_{\alpha_2} + E_{-\alpha_3}$ and $E_{\alpha_3} + E_{-\alpha_2}$ are
non-compact. Taking into account also the additional generators in \autoref{l1+l2stabsl3C}, the resulting stabilizing algebras are respectively $[
\mathfrak{su}(2)\oplus \mathfrak{so}(2)]\ltimes \mathbb{R}^{(2,2)}$ and $[
\mathfrak{sl}(2,\mathbb{R})\oplus \mathfrak{so}(2)]\ltimes \mathbb{R}^{(2,2)}
$, which correspond to the two rank-2 orbits of the theory. From comparison with the literature \cite{Borsten:2011ai}, the former orbit is 1/2-BPS, and the latter one is non-supersymmetric (non-BPS).

In the previous section, we have shown that in the $\mathcal{N}=2$, $D=5$ Maxwell-Einstein supergravity based
on $J_{3}^{\mathbb{R}}$ the splitting of the 2-charge
configuration into two different rank-2 orbits is due to the fact that the
conjunction stabilizers connect two long-weight vectors to the same short-weight vector.
By looking at \autoref{l1+l2stabsl3C} and \autoref{3bar3sl3cdynkintree}, we
notice that in $\mathcal{N}=2$, $D=5$ Maxwell-Einstein supergravity based
on $J_{3}^{\mathbb{C}}$ (and analogously for theories based on $J_{3}^{\mathbb{H}}$ and $J_{3}^{\mathbb{O}}$) something very similar happens, namely the
conjunction stabilizers connect two real-weight vectors, say $\ket{\Lambda_{1}}$ and $\ket{\Lambda_{2}}$  to the complex-weight vectors $\ket{\Sigma_1}$ and $\ket{\Sigma_2}$. By projecting on the restricted-root algebra, both
the complex weights $\Sigma_1$ and $\Sigma_2$ are projected on the same short weight of the $\mathbf{6}$ of $\mathfrak{sl}(3,\mathbb{R})$. More generally, by considering the
action of the Cartan involution on the complex weights,
\begin{flalign}
&\theta \Sigma_{1}=-\Sigma_{2} \qquad  \theta \Sigma_{3}=-\Sigma_{4} \qquad \theta \Sigma_{5}=-\Sigma_{6} \quad ,
\end{flalign}
one can see that each of the three pairs of complex weights of the $\mathbf{(\bar{3},3)}$ of $\mathfrak{sl}(3,\mathbb{C})$ connected by $\theta$ is
projected onto each of the three short weights of the $\mathbf{6}$ of $\mathfrak{sl}(3,\mathbb{R})$.

In the previous section we have also shown that the rank-2 non-supersymmetric $\ket{\Lambda_1} - \ket{\Lambda_2}$ orbit of $\mathcal{N}=2$,
$D=5$ Maxwell-Einstein supergravity based
on $J_{3}^{\mathbb{R}}$ can be obtained as the orbit of a single short-weight vector. We now show that this naturally generalizes to the magic theory based on $J_{3}^{\mathbb{C}}$ (and analogously to the supergravities based on $J_{3}^{\mathbb{H}}$ and $J_{3}^{\mathbb{O}}$) as follows. The
reality properties of the algebra imply that we must consider real
combinations of weights, because this corresponds to the physically meaningful real electric charges of the extremal BH. We thus define the real state $\ket{\Sigma_{1}}+\ket{\Sigma_{2}}$, and we list in \autoref{sigma1sigma2ofsl3C} the
corresponding stabilizers. One finds that the stabilizing algebra is $(\mathfrak{sl}(2,\mathbb{R})\oplus \mathfrak{so}(2))\ltimes \mathbb{R}^{(2,2)}
$, which is the same as that of $\ket{\Lambda_{1}}-\ket{\Lambda_{2}}$, as expected.

\begin{table}[t!]
\renewcommand{\arraystretch}{1.5}
\begin{center}
\begin{tabular}{|c||c|c|c|}
\hline
Common &2-conj. & $\ket{\Lambda_{3}}+\ket{\Sigma_{1}}+\ket{\Sigma_{2}}$   &$\ket{\Lambda_{1}}-(\ket{\Sigma_{1}}+\ket{\Sigma_{2}})$ \\
\hline
\hline
$H_{\alpha_{2}}+H_{\alpha_{3}}$&&$E_{\alpha_{3}}-E_{-\alpha_{2}}$ &$E_{\alpha_{3}}-E_{-\alpha_{2}}$\\
$H_{\alpha_{1}}-H_{\alpha_{3}}-H_{\alpha_{4}}$&\multirow{-2}{*}{\rotatebox{90}{$\Sigma_{1},\Sigma_{2}$}}&$E_{-\alpha_{2}}-E_{-\alpha_{3}}$&$E_{-\alpha_{2}}-E_{-\alpha_{3}}$\\ \cline{2-4}
&&$E_{\alpha_{1}+\alpha_{2}}-E_{-\alpha_{4}}$ & $E_{\alpha_{1}+\alpha_{2}}+E_{-\alpha_{4}}$\\
&\multirow{-2}{*}{\rotatebox{90}{$\Sigma_{1},\Lambda_{3}$}}&$E_{\alpha_{4}}-E_{-\alpha_{1}-\alpha_{2}}$ &$E_{\alpha_{4}}+E_{-\alpha_{1}-\alpha_{2}}$\\ \cline{2-4}
&&$E_{\alpha_{1}}-E_{-\alpha_{3}-\alpha_{4}}$  &$E_{\alpha_{1}}+E_{-\alpha_{3}-\alpha_{4}}$\\
&\multirow{-2}{*}{\rotatebox{90}{$\Sigma_{2},\Lambda_{3}$}}&$E_{\alpha_{3}+\alpha_{4}}-E_{-\alpha_{1}}$  &$E_{\alpha_{3}+\alpha_{4}}+E_{-\alpha_{1}}$\\
\hline
\end{tabular}

\caption{The generators of the $\mathfrak{su}(1,2)$ stabilizing algebra of the bound states $\ket{\Lambda_{3}}+\ket{\Sigma_{1}}+\ket{\Sigma_{2}}$ and $\ket{\Lambda_{3}}-(\ket{\Sigma_{1}}+\ket{\Sigma_{2}})$. The 2-conjunction stabilizers connect the two weights in the second column and annihilate the third weight. \label{l3+s1+s2stabsl3C}}
\end{center}
\end{table}

\begin{table}[t]
\renewcommand{\arraystretch}{1.5}
\par
\begin{center}
\begin{tabular}{|c|c|c|}
\hline
\multicolumn{2}{|c|}{State} & Stabilizer \\ \hline\hline
\multirow{-1}{*}{\rotatebox{90}{1-s}} & $\ket{\Lambda_{1}}$ & $(\mathfrak{sl}(2,
\mathbb{C})\oplus \mathfrak{so}(2))\ltimes \mathbb{R}^{(2,2)}$ \\ \hline
& $\ket{\Lambda_{1}}+\ket{\Lambda_{2}}$ & $(\mathfrak{su}(2)\oplus \mathfrak{so}
(2))\ltimes \mathbb{R}^{(2,2)}$ \\
& $\ket{\Lambda_{1}}-\ket{\Lambda_{2}}$ & $(\mathfrak{sl}(2,\mathbb{R})\oplus \mathfrak{
so}(2))\ltimes \mathbb{R}^{(2,2)}$ \\
\multirow{-3}{*}{\rotatebox{90}{2-s}} & $\ket{\Sigma_{1}}+\ket{\Sigma_{2}}$ & $(
\mathfrak{sl}(2,\mathbb{R})\oplus \mathfrak{so}(2))\ltimes \mathbb{R}^{(2,2)}
$ \\ \hline
& $\ket{\Lambda_{1}}+\ket{\Lambda_{2}}+\ket{\Lambda_{3}}$ & $\mathfrak{su}(3)$ \\
& $\ket{\Lambda_{1}}+\ket{\Lambda_{2}}-\ket{\Lambda_{3}}$ & $\mathfrak{sl}(1,2)$ \\
& $\ket{\Lambda_{3}}+\ket{\Sigma_{1}}+\ket{\Sigma_{2}}$ & $\mathfrak{sl}(1,2)$ \\
\multirow{-4}{*}{\rotatebox{90}{3-s}} & $\ket{\Lambda_{1}}-(\ket{\Sigma_{1}}+\ket{\Sigma_{2}})$
& $\mathfrak{sl}(1,2)$ \\ \hline
\end{tabular}
\end{center}
\caption{Stabilizers in the $\mathbf{(\bar{3},3)}$ of $\mathfrak{sl}(3,\mathbb{C})$. }
\label{3bar3summary}
\end{table}

\subsubsection{3-charge orbits}

We can now proceed to show that our prescription works for the case of the
3-charge (\ie rank-3 \cite{Ferrar,Krutelevich}) ``large'' U-duality orbits. In \autoref{l1+l2+l3stabsl3C} we list the stabilizers
on the $\ket{\Lambda_1} +\ket{ \Lambda_2} +\ket{ \Lambda_3}$ and $\ket{\Lambda_1} +\ket{ \Lambda_2} -
\ket{\Lambda_3}$ bound states. The reader can check that in the first case all
generators are compact, while in the second case there are four compact and
four non-compact generators, resulting in the algebras $\mathfrak{su}(3)$
and $\mathfrak{su}(1,2)$, respectively. From comparison with literature \cite{Ferrara:2006xx,Borsten:2011ai}, we recognize the first as the 1/2-BPS
``large'' orbit and the second as the non-supersymmetric (non-BPS) ``large'' one. By trading a real combination of two complex-weight vectors for the difference of two real-weight vectors, in \autoref{l3+s1+s2stabsl3C} we list the stabilizers of considered bound states $\ket{\Lambda_{3}}
\pm (\ket{\Sigma_{1}}+\ket{\Sigma_{2}}) $, yielding the algebra $\mathfrak{su}(1,2)$ which
as expected corresponds to the non-supersymmetric 3-charge ``large'' orbit.

We summarize the stabilizers of all the bound states in the $\mathbf{(\bar{3},3)}$ of $\mathfrak{sl}(3,\mathbb{C})$ in \autoref{3bar3summary}, matching the results reported in Table II of \cite{Borsten:2011ai} (cfr. also Refs. therein).

In the next subsection we will perform the same analysis for the four-dimensional
theory, showing that again all  extremal BH orbits are obtained
as bound states of real weights, thus providing a natural explanation for the
splitting of the charge configurations exactly as in all other cases analyzed above.

\subsection{\label{4dimtheoryonCorbits}$D=4$}

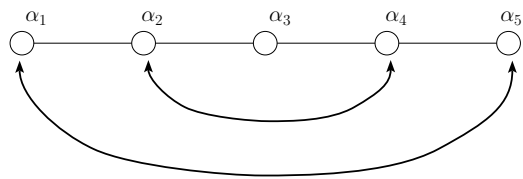
\begin{figure}[h!]
\centering
\scalebox{0.4} {\
\begin{pspicture}(0,-4.674844)(16.8,1.7148438)
\psline[linewidth=0.02cm](0.6,-0.25796875)(16.2,-0.25796875)
\pscircle[linewidth=0.02,dimen=outer,fillstyle=solid](0.4,-0.25796875){0.4}
\pscircle[linewidth=0.02,dimen=outer,fillstyle=solid](16.4,-0.25796875){0.4}
\pscircle[linewidth=0.02,dimen=outer,fillstyle=solid](8.4,-0.25796875){0.4}
\pscircle[linewidth=0.02,dimen=outer,fillstyle=solid](12.4,-0.25796875){0.4}
\pscircle[linewidth=0.02,dimen=outer,fillstyle=solid](4.4,-0.25796875){0.4}
\psbezier[linewidth=0.06,arrowsize=0.05291667cm 4.0,arrowlength=1.4,arrowinset=0.2]{<->}(0.32453123,-0.83484375)(0.32453123,-2.0071514)(1.5156925,-3.1833103)(2.7545307,-3.765613)(3.993369,-4.3479156)(7.075781,-4.6448436)(8.307333,-4.6448436)(9.538885,-4.6448436)(12.474531,-4.6448436)(14.094531,-3.765613)(15.714531,-2.886382)(16.52453,-2.3002284)(16.52453,-0.83484375)
\psbezier[linewidth=0.06,arrowsize=0.05291667cm 4.0,arrowlength=1.4,arrowinset=0.2]{<->}(4.558437,-0.81576675)(4.558437,-1.440098)(5.1441727,-2.0664806)(5.7533517,-2.3765953)(6.3625307,-2.68671)(7.878258,-2.8448439)(8.541485,-2.8448439)(9.204712,-2.8448439)(10.524531,-2.8448439)(11.3296175,-2.3765953)(12.134704,-1.9083468)(12.524531,-1.596181)(12.524531,-0.8157669)
\rput(0.87640625,0.5651562){\huge $\alpha_1$}
\rput(8.8925,0.5651562){\huge $\alpha_3$}
\rput(12.710468,0.5651562){\huge $\alpha_4$}
\rput(16.48953,0.5651562){\huge $\alpha_5$}
\rput(4.6865625,0.5651562){\huge $\alpha_2$}
\end{pspicture}
}
\caption{The Tits-Satake diagram of the algebra $\mathfrak{su}(3,3)$.}
\label{su33dynkindiagram}
\end{figure}

The $\mathcal{N}=2$, $D=4$ Maxwell-Einstein supergravity theory based on $J_{3}^{\mathbb{C}}$ (coupled to 9 vector multiplets) has
U-duality symmetry group $SU(3,3)$, and we draw in \autoref{su33dynkindiagram}
the Tits-Satake diagram of the corresponding Lie algebra $\mathfrak{su}(3,3)$.
From the diagram, one deduces that the Cartan involution acts on the simple
roots as
\begin{equation}
\theta \alpha_1 =-\alpha_5 \qquad \theta \alpha_2 =-\alpha_4 \qquad \theta
\alpha_3 =-\alpha_3 \quad ,  \label{cartansimplerootssu33}
\end{equation}
which leads to the action
\begin{eqnarray}
& & \theta E_{\alpha_1 } =-E_{-\alpha_5} \qquad \theta E_{\alpha_2}
=-E_{-\alpha_4} \qquad \theta E_{\alpha_3} =-E_{-\alpha_3}  \notag \\
& & \theta H_{\alpha_1} = -H_{\alpha_5} \qquad \theta H_{\alpha_2}
=-H_{\alpha_4} \qquad \theta H_{\alpha_3} =-H_{\alpha_3}
\label{cartaninvgeneratorssu33}
\end{eqnarray}
on the corresponding simple-root generators. From \autoref{cartansimplerootssu33} one
obtains the restricted roots
\begin{equation}
(\alpha_1 )_R = \tfrac{1}{2} (\alpha_1 + \alpha_5 ) \quad (\alpha_2 )_R =
\tfrac{1}{2} (\alpha_2 + \alpha_4 ) \quad (\alpha_3 )_R = \alpha_3 \quad ,
\end{equation}
which are nothing but the simple roots of the restricted-root algebra $\mathfrak{sp}(6,\mathbb{R})$, namely the U-duality Lie algebra of the $\mathcal{N}=2$, $D=4$ Maxwell-Einstein supergravity theory based on $J_{3}^{\mathbb{R}}$.

\begin{figure}[h]
\centering
\scalebox{0.65} 
{\
\begin{pspicture}(0,-10.301927)(14.2890625,10.301927)
\usefont{T1}{ppl}{m}{n}
\rput{35.35122}(-0.6673004,-6.4997444){\rput(9.834531,-4.2810936){$\alpha_{5}$}}
\psline[linewidth=0.02cm,fillcolor=black,dotsize=0.07055555cm 2.0]{*-*}(11.57,-3.3310938)(8.37,-5.5310936)
\psline[linewidth=0.02cm,fillcolor=black,dotsize=0.07055555cm 2.0]{*-*}(3.77,-1.1310937)(6.97,-3.3310938)
\usefont{T1}{ppl}{m}{n}
\rput{-32.069756}(1.9137675,2.5450494){\rput(5.3545313,-2.0410938){$\alpha_{1}$}}
\usefont{T1}{ppl}{m}{n}
\rput{35.35122}(1.239886,-4.604544){\rput(7.8145313,-0.34109375){$\alpha_{5}$}}
\psline[linewidth=0.02cm,fillcolor=black,dotsize=0.07055555cm 2.0]{*-*}(10.17,1.0689063)(6.97,-1.1310937)
\psline[linewidth=0.02cm,fillcolor=black,dotsize=0.07055555cm 2.0]{*-*}(6.97,1.0689063)(10.17,-1.1310937)
\usefont{T1}{ppl}{m}{n}
\rput{-32.069756}(1.6062813,4.875176){\rput(9.254531,-0.34109375){$\alpha_{1}$}}
\psline[linewidth=0.02cm,fillcolor=black,dotsize=0.07055555cm 2.0]{*-*}(6.97,-1.1310937)(6.97,-3.3310938)
\usefont{T1}{ppl}{m}{n}
\rput{90.38534}(4.493472,-9.177103){\rput(6.7745314,-2.3410938){$\alpha_{3}$}}
\usefont{T1}{ppl}{m}{n}
\rput{35.35122}(2.3032286,-2.6213405){\rput(5.2345314,2.3189063){$\alpha_{5}$}}
\psline[linewidth=0.02cm,fillcolor=black,dotsize=0.07055555cm 2.0]{*-*}(6.97,3.2689064)(3.77,1.0689063)
\psline[linewidth=0.02cm,fillcolor=black,dotsize=0.07055555cm 2.0]{*-*}(8.37,-5.5310936)(6.97,-7.731094)
\usefont{T1}{ppl}{m}{n}
\rput{54.790813}(-2.2867286,-8.901338){\rput(7.414531,-6.6410937){$\alpha_{4}$}}
\psline[linewidth=0.02cm,fillcolor=black,dotsize=0.07055555cm 2.0]{*-*}(3.77,1.0689063)(6.97,-1.1310937)
\usefont{T1}{ppl}{m}{n}
\rput{-32.069756}(0.44478738,2.6338518){\rput(4.7745314,0.55890626){$\alpha_{1}$}}
\psline[linewidth=0.02cm,fillcolor=black,dotsize=0.07055555cm 2.0]{*-*}(6.97,-3.3310938)(8.37,-5.5310936)
\usefont{T1}{ppl}{m}{n}
\rput{-58.967655}(7.550945,4.6012936){\rput(7.8145313,-4.3610935){$\alpha_{2}$}}
\psline[linewidth=0.02cm,fillcolor=black,dotsize=0.07055555cm 2.0]{*-*}(6.97,-7.731094)(6.97,-9.931094)
\usefont{T1}{ppl}{m}{n}
\rput{90.38534}(-2.1063786,-15.82149){\rput(6.7745314,-8.941093){$\alpha_{3}$}}
\usefont{T1}{ppl}{m}{n}
\rput{35.35122}(3.3179893,-1.4056844){\rput(3.8345313,4.518906){$\alpha_{5}$}}
\psline[linewidth=0.02cm,fillcolor=black,dotsize=0.07055555cm 2.0]{*-*}(5.57,5.4689064)(2.37,3.2689064)
\psline[linewidth=0.02cm,fillcolor=black,dotsize=0.07055555cm 2.0]{*-*}(6.97,-3.3310938)(5.57,-5.5310936)
\usefont{T1}{ppl}{m}{n}
\rput{54.790813}(-1.0820243,-6.8259034){\rput(6.014531,-4.441094){$\alpha_{4}$}}
\psline[linewidth=0.02cm,fillcolor=black,dotsize=0.07055555cm 2.0]{*-*}(8.37,5.4689064)(11.57,3.2689064)
\usefont{T1}{ppl}{m}{n}
\rput{-32.069756}(-0.88856196,5.9945707){\rput(9.954532,4.558906){$\alpha_{1}$}}
\psline[linewidth=0.02cm,fillcolor=black,dotsize=0.07055555cm 2.0]{*-*}(5.57,-5.5310936)(6.97,-7.731094)
\usefont{T1}{ppl}{m}{n}
\rput{-58.967655}(8.757804,2.3358147){\rput(6.414531,-6.561094){$\alpha_{2}$}}
\psline[linewidth=0.02cm,fillcolor=black,dotsize=0.07055555cm 2.0]{*-*}(10.17,1.0689063)(10.17,-1.1310937)
\usefont{T1}{ppl}{m}{n}
\rput{90.38534}(9.914943,-10.162234){\rput(9.974531,-0.14109375){$\alpha_{3}$}}
\psline[linewidth=0.02cm,fillcolor=black,dotsize=0.07055555cm 2.0]{*-*}(3.77,-1.1310937)(2.37,-3.3310938)
\usefont{T1}{ppl}{m}{n}
\rput{54.790813}(-0.63950646,-3.279775){\rput(2.8145313,-2.2410936){$\alpha_{4}$}}
\psline[linewidth=0.02cm,fillcolor=black,dotsize=0.07055555cm 2.0]{*-*}(5.57,5.4689064)(6.97,3.2689064)
\usefont{T1}{ppl}{m}{n}
\rput{-58.967655}(-0.6678368,7.6650743){\rput(6.414531,4.438906){$\alpha_{2}$}}
\psline[linewidth=0.02cm,fillcolor=black,dotsize=0.07055555cm 2.0]{*-*}(3.77,1.0689063)(3.77,-1.1310937)
\usefont{T1}{ppl}{m}{n}
\rput{90.38534}(3.4719014,-3.76238){\rput(3.5745313,-0.14109375){$\alpha_{3}$}}
\usefont{T1}{ppl}{m}{n}
\rput{35.35122}(0.69196266,-2.7651465){\rput(4.6545315,-0.28109375){$\alpha_{5}$}}
\psline[linewidth=0.02cm,fillcolor=black,dotsize=0.07055555cm 2.0]{*-*}(6.97,1.0689063)(3.77,-1.1310937)
\psline[linewidth=0.02cm,fillcolor=black,dotsize=0.07055555cm 2.0]{*-*}(11.57,3.2689064)(10.17,1.0689063)
\usefont{T1}{ppl}{m}{n}
\rput{54.790813}(6.2583303,-7.789663){\rput(10.6145315,2.1589062){$\alpha_{4}$}}
\psline[linewidth=0.02cm,fillcolor=black,dotsize=0.07055555cm 2.0]{*-*}(2.37,-3.3310938)(5.57,-5.5310936)
\usefont{T1}{ppl}{m}{n}
\rput{-32.069756}(2.8682237,1.4660028){\rput(3.9545312,-4.2410936){$\alpha_{1}$}}
\psline[linewidth=0.02cm,fillcolor=black,dotsize=0.07055555cm 2.0]{*-*}(10.17,-1.1310937)(11.57,-3.3310938)
\usefont{T1}{ppl}{m}{n}
\rput{-58.967655}(7.2161465,8.40915){\rput(11.014531,-2.1610937){$\alpha_{2}$}}
\psline[linewidth=0.02cm,fillcolor=black,dotsize=0.07055555cm 2.0]{*-*}(6.97,3.2689064)(6.97,1.0689063)
\usefont{T1}{ppl}{m}{n}
\rput{90.38534}(8.893373,-4.747512){\rput(6.7745314,2.0589063){$\alpha_{3}$}}
\psline[linewidth=0.02cm,fillcolor=black,dotsize=0.07055555cm 2.0]{*-*}(8.37,5.4689064)(6.97,3.2689064)
\usefont{T1}{ppl}{m}{n}
\rput{54.790813}(6.700848,-4.2435346){\rput(7.414531,4.3589063){$\alpha_{4}$}}
\psline[linewidth=0.02cm,fillcolor=black,dotsize=0.07055555cm 2.0]{*-*}(2.37,3.2689064)(3.77,1.0689063)
\usefont{T1}{ppl}{m}{n}
\rput{-58.967655}(-0.333039,3.8572178){\rput(3.2145312,2.2389061){$\alpha_{2}$}}
\psline[linewidth=0.02cm,fillcolor=black,dotsize=0.07055555cm 2.0]{*-*}(6.97,9.868906)(6.97,7.668906)
\usefont{T1}{ppl}{m}{n}
\rput{90.38534}(15.493223,1.8968751){\rput(6.7745314,8.658906){$\alpha_{3}$}}
\usefont{T1}{ppl}{m}{n}
\rput{35.35122}(0.34745985,-5.2840877){\rput(8.434531,-2.0810938){$\alpha_{5}$}}
\psline[linewidth=0.02cm,fillcolor=black,dotsize=0.07055555cm 2.0]{*-*}(10.17,-1.1310937)(6.97,-3.3310938)
\psline[linewidth=0.02cm,fillcolor=black,dotsize=0.07055555cm 2.0]{*-*}(6.97,7.668906)(5.57,5.4689064)
\usefont{T1}{ppl}{m}{n}
\rput{54.790813}(7.9055524,-2.1681){\rput(6.014531,6.558906){$\alpha_{4}$}}
\psline[linewidth=0.02cm,fillcolor=black,dotsize=0.07055555cm 2.0]{*-*}(6.97,3.2689064)(10.17,1.0689063)
\usefont{T1}{ppl}{m}{n}
\rput{-32.069756}(0.06589412,4.9155235){\rput(8.554531,2.3589063){$\alpha_{1}$}}
\psline[linewidth=0.02cm,fillcolor=black,dotsize=0.07055555cm 2.0]{*-*}(6.97,7.668906)(8.37,5.4689064)
\usefont{T1}{ppl}{m}{n}
\rput{-58.967655}(-1.8746959,9.930553){\rput(7.8145313,6.6389065){$\alpha_{2}$}}
\usefont{T1}{ppl}{m}{n}
\rput(6.940625,9.978907){\psframebox[linewidth=0.02,fillstyle=solid]{0 0 1 0 0}}
\usefont{T1}{ppl}{m}{n}
\rput(7.010625,7.578906){\psframebox[linewidth=0.02,fillstyle=solid]{0 1 -1 1 0}}
\usefont{T1}{ppl}{m}{n}
\rput(5.606094,5.3789062){\psframebox[linewidth=0.02,fillstyle=solid]{0 1 0 -1 1}}
\usefont{T1}{ppl}{m}{n}
\rput(2.4060938,3.1789062){\psframebox[linewidth=0.02,fillstyle=solid]{0 1 0 0 -1}}
\usefont{T1}{ppl}{m}{n}
\rput(8.402657,5.3789062){\psframebox[linewidth=0.02,fillstyle=solid]{1 -1 0 1 0}}
\usefont{T1}{ppl}{m}{n}
\rput(6.868125,3.1789062){\psframebox[linewidth=0.02,fillstyle=solid]{1 -1 1 -1 1}}
\usefont{T1}{ppl}{m}{n}
\rput(11.41375,3.1789062){\psframebox[linewidth=0.02,fillstyle=solid]{-1 0 0 1 0}}
\usefont{T1}{ppl}{m}{n}
\rput(3.668125,0.9789063){\psframebox[linewidth=0.02,fillstyle=solid]{1 -1 1 0 -1}}
\usefont{T1}{ppl}{m}{n}
\rput(6.798125,0.9789063){\psframebox[linewidth=0.02,fillstyle=solid]{1 0 -1 0 1}}
\usefont{T1}{ppl}{m}{n}
\rput(10.079219,0.9789063){\psframebox[linewidth=0.02,fillstyle=solid]{-1 0 1 -1 1}}
\usefont{T1}{ppl}{m}{n}
\rput(3.668125,-1.2210938){\psframebox[linewidth=0.02,fillstyle=solid]{1 0 -1 1 -1}}
\usefont{T1}{ppl}{m}{n}
\rput(6.8792186,-1.2210938){\psframebox[linewidth=0.02,fillstyle=solid]{-1 0 1 0 -1}}
\usefont{T1}{ppl}{m}{n}
\rput(10.079219,-1.2210938){\psframebox[linewidth=0.02,fillstyle=solid]{-1 1 -1 0 1}}
\usefont{T1}{ppl}{m}{n}
\rput(7.010625,-10.021093){\psframebox[linewidth=0.02,fillstyle=solid]{0 0 -1 0 0}}
\usefont{T1}{ppl}{m}{n}
\rput(6.880625,-7.6210938){\psframebox[linewidth=0.02,fillstyle=solid]{0 -1 1 -1 0}}
\usefont{T1}{ppl}{m}{n}
\rput(8.2760935,-5.6210938){\psframebox[linewidth=0.02,fillstyle=solid]{0 -1 0 1 -1}}
\usefont{T1}{ppl}{m}{n}
\rput(5.68375,-5.6210938){\psframebox[linewidth=0.02,fillstyle=solid]{-1 1 0 -1 0}}
\usefont{T1}{ppl}{m}{n}
\rput(2.4026563,-3.4210937){\psframebox[linewidth=0.02,fillstyle=solid]{1 0 0 -1 0}}
\usefont{T1}{ppl}{m}{n}
\rput(11.606093,-3.4210937){\psframebox[linewidth=0.02,fillstyle=solid]{0 -1 0 0 1}}
\usefont{T1}{ppl}{m}{n}
\rput(6.9492188,-3.4210937){\psframebox[linewidth=0.02,fillstyle=solid]{-1 1 -1 1 -1}}
\usefont{T1}{ptm}{m}{n}
\rput(8.134531,9.978907){$\Lambda_{1}$}
\usefont{T1}{ptm}{m}{n}
\rput(4.384531,5.3789062){$\Sigma_{1}$}
\usefont{T1}{ptm}{m}{n}
\rput(8.334531,7.578906){$\Lambda_{2}$}
\usefont{T1}{ptm}{m}{n}
\rput(8.134531,3.1789062){$\Lambda_{3}$}
\usefont{T1}{ptm}{m}{n}
\rput(8.134531,0.9789063){$\Lambda_{4}$}
\usefont{T1}{ptm}{m}{n}
\rput(8.334531,-3.4210937){$\Lambda_{6}$}
\usefont{T1}{ptm}{m}{n}
\rput(8.134531,-7.6210938){$\Lambda_{7}$}
\usefont{T1}{ptm}{m}{n}
\rput(8.134531,-1.2210938){$\Lambda_{5}$}
\usefont{T1}{ptm}{m}{n}
\rput(8.334531,-10.021093){$\Lambda_{8}$}
\usefont{T1}{ptm}{m}{n}
\rput(9.784532,5.3789062){$\Sigma_{2}$}
\usefont{T1}{ptm}{m}{n}
\rput(1.1845312,3.1789062){$\Sigma_{3}$}
\usefont{T1}{ptm}{m}{n}
\rput(2.3845313,0.9789063){$\Sigma_{5}$}
\usefont{T1}{ptm}{m}{n}
\rput(12.784532,3.1789062){$\Sigma_{4}$}
\usefont{T1}{ptm}{m}{n}
\rput(11.384531,0.9789063){$\Sigma_{6}$}
\usefont{T1}{ptm}{m}{n}
\rput(2.3845313,-1.2210938){$\Sigma_{7}$}
\usefont{T1}{ptm}{m}{n}
\rput(1.1845312,-3.4210937){$\Sigma_{9}$}
\usefont{T1}{ptm}{m}{n}
\rput(11.384531,-1.2210938){$\Sigma_{8}$}
\usefont{T1}{ptm}{m}{n}
\rput(12.894531,-3.4210937){$\Sigma_{10}$}
\usefont{T1}{ptm}{m}{n}
\rput(4.2945313,-5.6210938){$\Sigma_{11}$}
\usefont{T1}{ptm}{m}{n}
\rput(9.694531,-5.6210938){$\Sigma_{12}$}
\end{pspicture}
}
\caption{The weights of the $\mathbf{20}$ of $\mathfrak{su}(3,3)$.}
\label{20su(3,3)dyntree}
\end{figure}
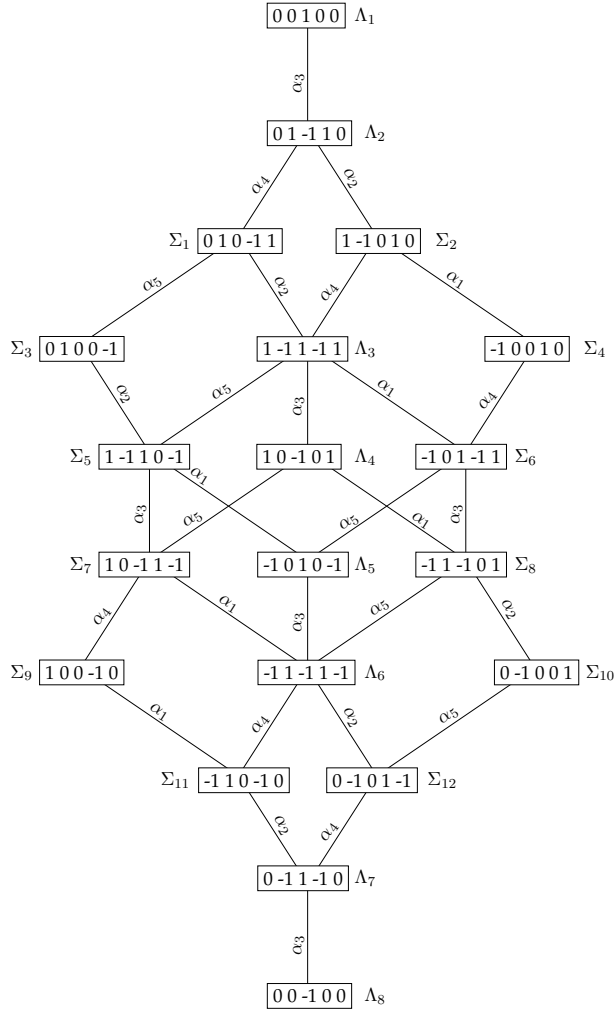

\begin{table}[h!]
\renewcommand{\arraystretch}{1.2}
\par
\begin{center}
\scalebox{0.85}{
\begin{tabular}{|c|c|c|c|c|}
\hline
height&$\ket{\Lambda_{1}}$&$\ket{\Lambda_{4}}$&$\ket{\Lambda_{6}}$&$\ket{\Lambda_{7}}$\\
\hline\hline
5&$E_{\alpha_{1}+\alpha_{2}+\alpha_{3}+\alpha_{4}+\alpha_{5}}$
&$E_{\alpha_{1}+\alpha_{2}+\alpha_{3}+\alpha_{4}+\alpha_{5}}$&&\\
\hline
4&\begin{tabular}{c}
$E_{\alpha_{1}+\alpha_{2}+\alpha_{3}+\alpha_{4}}$\\
$ E_{\alpha_{2}+\alpha_{3}+\alpha_{4}+\alpha_{5}}$
\end{tabular}&
\begin{tabular}{c}
$E_{\alpha_{1}+\alpha_{2}+\alpha_{3}+\alpha_{4}}$\\
$ E_{\alpha_{2}+\alpha_{3}+\alpha_{4}+\alpha_{5}}$
\end{tabular}&
\begin{tabular}{c}
$E_{\alpha_{1}+\alpha_{2}+\alpha_{3}+\alpha_{4}}$\\
$E_{\alpha_{2}+\alpha_{3}+\alpha_{4}+\alpha_{5}}$
\end{tabular}
&\\

\hline

3&
\begin{tabular}{c}
$E_{\alpha_{1}+\alpha_{2}+\alpha_{3}}$\\
$E_{\alpha_{2}+\alpha_{3}+\alpha_{4}}$\\
$E_{\alpha_{3}+\alpha_{4}+\alpha_{5}}$
\end{tabular}&
\begin{tabular}{c}
$E_{\alpha_{1}+\alpha_{2}+\alpha_{3}}$\\
$ E_{\alpha_{3}+\alpha_{4}+\alpha_{5}}$
\end{tabular}&
\begin{tabular}{c}
$E_{\alpha_{2}+\alpha_{3}+\alpha_{4}}$
\end{tabular}&
\begin{tabular}{c}
$E_{\alpha_{1}+\alpha_{2}+\alpha_{3}}$\\
$E_{\alpha_{3}+\alpha_{4}+\alpha_{5}}$
\end{tabular}\\

\hline

2&
\begin{tabular}{c}
$E_{\alpha_{1}+\alpha_{2}}$\\
$E_{\alpha_{2}+\alpha_{3}}$\\
$E_{\alpha_{3}+\alpha_{4}}$\\
$E_{\alpha_{4}+\alpha_{5}}$
\end{tabular}&
\begin{tabular}{c}
$E_{\alpha_{1}+\alpha_{2}}$\\
$ E_{\alpha_{4}+\alpha_{5}}$
\end{tabular}
&
\begin{tabular}{c}
$E_{\alpha_{1}+\alpha_{2}}$\\
$E_{\alpha_{2}+\alpha_{3}}$\\
$E_{\alpha_{3}+\alpha_{4}}$\\
$E_{\alpha_{4}+\alpha_{5}}$
\end{tabular}
&
\begin{tabular}{c}
$E_{\alpha_{2}+\alpha_{3}}$\\
$E_{\alpha_{3}+\alpha_{4}}$
\end{tabular}
\\

\hline
1&
\begin{tabular}{c}
$E_{\alpha_{1}}$\\
$E_{\alpha_{2}}$\\
$E_{\alpha_{3}}$\\
$E_{\alpha_{4}}$\\
$E_{\alpha_{5}}$
\end{tabular}&
\begin{tabular}{c}
$E_{\alpha_{1}}$\\
$E_{\alpha_{2}}$\\
$E_{\alpha_{4}}$\\
$E_{\alpha_{5}}$
\end{tabular}&
\begin{tabular}{c}
$E_{\alpha_{2}}$\\
$E_{\alpha_{4}}$
\end{tabular}&
\begin{tabular}{c}
$E_{\alpha_{1}}$\\
$E_{\alpha_{3}}$\\
$E_{\alpha_{5}}$
\end{tabular}
\\

\hline

0&
\begin{tabular}{c}
$H_{\alpha_{1}}$\\
$H_{\alpha_{2}}$\\
$H_{\alpha_{4}}$\\
$H_{\alpha_{5}}$
\end{tabular}&

\begin{tabular}{c}
$H_{\alpha_{2}}$\\
$H_{\alpha_{4}}$\\
$H_{\alpha_{1}}+H_{\alpha_{3}}$\\
$H_{\alpha_{3}}+H_{\alpha_{5}}$
\end{tabular}&
\begin{tabular}{c}
$H_{\alpha_{1}}+H_{\alpha_{2}}$\\
$H_{\alpha_{2}}+H_{\alpha_{3}}$\\
$H_{\alpha_{3}}+H_{\alpha_{4}}$\\
$H_{\alpha_{4}}+H_{\alpha_{5}}$
\end{tabular}&
\begin{tabular}{c}
$H_{\alpha_{1}}$\\
$H_{\alpha_{5}}$\\
$H_{\alpha_{2}}+H_{\alpha_{3}}$\\
$H_{\alpha_{3}}+H_{\alpha_{4}}$
\end{tabular}
\\

\hline

$-1$&
\begin{tabular}{c}
$E_{-\alpha_{1}}$\\
$E_{-\alpha_{2}}$\\
$E_{-\alpha_{4}}$\\
$E_{-\alpha_{5}}$
\end{tabular}&
\begin{tabular}{c}
$E_{-\alpha_{2}}$\\
$E_{-\alpha_{3}}$\\
$E_{-\alpha_{4}}$
\end{tabular}&
\begin{tabular}{c}
$E_{-\alpha_{1}}$\\
$E_{-\alpha_{3}}$\\
$E_{-\alpha_{5}}$
\end{tabular}&
\begin{tabular}{c}
$E_{-\alpha_{1}}$\\
$E_{-\alpha_{2}}$\\
$E_{-\alpha_{4}}$\\
$E_{-\alpha_{5}}$
\end{tabular}\\

\hline

$-2$&
\begin{tabular}{c}
$E_{-\alpha_{1}-\alpha_{2}}$\\
$E_{-\alpha_{4}-\alpha_{5}}$
\end{tabular}&
\begin{tabular}{c}
$E_{-\alpha_{2}-\alpha_{3}}$\\
$E_{-\alpha_{3}-\alpha_{4}}$
\end{tabular}&
\begin{tabular}{c}
$E_{-\alpha_{1}-\alpha_{2}}$\\
$E_{-\alpha_{2}-\alpha_{3}}$\\
$ E_{-\alpha_{3}-\alpha_{4}}$\\
$ E_{-\alpha_{4}-\alpha_{5}}$
\end{tabular}&
\begin{tabular}{c}
$E_{-\alpha_{1}-\alpha_{2}}$\\
$E_{-\alpha_{2}-\alpha_{3}}$\\
$ E_{-\alpha_{3}-\alpha_{4}}$\\
$ E_{-\alpha_{4}-\alpha_{5}}$
\end{tabular}\\

\hline

$-3$&&
\begin{tabular}{c}
$E_{-\alpha_{1}-\alpha_{2}-\alpha_{3}}$\\
$E_{-\alpha_{2}-\alpha_{3}-\alpha_{4}}$\\
$E_{-\alpha_{3}-\alpha_{4}-\alpha_{5}}$
\end{tabular}&
\begin{tabular}{c}
$E_{-\alpha_{1}-\alpha_{2}-\alpha_{3}}$\\
$E_{-\alpha_{3}-\alpha_{4}-\alpha_{5}}$
\end{tabular}&
\begin{tabular}{c}
$E_{-\alpha_{1}-\alpha_{2}-\alpha_{3}}$\\
$E_{-\alpha_{2}-\alpha_{3}-\alpha_{4}}$\\
$E_{-\alpha_{3}-\alpha_{4}-\alpha_{5}}$
\end{tabular}\\

\hline
$-4$&&
\begin{tabular}{c}
$E_{-\alpha_{1}-\alpha_{2}-\alpha_{3}-\alpha_{4}}$\\
$E_{-\alpha_{2}-\alpha_{3}-\alpha_{4}-\alpha_{5}}$
\end{tabular}&
\begin{tabular}{c}
$E_{-\alpha_{1}-\alpha_{2}-\alpha_{3}-\alpha_{4}}$\\
$E_{-\alpha_{2}-\alpha_{3}-\alpha_{4}-\alpha_{5}}$
\end{tabular}&
\begin{tabular}{c}
$E_{-\alpha_{1}-\alpha_{2}-\alpha_{3}-\alpha_{4}}$\\
$E_{-\alpha_{2}-\alpha_{3}-\alpha_{4}-\alpha_{5}}$
\end{tabular}\\

\hline
$-5$ &&&$E_{-\alpha_{1}-\alpha_{2}-\alpha_{3}-\alpha_{4}-\alpha_{5}}$
&$E_{-\alpha_{1}-\alpha_{2}-\alpha_{3}- \alpha_{4}-\alpha_{5}}$\\
\hline
\end{tabular}
}
\end{center}
\caption{Stabilizers of the real-weight vectors $\ket{\Lambda_1}$, $\ket{\Lambda_4}$, $\ket{\Lambda_6
}$ and $\ket{\Lambda_7}$ in the $\mathbf{20}$ of $\mathfrak{su}(3,3)$.}
\label{20su(3,3)singlestab}
\end{table}

The BH charges of the theory belong to the (rank-3 antisymmetric, self-dual) irrep. $\mathbf{20}$ of $\mathfrak{su}(3,3)$. We draw in \autoref{20su(3,3)dyntree} the Dynkin labels of the
weights of such representation, and as usual we label with $\Lambda$'s the real weights
and with $\Sigma$'s the complex ones. From the figure one sees that there are
eight real weights that, written in terms of the restricted roots, become the
long weights of the $\mathbf{14^\prime}$ of $\mathfrak{sp}(6,\mathbb{R})$.
In particular, the highest weight is
\begin{equation}
\Lambda_1 = \tfrac{1}{2} \alpha_1 + \alpha_2 + \tfrac{3}{2} \alpha_3 +
\alpha_4 + \tfrac{1}{2} \alpha_5 \quad ,
\end{equation}
which can be recast in terms of the restricted roots as $\Lambda_1 = (\alpha_R )_1
+ 2 (\alpha_R )_2 + \tfrac{3}{2} (\alpha_R )_3$, yielding the highest weight
of the $\mathbf{14^\prime}$ of $\mathfrak{sp}(6,\mathbb{R})$, as it can be seen
from \autoref{highestweightofthe14ofsp6}. There are twelve complex roots,
which form six pairs $\Sigma_{2i-1},\Sigma_{2i}$, $i=1,...,6$, where in each
pair the Cartan involution acts according to
\begin{equation}
\theta \Sigma_{2i-1} =-\Sigma_{2i}\quad .
\end{equation}
The projected weights $\tfrac{1}{2}(\Sigma_{2i-1} + \Sigma_{2i})$ become the
six short weights of the $\mathbf{14^\prime}$ of $\mathfrak{sp}(6,\mathbb{R})
$, when written in terms of the restricted roots.

\subsubsection{1-charge orbit}

Following our prescription, we want to obtain the orbits of extremal BH solutions
by considering bound states of only real-weight vectors of the $\mathbf{20}$. From
\autoref{20su(3,3)dyntree}, one deduces that there are four real-weight vectors that
are not connected by transformations of the algebra; in the chosen labelling, these are given by $\ket{\Lambda_1}$, $\ket{\Lambda_4}$, $\ket{\Lambda_6}$ and $\ket{\Lambda_7}$. Their stabilizers are listed in \autoref{20su(3,3)singlestab}; by
looking at each of its columns, one obtains that the stabilizing algebra of each of the real-weight vectors is $\mathfrak{sl}(3,\mathbb{C})\ltimes \mathbb{R}^{9}$, as expected for the
highest-weight, 1-charge (\ie rank-1 \cite{Ferrar,Krutelevich}) U-duality orbit \cite{Borsten:2011ai}. Considering for example the $\ket{\Lambda_1}$ stabilizers,
the semisimple part of the stabilizing algebra is generated by $E_{\pm
(\alpha_1 + \alpha_2 )}$, $E_{\pm (\alpha_4 + \alpha_5)}$, $E_{\pm \alpha_1}$, $E_{\pm \alpha_2}$, $E_{\pm \alpha_4}$, $E_{\pm \alpha_5}$ and the Cartan
generators $H_{\alpha_1}$, $H_{\alpha_2}$, $H_{\alpha_4}$ and $H_{\alpha_5}$, and from the action of the Cartan involution given in \autoref{cartaninvgeneratorssu33} one indeed obtains the real form $\mathfrak{sl}(3,\mathbb{C})$.

\begin{table}[t!]
\renewcommand{\arraystretch}{1.2}
\begin{center}
\begin{tabular}{|c|c|c|}
\hline
Common& $\ket{\Lambda_{1}}+\ket{\Lambda_{4}}$ Conjunction & $\ket{\Lambda_{1}}-\ket{\Lambda_{4}}$ Conjunction\\
\hline
\hline
$E_{\alpha_{1}+\alpha_{2}+\alpha_{3}+\alpha_{4}+\alpha_{5}}$&$E_{\alpha_{2}+\alpha_{3}+\alpha_{4}}-E_{-\alpha_{3}}$ &  $E_{\alpha_{2}+\alpha_{3}+\alpha_{4}}+E_{-\alpha_{3}}$\\
$E_{\alpha_{1}+\alpha_{2}+\alpha_{3}+\alpha_{4}}\quad E_{\alpha_{2}+\alpha_{3}+\alpha_{4}+\alpha_{5}}$&$E_{\alpha_{3}}-E_{-\alpha_{2}-\alpha_{3}-\alpha_{4}}$
&$E_{\alpha_{3}}+E_{-\alpha_{2}-\alpha_{3}-\alpha_{4}}$
\\
$E_{\alpha_{1}+\alpha_{2}+\alpha_{3}}\quad E_{\alpha_{3}+\alpha_{4}+\alpha_{5}}$&$E_{\alpha_{2}+\alpha_{3}}-E_{-\alpha_{3}-\alpha_{4}}$
& $E_{\alpha_{2}+\alpha_{3}}+E_{-\alpha_{3}-\alpha_{4}}$
\\
$E_{\alpha_{1}+\alpha_{2}}\quad E_{\alpha_{4}+\alpha_{5}}$&$E_{\alpha_{3}+\alpha_{4}}-E_{-\alpha_{2}-\alpha_{3}}$
& $E_{\alpha_{3}+\alpha_{4}}+E_{-\alpha_{2}-\alpha_{3}}$
\\
$E_{\alpha_{1}}\quad E_{\alpha_{2}}\quad E_{\alpha_{4}}\quad E_{\alpha_{5}}$&
&
\\
$H_{\alpha_{2}}\quad H_{\alpha_{4}}\quad H_{\alpha_{1}}-H_{\alpha_{5}}$&
&
\\
$E_{-\alpha_{2}}\quad E_{-\alpha_{4}}$&
&
\\
\hline
\end{tabular}

\caption{The stabilizers of the $\ket{\Lambda_{1}}+\ket{\Lambda_{4}}$ and $\ket{\Lambda_{1}}-\ket{\Lambda_{4}}$ bound states, yielding the algebras  $[\mathfrak{so}(1,4)\oplus \mathfrak{so}(2)]\ltimes (\mathbb{R}\times\mathbb{R}^{(4,2)})$ and $[\mathfrak{so}(2,3)\oplus \mathfrak{so}(2)]\ltimes (\mathbb{R}\times\mathbb{R}^{(4,2)})$, respectively. }\label{l1+l420su33stab}
\end{center}
\end{table}

\begin{table}[b!]
\renewcommand{\arraystretch}{1.3}
\begin{center}
\scalebox{0.9}{
\begin{tabular}{|c||c|c|c|}
\hline
Common &2-conj. & $\ket{\Lambda_{1}}+\ket{\Lambda_{4}}+\ket{\Lambda_{6}}$   &$\ket{\Lambda_{1}}+\ket{\Lambda_{4}}-\ket{\Lambda_{6}}$ \\
\hline
\hline
$ E_{\alpha_1 + \alpha_2 +\alpha_3 + \alpha_4}$& & $E_{\alpha_{2}+\alpha_{3}+\alpha_{4}}-E_{-\alpha_{3}}$& $E_{\alpha_{2}+\alpha_{3}+\alpha_{4}}-E_{-\alpha_{3}}$\\
$E_{\alpha_2 + \alpha_3 +\alpha_4 + \alpha_5}$ &  & $E_{\alpha_{2}+\alpha_{3}}-E_{-\alpha_{3}-\alpha_{4}}$ & $E_{\alpha_{2}+\alpha_{3}}-E_{-\alpha_{3}-\alpha_{4}}$\\
$E_{\alpha_1 + \alpha_2}$ & \multirow{-3}{*}{\rotatebox{90}{$\Lambda_{1},\Lambda_{4}$}}&$E_{\alpha_{3}+\alpha_{4}}-E_{-\alpha_{2}-\alpha_{3}}$ & $E_{\alpha_{3}+\alpha_{4}}-E_{-\alpha_{2}-\alpha_{3}}$\\
\cline{2-4}
 $E_{\alpha_4+ \alpha_5}$ & &$E_{\alpha_{1}+\alpha_{2}+\alpha_{3}+\alpha_{4}+\alpha_{5}}-E_{-\alpha_{3}}$ & $E_{\alpha_{1}+\alpha_{2}+\alpha_{3}+\alpha_{4}+\alpha_{5}}+ E_{-\alpha_{3}}$\\
$E_{\alpha_2}$& &$E_{\alpha_{1}+\alpha_{2}+\alpha_{3}}-E_{-\alpha_{3}-\alpha_{4}-\alpha_{5}}$&  $E_{\alpha_{1}+\alpha_{2}+\alpha_{3}}+E_{-\alpha_{3}-\alpha_{4}-\alpha_{5}}$\\
$E_{\alpha_4}$ & \multirow{-3}{*}{\rotatebox{90}{$\Lambda_{1},\Lambda_{6}$}}&$E_{\alpha_{3}+\alpha_{4}+\alpha_{5}}-E_{-\alpha_{1}-\alpha_{2}-\alpha_{3}}$&  $E_{\alpha_{3}+\alpha_{4}+\alpha_{5}} + E_{-\alpha_{1}-\alpha_{2}-\alpha_{3}}$\\
\cline{2-4}
$H_{\alpha_1} - H_{\alpha_5}$& &$E_{\alpha_{2}+\alpha_{3}+\alpha_{4}}-E_{\alpha_{1}+\alpha_{2}+\alpha_{3}+\alpha_{4}+\alpha_{5}}$&  $E_{\alpha_{2}+\alpha_{3}+\alpha_{4}} + E_{\alpha_{1}+\alpha_{2}+\alpha_{3}+\alpha_{4}+\alpha_{5}}$\\
$H_{\alpha_2} - H_{\alpha_4}$  &
&$E_{\alpha_{5}}-E_{-\alpha_{1}}$&  $E_{\alpha_{5}}+ E_{-\alpha_{1}}$\\
& \multirow{-3}{*}{\rotatebox{90}{$\Lambda_{4},\Lambda_{6}$}}&$E_{\alpha_{1}}-E_{-\alpha_{5}}$&  $E_{\alpha_{1}}+ E_{-\alpha_{5}}$\\
\hline
\end{tabular}
}
\caption{The stabilizers of the $\ket{\Lambda_{1}}+\ket{\Lambda_{4}}+\ket{\Lambda_{6}}$ and $\ket{\Lambda_{1}}+\ket{\Lambda_{4}}-\ket{\Lambda_{6}}$ bound states. The 2-conjunction stabilizers connect the two weight vectors whose weights are listed in the second  column, and annihilate the third weight vector. The stabilizing algebra is $\mathfrak{su}(3)\ltimes \mathbb{R}^{8}$ for the  $\ket{\Lambda_{1}}+\ket{\Lambda_{4}}+\ket{\Lambda_{6}}$ bound state, and $\mathfrak{su}(1,2)\ltimes \mathbb{R}^{8}$ for the  $\ket{\Lambda_{1}}+\ket{\Lambda_{2}}-\ket{\Lambda_{3}}$ bound state.} \label{l1+l4+l6stabsu33}

\end{center}
\end{table}

\subsubsection{2-charge orbits}

We now consider the ``small'' 2-charge (\ie rank-2 \cite{Ferrar,Krutelevich}) orbits as associated to the bound states $\ket{\Lambda_{1}}+\ket{\Lambda_{4}}$ and $\ket{\Lambda_{1}}-\ket{\Lambda_{4}}$. We list in \autoref{l1+l420su33stab} the generators that stabilize the two bound states. The
semisimple part of the stabilizing algebra is generated by the common
stabilizers $E_{\pm \alpha_2}$, $E_{\pm \alpha_4}$, $H_{\alpha_2}$, $H_{\alpha_4}$ and $H_{\alpha_1} - H_{\alpha_5}$, together with all the
conjunction stabilizers. From \autoref{cartaninvgeneratorssu33}, one deduces
the action of the Cartan involution on all these generators. In particular, $H_{\alpha_1} - H_{\alpha_5}$ is compact and generates $\mathfrak{so}(2)$.
The stabilizers $E_{\pm \alpha_2}$, $E_{\pm \alpha_4}$, $H_{\alpha_2}$, $H_{\alpha_4}$, $E_{\alpha_2 + \alpha_3 + \alpha_4} \pm E_{-\alpha_3}$ and $E_{\alpha_3} \pm E_{-\alpha_2 -\alpha_3 -\alpha_4}$ give four compact and
four non-compact generators, while the remaining two conjunction stabilizers
are compact in the $\ket{\Lambda_{1}}+\ket{\Lambda_{4}}$ case and non-compact in the $\ket{\Lambda_{1}}-\ket{\Lambda_{4}}$ case, so that one obtains the ten generators of $\mathfrak{so}(1,4)\sim \mathfrak{usp}\left( 2,2\right)$ and $\mathfrak{so}(2,3)\sim \mathfrak{sp}\left( 4,\mathbb{R})\right)$, respectively. Together with the
remaining stabilizers, this leads to the algebras $[\mathfrak{so}(1,4)\oplus
\mathfrak{so}(2)]\ltimes (\mathbb{R}\times\mathbb{R}^{(4,2)})$ and $[
\mathfrak{so}(2,3)\oplus \mathfrak{so}(2)]\ltimes (\mathbb{R}\times\mathbb{R}^{(4,2)})$, which are the stabilizing algebras for the 1/2-BPS and the
non-supersymmetric 2-charge orbits of this theory \cite{Borsten:2011ai}.

\begin{table}[t!]
\renewcommand{\arraystretch}{1.3}
\par
\begin{center}
\scalebox{0.7}{
\begin{tabular}{|c||c|c|c|c|}
\hline
Common & 2-c. &$\ket{\Lambda_1} + \ket{\Lambda_4} + \ket{\Lambda_6} + \ket{\Lambda_7}$ & $\ket{\Lambda_1} +\ket{ \Lambda_4} +\ket{ \Lambda_6} - \ket{\Lambda_7}$ & $\ket{\Lambda_1} +\ket{ \Lambda_4 }- \ket{\Lambda_6 }- \ket{\Lambda_7}$ \\
\hline
\hline
$H_{\alpha_2} - H_{\alpha_4}$ & &$E_{\alpha_2 + \alpha_3} - E_{-\alpha_3 - \alpha_4}$& $E_{\alpha_2 + \alpha_3} - E_{-\alpha_3 - \alpha_4}$ & $E_{\alpha_2 + \alpha_3} - E_{-\alpha_3 - \alpha_4}$\\
$H_{\alpha_1} - H_{\alpha_5}$ & \multirow{-2}{*}{\rotatebox{90}{$\Lambda_{1},\Lambda_{4}$}} & $E_{\alpha_3 + \alpha_4} - E_{-\alpha_2 - \alpha_3}$& $E_{\alpha_3 + \alpha_4} - E_{-\alpha_2 - \alpha_3}$ & $E_{\alpha_3 + \alpha_4} - E_{-\alpha_2 - \alpha_3}$
\\
& & $F^-_{\alpha_3} + F^-_{\alpha_2 + \alpha_3 + \alpha_4}$ & $F^+_{\alpha_3} - F^+_{\alpha_2 + \alpha_3 + \alpha_4}$ & $F^-_{\alpha_3} + F^-_{\alpha_2 + \alpha_3 + \alpha_4}$ \\
\cline{2-5}
&   &$E_{\alpha_1 + \alpha_2 + \alpha_3} - E_{-\alpha_3 - \alpha_4 - \alpha_5}$& $E_{\alpha_1+ \alpha_2 + \alpha_3} - E_{-\alpha_3 - \alpha_4 - \alpha_5}$ & $E_{\alpha_1 + \alpha_2 + \alpha_3} + E_{-\alpha_3 - \alpha_4 - \alpha_5}$\\
 & \multirow{-2}{*}{\rotatebox{90}{$\Lambda_{1},\Lambda_{6}$}}  & $E_{\alpha_3 + \alpha_4 + \alpha_5 } - E_{- \alpha_1 -\alpha_2 - \alpha_3}$& $E_{\alpha_3 + \alpha_4 + \alpha_5} - E_{-\alpha_1 -\alpha_2 - \alpha_3}$ & $E_{\alpha_3 + \alpha_4 + \alpha_5} + E_{-\alpha_1 -\alpha_2 - \alpha_3}$
\\
& & $F^-_{\alpha_3} + F^-_{\alpha_1 +\alpha_2 + \alpha_3 + \alpha_4 + \alpha_5}$ & $F^+_{\alpha_3} - F^+_{\alpha_1 + \alpha_2 + \alpha_3 + \alpha_4 + \alpha_5 }$ & $F^-_{\alpha_3} - F^-_{\alpha_1 +\alpha_2 + \alpha_3 + \alpha_4 +\alpha_5 }$ \\
\cline{2-5}
&   &$E_{\alpha_1 + \alpha_2 + \alpha_3 +\alpha_4} - E_{-\alpha_2 -\alpha_3 - \alpha_4 - \alpha_5}$& $E_{\alpha_1+ \alpha_2 + \alpha_3 + \alpha_4} + E_{-\alpha_2 -\alpha_3 - \alpha_4 - \alpha_5}$ & $E_{\alpha_1 + \alpha_2 + \alpha_3 + \alpha_4} + E_{-\alpha_2 -\alpha_3 - \alpha_4 - \alpha_5}$\\
 & \multirow{-2}{*}{\rotatebox{90}{$\Lambda_{1},\Lambda_{7}$}}  & $E_{\alpha_2 +\alpha_3 + \alpha_4 + \alpha_5 } - E_{- \alpha_1 -\alpha_2 - \alpha_3 -\alpha_4}$& $E_{\alpha_2+ \alpha_3 + \alpha_4 + \alpha_5} + E_{-\alpha_1 -\alpha_2 - \alpha_3 - \alpha_4}$ & $E_{\alpha_2+ \alpha_3 + \alpha_4 + \alpha_5} + E_{-\alpha_1 -\alpha_2 - \alpha_3 - \alpha_4}$
\\
& & $F^-_{\alpha_2+ \alpha_3 + \alpha_4} - F^-_{\alpha_1 +\alpha_2 + \alpha_3 + \alpha_4 + \alpha_5}$ & $F^+_{\alpha_2+ \alpha_3+ \alpha_4} - F^+_{\alpha_1 + \alpha_2 + \alpha_3 + \alpha_4 + \alpha_5 }$ & $F^-_{\alpha_2 + \alpha_3+ \alpha_4} + F^-_{\alpha_1 +\alpha_2 + \alpha_3 + \alpha_4 +\alpha_5 }$ \\
\cline{2-5}
&   &$E_{\alpha_1} - E_{-\alpha_5}$& $E_{\alpha_1} - E_{-\alpha_5}$ & $E_{\alpha_1} + E_{-\alpha_5}$\\
 & \multirow{-2}{*}{\rotatebox{90}{$\Lambda_{4},\Lambda_{6}$}}  & $E_{\alpha_5 } - E_{- \alpha_1}$& $E_{\alpha_5} - E_{-\alpha_1}$ & $E_{\alpha_5} + E_{-\alpha_1}$
\\
& & $F^-_{\alpha_2+ \alpha_3 + \alpha_4} - F^-_{\alpha_1 +\alpha_2 + \alpha_3 + \alpha_4 + \alpha_5}$ & $F^+_{\alpha_2+ \alpha_3+ \alpha_4} - F^+_{\alpha_1 + \alpha_2 + \alpha_3 + \alpha_4 + \alpha_5 }$ & $F^-_{\alpha_2 + \alpha_3+ \alpha_4} + F^-_{\alpha_1 +\alpha_2 + \alpha_3 + \alpha_4 +\alpha_5 }$ \\
\cline{2-5}
&   &$E_{\alpha_1 + \alpha_2 } - E_{ - \alpha_4 - \alpha_5}$& $E_{\alpha_1+ \alpha_2 } + E_{ - \alpha_4 - \alpha_5}$ & $E_{\alpha_1 + \alpha_2 } + E_{ - \alpha_4 - \alpha_5}$\\
 & \multirow{-2}{*}{\rotatebox{90}{$\Lambda_{4},\Lambda_{7}$}}  & $E_{ \alpha_4 + \alpha_5 } - E_{- \alpha_1 -\alpha_2 }$& $E_{ \alpha_4 + \alpha_5} + E_{-\alpha_1 -\alpha_2 }$ & $E_{\alpha_4 + \alpha_5} + E_{-\alpha_1 -\alpha_2 }$
\\
& & $F^-_{\alpha_3} + F^-_{\alpha_1 +\alpha_2 + \alpha_3 + \alpha_4 + \alpha_5}$ & $F^+_{\alpha_3} - F^+_{\alpha_1 + \alpha_2 + \alpha_3 + \alpha_4 + \alpha_5 }$ & $F^-_{\alpha_3} - F^-_{\alpha_1 +\alpha_2 + \alpha_3 + \alpha_4 +\alpha_5 }$ \\
\cline{2-5}
 & &$E_{\alpha_2 } - E_{ - \alpha_4}$& $E_{\alpha_2 } + E_{- \alpha_4}$ & $E_{\alpha_2 } - E_{ - \alpha_4}$\\
 & \multirow{-2}{*}{\rotatebox{90}{$\Lambda_{6},\Lambda_{7}$}} & $E_{ \alpha_4} - E_{-\alpha_2 }$& $E_{ \alpha_4} + E_{-\alpha_2 }$ & $E_{ \alpha_4} - E_{-\alpha_2 }$
\\
& & $F^-_{\alpha_3} + F^-_{\alpha_2 + \alpha_3 + \alpha_4}$ & $F^+_{\alpha_3} - F^+_{\alpha_2 + \alpha_3 + \alpha_4}$ & $F^-_{\alpha_3} + F^-_{\alpha_2 + \alpha_3 + \alpha_4}$ \\
\hline
\end{tabular}
}
\end{center}
\caption{The stabilizers of the 4-charge orbits of the  $\mathbf{20}$ of $\mathfrak{su}(3,3)$. In the second column we
list the pair of states for which the operators in the first two lines of each row
are  2-conjunction stabilizers. In any column, there are only two independent
generators among those in the third  line of each row. The stabilizing
algebras are $\mathfrak{su}(3)\oplus \mathfrak{su}(3)$, $\mathfrak{sl}(3,\mathbb{C})$ and $\mathfrak{su}(1,2)\oplus \mathfrak{su}(1,2) $. }
\label{fourchargeorbits20su33}
\end{table}

\subsubsection{3-charge orbits}

We then move to the ``small'' 3-charge (\ie rank-3 \cite{Ferrar,Krutelevich}) orbits, corresponding to the bound states $\ket{\Lambda_{1}}+\ket{\Lambda_{4}}+\ket{ \Lambda_{6}}$ and $\ket{\Lambda_{1}}+\ket{\Lambda_{4}} -
\ket{\Lambda_{6}}$. The stabilizers of the bound states are listed in \autoref{l1+l4+l6stabsu33}. In the third and fourth column of the table,
we list the conjunction stabilizers for two of the weight vectors (whose weights are given in the second column) that annihilate
the third. The semisimple part of the algebra is generated by the two
Cartan among the common stabilizers and by the conjunction stabilizers in the
second and third line of each row. One can check that in the $\ket{\Lambda_{1}}+\ket{\Lambda_{4}}+\ket{ \Lambda_{6}}$ case all these generators are compact,
while in the $\ket{\Lambda_{1}}+\ket{\Lambda_{4}} - \ket{\Lambda_{6}}$ case there are four compact
and four non-compact generators. Both in the third and fourth column, among the three conjunction stabilizers in
the first line of each row only two are independent. The resulting
stabilizing algebras are $\mathfrak{su}(3)\ltimes \mathbb{R}^{8}$ for the $\ket{\Lambda_{1}}+\ket{\Lambda_{4}}+\ket{\Lambda_{6}}$ bound state, and $\mathfrak{su}(1,2)\ltimes \mathbb{R}^{8}$ for the $\ket{\Lambda_{1}}+\ket{\Lambda_{4}}-\ket{\Lambda_{6}}$
bound state. From comparison with the literature \cite{Borsten:2011ai}, the former corresponds to the 1/2-BPS orbit, while the latter is non-supersymmetric (non-BPS).

\begin{table}[h!]
\renewcommand{\arraystretch}{1.2}
\par
\begin{center}
\scalebox{0.9}{
\begin{tabular}{|c|c|c|c|}
\hline
height &$\ket{\Sigma_{11}}$&$\ket{\Sigma_{12}}$&$\ket{\Lambda_{8}}$\\
\hline\hline

4&\begin{tabular}{c}
$ E_{\alpha_{2}+\alpha_{3}+\alpha_{4}+\alpha_{5}}$
\end{tabular}&
\begin{tabular}{c}
$E_{\alpha_{1}+\alpha_{2}+\alpha_{3}+\alpha_{4}}$
\end{tabular}

&\\

\hline

3&\begin{tabular}{c}
$E_{\alpha_{1}+\alpha_{2}+\alpha_{3}}$\\
$E_{\alpha_{2}+\alpha_{3}+\alpha_{4}}$
\end{tabular}&
\begin{tabular}{c}
$E_{\alpha_{3}+\alpha_{4}+\alpha_{5}}$\\
$E_{\alpha_{2}+\alpha_{3}+\alpha_{4}}$
\end{tabular}
&\\

\hline

2&\begin{tabular}{c}
$E_{\alpha_{1}+\alpha_{2}}$\\
$E_{\alpha_{2}+\alpha_{3}}$
\end{tabular}&
\begin{tabular}{c}
$E_{\alpha_{3}+\alpha_{4}}$\\
$E_{\alpha_{4}+\alpha_{5}}$
\end{tabular}
&
\begin{tabular}{c}
$E_{\alpha_{1}+\alpha_{2}}$\\
$E_{\alpha_{4}+\alpha_{5}}$
\end{tabular}\\

\hline
1&\begin{tabular}{c}
$E_{\alpha_{2}}$\\
$E_{\alpha_{3}}$\\
$E_{\alpha_{5}}$
\end{tabular}&
\begin{tabular}{c}
$E_{\alpha_{1}}$\\
$E_{\alpha_{3}}$\\
$E_{\alpha_{4}}$
\end{tabular}
&
\begin{tabular}{c}
$E_{\alpha_{1}}$\\
$E_{\alpha_{2}}$\\
$E_{\alpha_{4}}$\\
$E_{\alpha_{5}}$
\end{tabular}
\\

\hline

0&\begin{tabular}{c}
$H_{\alpha_{3}}$\\
$H_{\alpha_{5}}$\\
$H_{\alpha_{1}}+H_{\alpha_{2}}$\\
$H_{\alpha_{2}}+H_{\alpha_{4}}$
\end{tabular}
&
\begin{tabular}{c}
$H_{\alpha_{1}}$\\
$H_{\alpha_{3}}$\\
$H_{\alpha_{2}}+H_{\alpha_{4}}$\\
$H_{\alpha_{4}}+H_{\alpha_{5}}$
\end{tabular}
&
\begin{tabular}{c}
$H_{\alpha_{1}}$\\
$H_{\alpha_{2}}$\\
$H_{\alpha_{4}}$\\
$H_{\alpha_{5}}$
\end{tabular}\\

\hline

$-1$ &
\begin{tabular}{c}
$E_{-\alpha_{1}}$\\
$E_{-\alpha_{3}}$\\
$E_{-\alpha_{4}}$\\
$E_{-\alpha_{5}}$
\end{tabular}&

\begin{tabular}{c}
$E_{-\alpha_{1}}$\\
$E_{-\alpha_{2}}$\\
$E_{-\alpha_{3}}$\\
$E_{-\alpha_{5}}$
\end{tabular}
&
\begin{tabular}{c}
$E_{-\alpha_{1}}$\\
$E_{-\alpha_{2}}$\\
$E_{-\alpha_{3}}$\\
$E_{-\alpha_{4}}$\\
$E_{-\alpha_{5}}$
\end{tabular}\\

\hline

$-2$&
\begin{tabular}{c}
$E_{-\alpha_{1}-\alpha_{2}}$\\
$E_{-\alpha_{3}-\alpha_{4}}$\\
$E_{-\alpha_{4}-\alpha_{5}}$
\end{tabular}&
\begin{tabular}{c}
$E_{-\alpha_{1}-\alpha_{2}}$\\
$E_{-\alpha_{2}-\alpha_{3}}$\\
$E_{-\alpha_{4}-\alpha_{5}}$
\end{tabular}
&\begin{tabular}{c}
$E_{-\alpha_{1}-\alpha_{2}}$\\
$E_{-\alpha_{2}-\alpha_{3}}$\\
$ E_{-\alpha_{3}-\alpha_{4}}$\\
$ E_{-\alpha_{4}-\alpha_{5}}$
\end{tabular}\\

\hline

$-3$&
\begin{tabular}{c}
$E_{-\alpha_{1}-\alpha_{2}-\alpha_{3}}$\\
$E_{-\alpha_{2}-\alpha_{3}-\alpha_{4}}$\\
$E_{-\alpha_{3}-\alpha_{4}-\alpha_{5}}$
\end{tabular}&
\begin{tabular}{c}
$E_{-\alpha_{1}-\alpha_{2}-\alpha_{3}}$\\
$E_{-\alpha_{2}-\alpha_{3}-\alpha_{4}}$\\
$E_{-\alpha_{3}-\alpha_{4}-\alpha_{5}}$
\end{tabular}
&\begin{tabular}{c}
$E_{-\alpha_{1}-\alpha_{2}-\alpha_{3}}$\\
$E_{-\alpha_{2}-\alpha_{3}-\alpha_{4}}$\\
$E_{-\alpha_{3}-\alpha_{4}-\alpha_{5}}$
\end{tabular}\\

\hline
$-4$ &\begin{tabular}{c}
$E_{-\alpha_{1}-\alpha_{2}-\alpha_{3}-\alpha_{4}}$\\
$E_{-\alpha_{2}-\alpha_{3}-\alpha_{4}-\alpha_{5}}$
\end{tabular}&
\begin{tabular}{c}
$E_{-\alpha_{1}-\alpha_{2}-\alpha_{3}-\alpha_{4}}$\\
$E_{-\alpha_{2}-\alpha_{3}-\alpha_{4}-\alpha_{5}}$
\end{tabular}
&\begin{tabular}{c}
$E_{-\alpha_{1}-\alpha_{2}-\alpha_{3}-\alpha_{4}}$\\
$E_{-\alpha_{2}-\alpha_{3}-\alpha_{4}-\alpha_{5}}$
\end{tabular}\\

\hline
$-5$ &$E_{-\alpha_{1}-\alpha_{2}-\alpha_{3}-\alpha_{4}-\alpha_{5}}$
&$E_{-\alpha_{1}-\alpha_{2}-\alpha_{3}-\alpha_{4}-\alpha_{5}}$
&$E_{-\alpha_{1}-\alpha_{2}-\alpha_{3}-\alpha_{4}-\alpha_{5}}$\\
\hline
\end{tabular}
}
\end{center}
\caption{The stabilizers of the complex-weight vectors $\ket{\Sigma_{11}}$, $\ket{\Sigma_{12}}$ and of the lowest-weight vector $\Lambda_8$ of the $\mathbf{20}$ of $\mathfrak{su}(3,3)$.}
\label{20su(3,3)sigmaandlambda8stab}
\end{table}

\subsubsection{4-charge orbits}

Moving to the 4-charge (\ie rank-4 \cite{Ferrar,Krutelevich}) ``large'' orbits, the only common stabilizers of the states $\ket{\Lambda_1}$, $\ket{\Lambda_4}$, $\ket{\Lambda_6}$ and $\ket{\Lambda_7}$ are the Cartan
generators $H_{\alpha_2} - H_{\alpha_4}$ and $H_{\alpha_1} - H_{\alpha_5}$,
which are both compact. Considering the pair $\ket{\Lambda_1} ,\ket{\Lambda_4}$, only
the third and fourth conjunction stabilizers in the second column of \autoref{l1+l420su33stab} annihilate both $\ket{\Lambda_6}$ and $\ket{\Lambda_7}$, while the sum
of the first two is a conjunction stabilizer for the pair $\ket{\Lambda_6}
,\ket{\Lambda_7}$. Repeating the same analysis for any other pair of weight vectors, one
derives the generators that are listed in \autoref{fourchargeorbits20su33}.
For each of the third, fourth and fifth column, only two of the generators in the last line of each row are independent. It
is easy to check the compactness of the stabilizers, and one can show that
they generate $\mathfrak{su}(3)\oplus \mathfrak{su}(3)$, $\mathfrak{sl}(3,\mathbb{C})$ and $\mathfrak{su}(1,2)\oplus \mathfrak{su}(1,2) $. By comparing with the literature \cite{Bellucci:2006xz,Borsten:2011ai}, the first
case is the stabilizer of the ``large'' time-like 1/2-BPS orbit, the second one is the stabilizer of the ``large'' space-like (dyonic) non-supersymmetric (non-BPS) orbit, and the third one stabilizes the ``large''
time-like non-supersymmetric (non-BPS) orbit.

Exactly as for the $\mathcal{N}=2$, $D=4$ Maxwell-Einstein supergravity theory based on $J_{3}^{\mathbb{R}}$ treated in \autoref{4dimtheoryorbits}, it can be checked that in the 4-charge dyonic orbit there are both conjunction stabilizers on
real and on complex-weight vectors, that change their compactness with respect to the
supersymmetric ``large'' orbit. Instead, the generators that change their compactness in the
non-supersymmetric (non-BPS) 4-charge $\ket{\Lambda_1} + \ket{\Lambda_4} -\ket{\Lambda_6} -\ket{\Lambda_7}$ orbit (compared
the the supersymmetric one) are only conjunction stabilizers on complex-weight vectors.

\begin{table}[t!]
\renewcommand{\arraystretch}{1.2}
\begin{center}
\begin{tabular}{|c|}
\hline
{$\ket{\Lambda_{1}}+\ket{\Lambda_{8}}$ Stabilizers (Common)}\\
\hline
\hline
$E_{\alpha_{1}+\alpha_{2}}\quad E_{\alpha_{4}+\alpha_{5}}$\\
$E_{\alpha_{1}}\quad E_{\alpha_{2}}\quad E_{\alpha_{4}}\quad E_{\alpha_{5}}$\\
$H_{\alpha_{1}}\quad H_{\alpha_{2}}\quad H_{\alpha_{4}}\quad H_{\alpha_{5}}$\\
$E_{-\alpha_{1}}\quad E_{-\alpha_{2}}\quad E_{-\alpha_{4}}\quad E_{-\alpha_{5}}$\\
$E_{-\alpha_{1}-\alpha_{2}}\quad E_{-\alpha_{4}-\alpha_{5}}$\\
\hline
\end{tabular}

\caption{The stabilizers of the bound state $\ket{\Lambda_{1}}+\ket{\Lambda_{8}}$, forming the algebra $\mathfrak{sl}(3,\mathbb{C})$.}\label{dyonic20su33stab}
\end{center}
\end{table}

\begin{table}[t!]
\renewcommand{\arraystretch}{1.5}
\begin{center}
\begin{tabular}{|c|c|c|}
\hline
\multicolumn{2}{|c|}{{States}}&{stabilizer}\\ \hline\hline
\multirow{-1}{*}{\rotatebox{90}{1-s}}&$\ket{\Lambda_{1}}$&$\mathfrak{sl}(3,\mathbb{C})\ltimes \mathbb{R}^{9}$\\ \hline
&$\ket{\Lambda_{1}}+\ket{\Lambda_{4}}$&$[\mathfrak{so}(1,4)\oplus \mathfrak{so}(2)]\ltimes (\mathbb{R}\times\mathbb{R}^{(4,2)})$\\
&$\ket{\Lambda_{1}}-\ket{\Lambda_{4}}$&$[\mathfrak{so}(2,3)\oplus \mathfrak{so}(2)]\ltimes (\mathbb{R}\times\mathbb{R}^{(4,2)})$\\
\multirow{-3}{*}{\rotatebox{90}{2-s}}&$\ket{\Sigma_{11}}+\ket{\Sigma_{12}}$
&$[\mathfrak{so}(2,3)\oplus \mathfrak{so}(2)]\ltimes (\mathbb{R}\times\mathbb{R}^{(4,2)})$\\ \hline
&$\ket{\Lambda_{1}}+\ket{\Lambda_{4}}+\ket{\Lambda_{6}}$&$\mathfrak{su}(3)\ltimes \mathbb{R}^{8}$\\
&$\ket{\Lambda_{1}}+\ket{\Lambda_{4}}-\ket{\Lambda_{6}}$ &$\mathfrak{su}(1,2)\ltimes \mathbb{R}^{8}$\\
&$\ket{\Lambda_{1}}+\ket{\Sigma_{11}}+\ket{\Sigma_{12}}$ &$\mathfrak{su}(1,2)\ltimes \mathbb{R}^{8}$\\
\multirow{-4}{*}{\rotatebox{90}{3-s}}&$\ket{\Lambda_{1}}-(\ket{\Sigma_{11}}+\ket{\Sigma_{12}})$
&$\mathfrak{su}(1,2)\ltimes \mathbb{R}^{8}$\\ \hline

&$\ket{\Lambda_{1}}+\ket{\Lambda_{4}}+\ket{\Lambda_{6}}+\ket{\Lambda_{7}}$ &$\mathfrak{su}(3)\oplus \mathfrak{su}(3)$\\
&$\ket{\Lambda_{1}}+\ket{\Lambda_{4}}+\ket{\Lambda_{6}}-\ket{\Lambda_{7}}$ &$\mathfrak{sl}(3,\mathbb{C})$\\
&$\ket{\Lambda_{1}}+\ket{\Lambda_{4}}-\ket{\Lambda_{6}}-\ket{\Lambda_{7}}$ &$\mathfrak{su}(1,2)\oplus \mathfrak{su}(1,2)$\\
&$\ket{\Lambda_{1}}+\ket{\Lambda_{4}}+\ket{\Sigma_{11}}+\ket{\Sigma_{12}}$ &$\mathfrak{sl}(3,\mathbb{C})$\\
&$\ket{\Lambda_{1}}+\ket{\Lambda_{4}}-(\ket{\Sigma_{11}}+\ket{\Sigma_{12}})$ &$\mathfrak{sl}(3,\mathbb{C})$\\
&$\ket{\Lambda_{1}}-\ket{\Lambda_{4}}+\ket{\Sigma_{11}}+\ket{\Sigma_{12}}$ &$\mathfrak{su}(1,2)\oplus \mathfrak{su}(1,2)$\\
\multirow{-7}{*}{\rotatebox{90}{4-s}}&$\ket{\Lambda_{1}}+\ket{\Lambda_{8}}$&$\mathfrak{sl}(3,\mathbb{C})$\\

\hline
\end{tabular}

\caption{Stabilizers of the different bound states in the ${\bf 20}$ of $\mathfrak{su}(3,3)$.}\label{20su33summary}
\end{center}
\end{table}

The analogy with the theory based on $J_{3}^{\mathbb{R}}$ discussed in the previous section can be further carried out. Indeed, one can show that the
non-supersymmetric
orbits can also be derived as bound states involving real combinations of complex-weight vectors. Moreover, the 4-charge dyonic orbit can also be
derived as a bound state of the highest-weight vector and the lowest-weight vector. In \autoref{20su(3,3)sigmaandlambda8stab}, we list the stabilizers of two
independent complex-weight vectors, that (without any loss of generality) we take to be $\ket{\Sigma_{11}}$ and $\ket{\Sigma_{12}}$, and of the
lowest-weight vector $\ket{\Lambda_8}$. By following the procedure established in the treatment above, the stabilizers and their compactness in the various cases can be determined, and it can be shown that
the signature of the resulting real form of the stabilizing algebra is
consistent with the general interpretation that the sum $\ket{\Sigma_{11}} +
\ket{\Sigma_{12}}$ is equivalent to the difference of two real-weight vectors. As far as
the $\ket{\Lambda_1} +\ket{\Lambda_8}$ orbit is concerned, we list in \autoref{dyonic20su33stab} its stabilizers, which form the algebra $\mathfrak{sl}(3,
\mathbb{C})$. As for the same realization of the representative of the 4-charge dyonic non-supersymmetric space-like orbit of the $\mathbf{14}^{\prime }$ of $\mathfrak{sp}(6,\mathbb{R})$, also in this case there are no conjunction stabilizers, but only common stabilizers.

To summarize, we list in \autoref{20su33summary} the stabilizing algebras for the various bound states of the ${\bf 20}$ of $\mathfrak{su}(3,3)$ which have been realized above, also including the aforementioned bound states involving real combinations of complex-weight vectors. The results match the ones reported in Table VI of \cite{Borsten:2011ai}.

\vskip 1cm

A completely analogous analysis, based on the  projection to the restricted-root subalgebra of the U-duality Lie algebra, can be performed for the other two magic theories, based $J_{3}^{\mathbb{H}}$ and $J_{3}^{\mathbb{O}}$, in $D=5$ and $D=4$. We leave it to the reader to verify that in all cases the results that one obtains match what is already known in the literature.

\section{\label{sectioninfinieseries}$\mathcal{N}=4$ and $\mathcal{N}=2$ supergravities based on $\mathbb{R}\oplus \mathbf{\Gamma }_{m-1,n-1}$}

In this section we consider the $\mathcal{N}=4$ and $\mathcal{N}=2$ supergravity theories based on the infinite sequences of semisimple rank-3 Jordan algebras $\mathbb{R}\oplus \mathbf{\Gamma }_{m-1,n-1}$, where $\mathbf{\Gamma }_{m-1,n-1}$ is the Clifford algebra of $O(m-1,n-1)$ \cite{JNW}. In the $\mathcal{N}=4$ case (corresponding to $m=6$), the five-dimensional theory coupled to $n-1$ vector multiplets contains $n+4$ vectors (where 5 of them belong to the gravity multiplet) transforming in the ${\bf (4+n)}$ of $SO(5,n-1)$ and an additional vector (dual of the universal 2-form in the gravity multiplet) which is a singlet of of $SO(5,n-1)$, together with
$5(n-1)+1$ real scalars parametrizing the coset manifold $\mathbb{R}^+ \times SO(5,n-1)/[SO(5) \times SO(n-1)]$. The four-dimensional theory coupled to $n$ vector multiplets contains $n+6$ vectors, that together with their magnetic duals transform in the ${\bf (6+n,2)}$ of $SO(6,n)\times SL(2,\mathbb{R})$, and $6n+2$ real scalars parametrizing the coset manifold $SO(6,n)/[SO(6)\times SO(n)] \times SL(2,\mathbb{R})/SO(2)$. On the other hand, in
 the $\mathcal{N}=2$ case (corresponding to $m=2$), the five-dimensional theory coupled to $n$ vector multiplets contains $n+1$ vectors transforming in the ${\bf 1 \oplus n}$ of $SO(1,n-1)$, and $n$ real scalars parametrizing the coset manifold $\mathbb{R}^+ \times SO(1,n-1)/SO(n-1)$, while the four-dimensional theory coupled to $n+1$ vector multiplets contains $n+2$ vectors,\footnote{Note that, in the $D=5\rightarrow 4$ reduction, the $D=5$ graviphoton
becomes the Maxwell vector field of the $D=4$ axio-dilatonic vector
multiplet, while the Kaluza-Klein vector becomes the $D=4$ graviphoton.} that together with their magnetic duals transform in the ${\bf (2+n,2)}$ of $SO(2,n)\times SL(2,\mathbb{R})$, and $n+1$ complex scalars parametrizing the special K\"{a}hler coset\,\footnote{Actually, after \cite{Ferrara:1989py}, this is the unique example of
non-irreducible (\textit{i.e.}, product) special K\"{a}hler manifold.}  $SO(2,n)/[SO(2)\times SO(n)] \times SL(2,\mathbb{R})/SO(2)$.

A complete classification of all the extremal BH orbits of these theories can be found in \cite{Cerchiai:2010xv} and \cite{Borsten:2011ai}. Here we want to show that the same results can be derived by using the methods applied in the   previous section for the magic supergravity theory based on $J_{3}^{\mathbb{C}}$. We thus want to obtain the orbits from the stabilizers of suitable bound states of real-weight vectors of the representation of the BH charge, where the reality properties of the weights are derived from the Tits-Satake diagram that defines the real form of (the semisimple part of) the symmetry algebra.
 In the first subsection, we will discuss the five-dimensional case, while in the second subsection we will perform a more detailed analysis in four dimensions for the particular case of the $\mathcal{N}=2$ theory coupled to 4 vector multiplets.

\subsection{\label{5diminfiniteseriesorbits}$D=5$}

In both $\mathcal{N}=4$ and $\mathcal{N}=2$  theories, the BH charges are $(Q, Q_M )$, where $M$ is a vector index of either $SO(5,n-1)$ or $SO(1,n-1)$. The BH orbits are characterized by the cubic expression $Q Q^M Q_M$, which is an invariant of $\mathbb{R}^+ \times SO(5,n-1)$ or of $\mathbb{R}^+ \times SO(1,n-1)$, because $Q$ has weight $+2$ and $Q_M$ has weight $-1$ under $\mathbb{R}^+$ \cite{Borsten:2011ai}. We want to determine the stabilizing algebra of each orbit by considering bound states of suitably chosen weight vectors. In particular, as far as the simple part of the symmetry algebra is concerned, we have to consider bound states of real-weight vectors $\ket{\Lambda_i}$, where the reality properties are derived from the corresponding Tits-Satake diagram. Together with those, we can also consider the state $\ket{\Lambda}$, corresponding to the $\mathbb{R}^+$ charge $Q$.

In \cite{Bergshoeff:2014lxa} it was shown that for an orthogonal algebra $\mathfrak{so}(p,q)$\,\footnote{Without loss of generality we can take $p+q$ even, so that all the weights of the vector representation have the same length.} the real-weight vectors of the ${\bf p+q}$ representation simply identify the light-like components of the charge $Q_M$. Choosing $p \leq q$, this means that there are $2p$ real weights in the representation, because one can make a choice of basis such that there are $2p$ light-like directions. If $p \neq 0$, the highest weight is always real, and we denote it with $\Lambda_+$, where $+ \equiv x+t$ with $x$ and $t$ two `space' and `time' directions of the vector representation of $\mathfrak{so}(p,q)$. Under transformations of the algebra, the only weight that cannot be reached from $\ket{\Lambda_+}$ is the lowest-weight vector $\ket{\Lambda_-}$, where $- \equiv x-t$.
The outcome of this analysis is that we have to consider bound states of $\ket{\Lambda}$, $\ket{\Lambda_+}$ and $\ket{\Lambda_-}$.

\subsubsection{1-charge orbits}

There are two different 1-charge orbits, corresponding to the weight vectors $\ket{\Lambda}$ and $\ket{\Lambda_+}$. The stabilizing algebras are $\mathfrak{so}(5,n-1)$ and $(\mathfrak{so}(1,1)\oplus \mathfrak{so}(4,n-2)) \ltimes \mathbb{R}^{n+2}$ in the $\mathcal{N}=4$ case, and $\mathfrak{so}(1,n-1)$ and $(\mathfrak{so}(1,1)\oplus \mathfrak{so}(n-2)) \ltimes \mathbb{R}^{n-2}$ in the  $\mathcal{N}=2$ case. Comparing with Table III and IV of \cite{Borsten:2011ai}, one recognizes these as the rank-$1a$ and rank-$1c$ orbits of that paper. Both orbits are 1/2-BPS in both theories.

\subsubsection{2-charge orbits}

The independent 2-charge orbits correspond to the bound states $\ket{\Lambda}+ \ket{\Lambda_+}$, $\ket{\Lambda}- \ket{\Lambda_+}$, $\ket{\Lambda_+}+ \ket{\Lambda_-}$ and  $\ket{\Lambda_+}- \ket{\Lambda_-}$. In the  $\mathcal{N}=4$ case, the first two bound states have the same stabilizing algebra $\mathfrak{so}(4,n-2) \ltimes \mathbb{R}^{n+2}$, while the third gives $\mathfrak{so}(4,n-1)$  and the fourth gives $\mathfrak{so}(5,n-2)$. From \cite{Borsten:2011ai}, we recognize these as the rank-$2c$ 1/4-BPS orbit, the rank-$2a$ 1/2-BPS orbit and the rank-$2b$ non-supersymmetric orbit.

Similarly, in the $\mathcal{N}=2$ case the same bound states have the stabilizers $\mathfrak{so}(n-2) \ltimes \mathbb{R}^{n-2}$, $\mathfrak{so}(n-1)$  and  $\mathfrak{so}(1,n-2)$. By comparing with Table IV of \cite{Borsten:2011ai} one realizes that the first orbit undergoes a further stratification in four different orbits $2c^\pm$ and $2d^\pm$. From the analysis of \cite{Cerchiai:2010xv} (see also \cite{Borsten:2011ai}), this corresponds to changing the sign of $\ket{\Lambda}$, which is not visible in our analysis. Two of these four orbits are 1/2-BPS and two are non-supersymmetric. Of the remaining two orbits, again the first is the rank-$2a$ 1/2-BPS orbit and the second is the rank-$2b$ non-supersymmetric orbit.

\subsubsection{3-charge orbits}

We now move to the 3-charge orbits.
The independent bound states are $\ket{\Lambda} + \ket{\Lambda_+} + \ket{\Lambda_-}$, $-\ket{\Lambda} + \ket{\Lambda_+} + \ket{\Lambda_-}$ and  $\ket{\Lambda} + \ket{\Lambda_+} - \ket{\Lambda_-}$.  In the  $\mathcal{N}=4$ case, the first two give the stabilizing algebra  $\mathfrak{so}(4,n-1)$  and the third gives $\mathfrak{so}(5,n-2)$. By comparing with Table III of  \cite{Borsten:2011ai} one can see that the first algebra gives  the 1/4-BPS rank-$3ab$ orbit\,\footnote{From point 3 of Theorem 3 of \cite{Borsten:2011ai}, the 1/4-BPS rank-3$ab$
orbit should better be called rank-3$a$ orbit in Table III therein.

The $\mathcal{N}=4\rightarrow \mathcal{N}=2$ splitting of orbits (for the
infinite Jordan sequence in $D=5$) is elucidated in Table 6 of \cite{Cerchiai:2010xv} (which however, with respect to the Tables II and IV of
\cite{Borsten:2011ai}, ``unifies'' the rank-3$a^{+}$ and rank-3$b^{-}$ orbits, as well as the rank-3$a^{-}$ and rank-3$b^{+}$ orbits).} and the second gives the non-supersymmetric rank-$3b$ orbit. In the  $\mathcal{N}=2$ case, the stabilizing algebras become $\mathfrak{so}(n-1)$  and the third gives $\mathfrak{so}(1,n-2)$, and by comparing with Table IV of \cite{Borsten:2011ai} one realizes that  the first stratifies into four different orbits (two 1/2-BPS and two non-supersymmetric). As in the case of the 2-charge orbit, from the analysis of \cite{Cerchiai:2010xv} (see also \cite{Borsten:2011ai}) this corresponds to flipping the sign of $\ket{\Lambda}$, but this cannot be detected in our analysis.

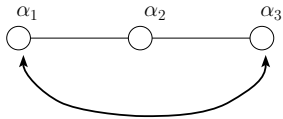
\begin{figure}[h!]
\centering
\scalebox{0.4} {\
\begin{pspicture}(0,-2.674844)(16.8,1.7148438)
\psline[linewidth=0.02cm](4.6,-0.25796875)(12.2,-0.25796875)
\pscircle[linewidth=0.02,dimen=outer,fillstyle=solid](8.4,-0.25796875){0.4}
\pscircle[linewidth=0.02,dimen=outer,fillstyle=solid](12.4,-0.25796875){0.4}
\pscircle[linewidth=0.02,dimen=outer,fillstyle=solid](4.4,-0.25796875){0.4}
\psbezier[linewidth=0.06,arrowsize=0.05291667cm 4.0,arrowlength=1.4,arrowinset=0.2]{<->}(4.558437,-0.81576675)(4.558437,-1.440098)(5.1441727,-2.0664806)(5.7533517,-2.3765953)(6.3625307,-2.68671)(7.878258,-2.8448439)(8.541485,-2.8448439)(9.204712,-2.8448439)(10.524531,-2.8448439)(11.3296175,-2.3765953)(12.134704,-1.9083468)(12.524531,-1.596181)(12.524531,-0.8157669)
\rput(8.8925,0.5651562){\huge $\alpha_2$}
\rput(12.710468,0.5651562){\huge $\alpha_3$}
\rput(4.6865625,0.5651562){\huge $\alpha_1$}
\end{pspicture}
}
\caption{The Tits-Satake diagram of the algebra $\mathfrak{so}(2,4)$.}
\label{so24dynkindiagram}
\end{figure}

\subsection{\label{4diminfiniteseriesorbits}$D=4$}

The BHs of the four-dimensional $\mathcal{N}=4$ and $\mathcal{N}=2$  theories based on the infinite sequences of semisimple rank-3 Jordan algebras $\mathbb{R}\oplus \mathbf{\Gamma }_{5,n-1}$ resp. $\mathbb{R}\oplus \mathbf{\Gamma }_{1,n-1}$ have electric and magnetic charges $Q_{aM}$, where $a$ is a doublet of $SL(2,\mathbb{R})$ and $M$ is a vector index of $SO(6,n)$ resp. $SO(2,n)$. In the present investigation, we will analyze in detail the particular case of the $\mathcal{N}=2$  theory with $n=4$, in which case the charge is in the representation ${\bf (6,2)}$ of the algebra\,\footnote{The motivation for choosing this algebra is
simply that this is the smallest \textit{non-split} relevant algebra, which
then allows to exploit all techniques for the slightly more involved case of
non-split U-duality algebras, already applied to the theories based on $J_{3}^{\mathbb{C}}$ in the treatment above. In fact, the smaller cases $n=1$
($ST^{2}$ model), $n=2$ ($STU$ model) and $n=3$ all exhibit split U-duality algebras.} $\mathfrak{so}(2,4)\oplus \mathfrak{sl}(2,\mathbb{R})\sim \mathfrak{su}(2,2)\oplus \mathfrak{sl}(2,\mathbb{R})$. The Tits-Satake diagram of $\mathfrak{so}(2,4)$ is drawn in \autoref{so24dynkindiagram}. Moreover, we denote with $\beta$ the positive root of $\mathfrak{sl}(2,\mathbb{R})$. From the diagram, and observing that $\mathfrak{sl}(2,\mathbb{R})$ is split, one obtains the action of the Cartan involution on the roots
\begin{equation}
\theta \alpha_1 = -\alpha_3  \qquad \quad \theta \alpha_2 = -\alpha_2 \qquad \quad \theta \beta  =-\beta
\quad . \label{Cartaninvolutionso24}
\end{equation}

\begin{figure}[b!]
\centering
\scalebox{0.65} 
{\
\begin{pspicture}(0,-6.5390625)(13.58,6.5790625)
\psline[linewidth=0.02cm,fillcolor=black,dotsize=0.07055555cm 2.0]{*-*}(9.436563,3.6046875)(11.836562,1.2046875)
\usefont{T1}{ppl}{m}{n}
\rput{-45.49773}(1.4651803,8.4520645){\rput(10.781094,2.4946876){$\alpha_{1}$}}
\psline[linewidth=0.02cm,fillcolor=black,dotsize=0.07055555cm 2.0]{*-*}(9.436563,6.0046873)(9.436563,3.6046875)
\usefont{T1}{ppl}{m}{n}
\rput{91.81141}(14.180749,-4.5015516){\rput(9.241094,4.6346874){$\alpha_{2}$}}
\psline[linewidth=0.02cm,fillcolor=black,dotsize=0.07055555cm 2.0]{*-*}(9.436563,3.6046875)(7.0365624,1.2046875)
\usefont{T1}{ppl}{m}{n}
\rput{43.80002}(4.0973134,-4.95457){\rput(8.181094,2.6346874){$\alpha_{3}$}}
\usefont{T1}{ppl}{m}{n}
\rput(9.9,6.3396873){\large $\Lambda^{+}_{1}$}
\psline[linewidth=0.02cm,fillcolor=black,dotsize=0.07055555cm 2.0]{*-*}(7.0365624,1.2046875)(9.436563,-1.1953125)
\usefont{T1}{ppl}{m}{n}
\rput{-45.49773}(2.4591649,6.0225806){\rput(8.381094,0.0946875){$\alpha_{1}$}}
\psline[linewidth=0.02cm,fillcolor=black,dotsize=0.07055555cm 2.0]{*-*}(9.436563,-1.1953125)(9.436563,-3.5953126)
\usefont{T1}{ppl}{m}{n}
\rput{91.81141}(6.9843473,-11.929142){\rput(9.241094,-2.5653124){$\alpha_{2}$}}
\psline[linewidth=0.02cm,fillcolor=black,dotsize=0.07055555cm 2.0]{*-*}(11.836562,1.2046875)(9.436563,-1.1953125)
\usefont{T1}{ppl}{m}{n}
\rput{43.80002}(3.103945,-7.2834897){\rput(10.581094,0.2346875){$\alpha_{3}$}}
\usefont{T1}{ppl}{m}{n}
\rput(6.5865626,1.3396875){\large $\Sigma_1^{+}$}
\usefont{T1}{ppl}{m}{n}
\rput(12.566719,1.3396875){\large $\Sigma_2^{+}$}
\usefont{T1}{ppl}{m}{n}
\rput(10.3,3.7396874){\large $\Lambda^{+}_{2}$}
\usefont{T1}{ppl}{m}{n}
\rput(10.3,-1.2603126){\large $\Lambda^{+}_{3}$}
\usefont{T1}{ppl}{m}{n}
\rput(9.7,-4.0603123){\large $\Lambda^{+}_{4}$}
\psline[linewidth=0.02cm,fillcolor=black,dotsize=0.07055555cm 2.0]{*-*}(3.8365624,1.4046875)(6.2365627,-0.9953125)
\usefont{T1}{ppl}{m}{n}
\rput{-45.49773}(1.3595203,3.8000808){\rput(5.1810937,0.2946875){$\alpha_{1}$}}
\psline[linewidth=0.02cm,fillcolor=black,dotsize=0.07055555cm 2.0]{*-*}(3.8365624,3.8046875)(3.8365624,1.4046875)
\usefont{T1}{ppl}{m}{n}
\rput{91.81141}(6.204834,-1.1738918){\rput(3.6410937,2.4346876){$\alpha_{2}$}}
\psline[linewidth=0.02cm,fillcolor=black,dotsize=0.07055555cm 2.0]{*-*}(3.8365624,1.4046875)(1.4365625,-0.9953125)
\usefont{T1}{ppl}{m}{n}
\rput{43.80002}(1.0164539,-1.6906948){\rput(2.5810938,0.4346875){$\alpha_{3}$}}
\usefont{T1}{ppl}{m}{n}
\rput(4.4,4.1396875){\large $\Lambda^{-}_{1}$}
\psline[linewidth=0.02cm,fillcolor=black,dotsize=0.07055555cm 2.0]{*-*}(1.4365625,-0.9953125)(3.8365624,-3.3953125)
\usefont{T1}{ppl}{m}{n}
\rput{-45.49773}(2.353505,1.3705963){\rput(2.7810938,-2.1053126){$\alpha_{1}$}}
\psline[linewidth=0.02cm,fillcolor=black,dotsize=0.07055555cm 2.0]{*-*}(3.8365624,-3.3953125)(3.8365624,-5.7953124)
\usefont{T1}{ppl}{m}{n}
\rput{91.81141}(-0.99156797,-8.601482){\rput(3.6410937,-4.7653127){$\alpha_{2}$}}
\psline[linewidth=0.02cm,fillcolor=black,dotsize=0.07055555cm 2.0]{*-*}(6.2365627,-0.9953125)(3.8365624,-3.3953125)
\usefont{T1}{ppl}{m}{n}
\rput{43.80002}(0.023085669,-4.0196147){\rput(4.981094,-1.9653125){$\alpha_{3}$}}
\usefont{T1}{ppl}{m}{n}
\rput(1.0865625,-0.8603125){\large $\Sigma_1^{-}$}
\usefont{T1}{ppl}{m}{n}
\rput(7.0667186,-0.8603125){\large $\Sigma_2^{-}$}
\usefont{T1}{ppl}{m}{n}
\rput(4.8,1.5396875){\large $\Lambda^{-}_{2}$}
\usefont{T1}{ppl}{m}{n}
\rput(4.8,-3.4603126){\large $\Lambda^{-}_{3}$}
\usefont{T1}{ppl}{m}{n}
\rput(4.2,-6.2603126){\large $\Lambda^{-}_{4}$}
\psline[linewidth=0.04cm,fillcolor=black,linestyle=dashed,dash=0.16cm 0.16cm,dotsize=0.07055555cm 2.0]{*-*}(3.8365624,-5.7953124)(9.436563,-3.5953126)
\usefont{T1}{ppl}{m}{n}
\rput{22.209826}(-1.2522349,-2.7411382){\rput(6.32,-4.5603123){\large $\beta$}}
\psline[linewidth=0.04cm,fillcolor=black,linestyle=dashed,dash=0.16cm 0.16cm,dotsize=0.07055555cm 2.0]{*-*}(3.8365624,-3.3953125)(9.436563,-1.1953125)
\usefont{T1}{ppl}{m}{n}
\rput{22.209826}(-0.34503603,-2.5630722){\rput(6.32,-2.1603124){\large $\beta$}}
\psline[linewidth=0.04cm,fillcolor=black,linestyle=dashed,dash=0.16cm 0.16cm,dotsize=0.07055555cm 2.0]{*-*}(1.4365625,-0.9953125)(7.0365624,1.2046875)
\usefont{T1}{ppl}{m}{n}
\rput{22.209826}(0.3840968,-1.477807){\rput(3.92,0.2396875){\large $\beta$}}
\psline[linewidth=0.04cm,fillcolor=black,linestyle=dashed,dash=0.16cm 0.16cm,dotsize=0.07055555cm 2.0]{*-*}(6.2365627,-0.9953125)(11.836562,1.2046875)
\usefont{T1}{ppl}{m}{n}
\rput{22.209826}(0.74022907,-3.2922049){\rput(8.72,0.2396875){\large $\beta$}}
\psline[linewidth=0.04cm,fillcolor=black,linestyle=dashed,dash=0.16cm 0.16cm,dotsize=0.07055555cm 2.0]{*-*}(3.8365624,1.4046875)(9.436563,3.6046875)
\usefont{T1}{ppl}{m}{n}
\rput{22.209826}(1.4693618,-2.20694){\rput(6.32,2.6396875){\large $\beta$}}
\psline[linewidth=0.04cm,fillcolor=black,linestyle=dashed,dash=0.16cm 0.16cm,dotsize=0.07055555cm 2.0]{*-*}(3.8365624,3.8046875)(9.436563,6.0046873)
\usefont{T1}{ppl}{m}{n}
\rput{22.209826}(2.376561,-2.0288737){\rput(6.32,5.0396876){\large $\beta$}}
\end{pspicture}
}
\caption{The weights of the ${\bf (6,2)}$ of  $\mathfrak{so}(2,4)\oplus \mathfrak{sl}(2,\mathbb{R})$. \label{weightdiagramof62ofso24sl2}}
\end{figure}
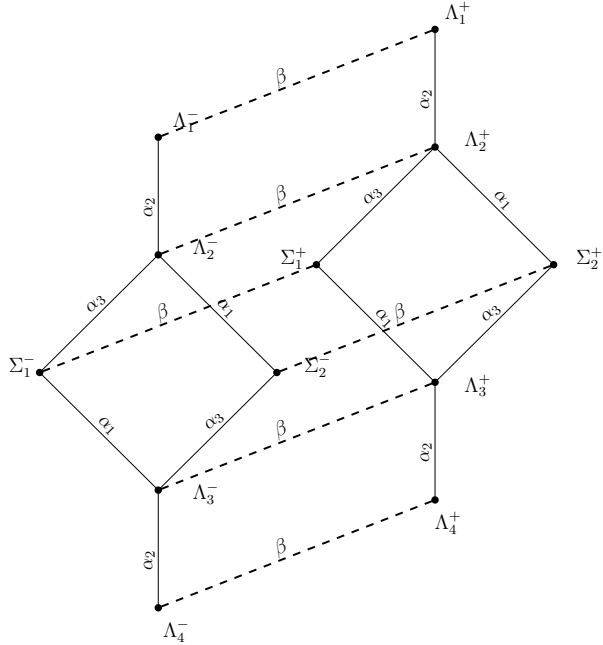

The highest weight of the ${\bf (6,2)}$ is the weight $\Lambda_1^+ = \tfrac{1}{2} \alpha_1 + \alpha_2  + \tfrac{1}{2} \alpha_3 + \tfrac{1}{2} \beta$, which is real. We draw in \autoref{weightdiagramof62ofso24sl2} all the weights of the representation, denoting with $\Lambda$ those that are real and with $\Sigma$ those that are complex. One gets in total eight real weights, four for each component of the $\mathfrak{sl}(2,\mathbb{R})$ doublet. This is precisely what we expect from the analysis of \cite{Bergshoeff:2014lxa} that we used in the previous subsection, that shows that for vector representations of orthogonal algebras the real-weight vectors correspond to components with indices along the light-like directions. Indeed,
given a vector of $\mathfrak{so}(2,4)$, one can choose four light-like and two space-like directions, and the light-like directions are associated to the real weights (while the two complex weights are associated to the two space-like directions).

\begin{table}[t!]
\renewcommand{\arraystretch}{1.2}
\par
\begin{center}
\resizebox{\textwidth}{!}{
\begin{tabular}{|c|c|c|c|c|}
\hline
height&$\ket{\Lambda^{+}_{1}}$&$\ket{\Lambda^{-}_{2}}$&$\ket{\Lambda^{-}_{3}}$&$\ket{\Lambda^{+}_{4}}$\\
\hline\hline
 3&$E_{\alpha_{1}+\alpha_{2}+\alpha_{3}}$&$E_{\alpha_{1}+\alpha_{2}+\alpha_{3}}$&&\\
 2&$E_{\alpha_{1}+\alpha_{2}}\quad E_{\alpha_{2}+\alpha_{3}}$&$E_{\alpha_{1}+\alpha_{2}}\quad E_{\alpha_{2}+\alpha_{3}}$&$E_{\alpha_{1}+\alpha_{2}}\quad  E_{\alpha_{2}+\alpha_{3}}$&\\
1&$E_{\alpha_{1}}\quad E_{\alpha_{2}}\quad E_{\alpha_{3}}\quad E_{\beta}$&$E_{\alpha_{1}}\quad E_{\alpha_{3}}$&$E_{\alpha_{2}}$&$E_{\alpha_{1}} \quad E_{\alpha_{3}}\quad E_{\beta} $\\
0&$H_{\alpha_{1}}\quad H_{\alpha_{3}}\quad H_{\alpha_{2}}-H_{\beta}$&$H_{\alpha_{1}}+H_{\alpha_{2}}\quad H_{\alpha_{3}}+H_{\alpha_{2}}\quad H_{\alpha_{2}}-H_{\beta}$&$H_{\alpha_{1}}+H_{\alpha_{2}}\quad H_{\alpha_{3}}+H_{\alpha_{2}}\quad H_{\alpha_{2}}+H_{\beta}$&$H_{\alpha_{1}}\quad H_{\alpha_{3}}\quad H_{\alpha_{2}}+H_{\beta}$\\
$-1$ &$E_{-\alpha_{1}}\quad E_{-\alpha_{3}} $&$E_{-\alpha_{2}} \quad E_{-\beta} $&$E_{-\alpha_{1}} \quad E_{-\alpha_{3}} \quad E_{-\beta} $&$E_{-\alpha_{1}} \quad E_{-\alpha_{2}} \quad E_{-\alpha_{3}}$\\
$ -2$ &&$E_{-\alpha_{1}-\alpha_{2}} \quad E_{-\alpha_{2}-\alpha_{3}} $ & $E_{-\alpha_{1}-\alpha_{2}} \quad E_{-\alpha_{2}-\alpha_{3}} $&$E_{-\alpha_{1}-\alpha_{2}} \quad E_{-\alpha_{2}-\alpha_{3}}$\\
$-3$ &&&$E_{-\alpha_{1}-\alpha_{2}-\alpha_{3}} $&$E_{-\alpha_{1}-\alpha_{2}-\alpha_{3}} $\\
\hline
\end{tabular}
}
\end{center}
\caption{The stabilizers of the weight vectors of the ${\mathbf (6,2)}$ of $\mathfrak{so}(2,4)\oplus \mathfrak{sl}(2,\mathbb{R})$ that are used in the paper. \label{stabilizersweightsso24sl2}}
\end{table}

\subsubsection{1-charge orbit}

We want to construct the relevant representatives of the duality orbits of multi-charge BH solutions as bound states of independent real-weight vectors. In particular, we choose the real-weight vectors
\begin{equation}
 \ket{\Lambda_1^+} \qquad \ket{\Lambda_2^-} \qquad \ket{\Lambda_3^-} \qquad \ket{\Lambda_4^+} \quad . \label{indeprealweightvectors26ofso24sl2}
 \end{equation}
 We list in \autoref{stabilizersweightsso24sl2} the generators that stabilize each of these weight vectors. In all cases,
we get the stabilizing algebra $[\mathfrak{so}(1,1)\oplus \mathfrak{so}(1,3)] \ltimes (\mathbb{R} \times \mathbb{R}^{4} )$, resulting in the rank-1 1/2-BPS orbit  of \cite{Borsten:2011ai} (set $n=4$ in Table VIII therein).

\begin{table}[t!]
\renewcommand{\arraystretch}{1.2}
\begin{center}
\begin{tabular}{|c|c|c|}
\hline
Common& $\ket{\Lambda_{1}^{+}}+\ket{\Lambda_{2}^{-}}$ Conjunction & $\ket{\Lambda_{1}^{+}}-\ket{\Lambda_{2}^{-}}$ Conjunction\\
\hline
\hline
$E_{\alpha_{1}+\alpha_{2}+\alpha_{3}}$&$E_{\alpha_{2}}-E_{-\beta}$ & $E_{\alpha_{2}} + E_{-\beta}$\\
$E_{\alpha_{1}+\alpha_{2}}$&$E_{\beta}-E_{-\alpha_{2}}$ & $E_{\beta}+E_{-\alpha_{2}}$\\
$  E_{\alpha_{2}+\alpha_{3}}$  & & \\
$E_{\alpha_{1}}$ & & \\
$ E_{\alpha_{3}}$& & \\
$H_{\alpha_{1}}- H_{\alpha_{3}}$ & & \\
$ H_{\alpha_{2}}- H_{\beta}$& & \\
\hline
\end{tabular}

\caption{The stabilizers of the bound states $\ket{\Lambda_{1}^{+}}+\ket{\Lambda_{2}^{-}}$ and $\ket{\Lambda_{1}^{+}}-\ket{\Lambda_{2}^{-}}$. In both cases the stabilizing algebra is $[\mathfrak{so}(2,1)\ltimes \mathbb{R}]\oplus [\mathfrak{so}%
(2)\ltimes (\mathbb{R}^{2}\oplus\mathbb{R}^{2})]$. \label{twochargelambda1+lambda2-} }
\end{center}
\end{table}

\begin{table}[b!]
\renewcommand{\arraystretch}{1.2}
\begin{center}
\begin{tabular}{|c|c|c|}
\hline
Common&$\ket{\Lambda_{1}^{+}}+\ket{\Lambda_{4}^{+}}$ Conjunction & $\ket{\Lambda_{1}^{+}}- \ket{\Lambda_{4}^{+}}$ Conjunction\\
\hline\hline
$E_{\alpha_{1}}$&$E_{\alpha_{1}+\alpha_{2}}-E_{-\alpha_{2}-\alpha_{3}}$ &$E_{\alpha_{1}+\alpha_{2}} + E_{-\alpha_{2}-\alpha_{3}}$\\
$E_{\alpha_{3}}$& $E_{\alpha_{2}+\alpha_{3}}-E_{-\alpha_{1}-\alpha_{2}}$ & $E_{\alpha_{2}+\alpha_{3}} +E_{-\alpha_{1}-\alpha_{2}}$\\
$E_{\beta}$ &$E_{\alpha_{1}+\alpha_{2}+\alpha_{3}}+E_{-\alpha_{2}}$ &$E_{\alpha_{1}+\alpha_{2}+\alpha_{3}} -E_{-\alpha_{2}}$\\
$H_{\alpha_{1}}$&$E_{\alpha_{2}}+E_{-\alpha_{1}-\alpha_{2}-\alpha_{3}}$ &$E_{\alpha_{2}} - E_{-\alpha_{1}-\alpha_{2} -\alpha_{3}}$\\
$H_{\alpha_{3}}$ & & \\
$E_{-\alpha_{1}}$ & & \\
$ E_{-\alpha_{3}}$ & & \\

\hline
\end{tabular}

\caption{The stabilizers of the $\ket{\Lambda_{1}^{+}}+\ket{\Lambda_{4}^{+}}$ and  $\ket{\Lambda_{1}^{+}}-\ket{\Lambda_{4}^{+}}$ bound states. The stabilizing algebra is $\mathfrak{so}(1,4)\oplus \mathbb{R}$ in the first case and  $\mathfrak{so}(2,3)\oplus \mathbb{R}$ in the second. \label{lambda1+lambda4+boundstates}}
\end{center}
\end{table}

\subsubsection{2-charge orbits}

There are four different 2-charge bound states that one can construct, that in terms of the real-weight vectors in \autoref{indeprealweightvectors26ofso24sl2} can be written as
\begin{equation}
\ket{\Lambda_{1}^{+}}+\ket{\Lambda_{2}^{-}} \quad \quad \ket{\Lambda_{1}^{+}}-\ket{\Lambda_{2}^{-}} \quad \quad \ket{\Lambda_{1}^{+}}+\ket{\Lambda_{4}^{+}} \quad \quad \ket{\Lambda_{1}^{+}}-\ket{\Lambda_{4}^{+}}.
\end{equation}
The stabilizers of all these bound states are listed in \autoref{twochargelambda1+lambda2-} and \autoref{lambda1+lambda4+boundstates}. From such tables, one deduces that the stabilizing algebra is $[\mathfrak{so}(2,1)\ltimes \mathbb{R}]\oplus [\mathfrak{so}(2)\ltimes (\mathbb{R}^{2}\oplus\mathbb{R}^{2})]$ for the $\ket{\Lambda_{1}^{+}}+\ket{\Lambda_{2}^{-}} $ and $\ket{\Lambda_{1}^{+}}-\ket{\Lambda_{2}^{-}}$ bound states, while it is $\mathfrak{so}(1,4) \oplus \mathbb{R}$ for $\ket{\Lambda_{1}^{+}}+\ket{\Lambda_{4}^{+}}$ and $\mathfrak{so}(2,3) \oplus \mathbb{R}$ for $\ket{\Lambda_{1}^{+}}-\ket{\Lambda_{4}^{+}}$. In particular, the real form of the semisimple part of the stabilizing algebra can be easily determined from  the Cartan involution in \autoref{Cartaninvolutionso24}.
 By comparing with Table VIII of \cite{Borsten:2011ai}, one gets that the $\ket{\Lambda_{1}^{+}}+\ket{\Lambda_{2}^{-}} $  and $\ket{\Lambda_{1}^{+}}-\ket{\Lambda_{2}^{-}} $ bound states correspond to the rank-$2c^+$ and rank-$2c^-$ orbits, where the first is 1/2-BPS and the second in non-supersymmetric, while  $\ket{\Lambda_{1}^{+}}+\ket{\Lambda_{4}^{+}}$ gives the 1/2-BPS rank-$2b$ orbit and $\ket{\Lambda_{1}^{+}}-\ket{\Lambda_{4}^{+}}$ gives the non-supersymmetric rank-$2a$ orbit.

\subsubsection{3-charge orbit}

There are three different 3-charge bound states that can be constructed, that we choose to be
\begin{equation}
\ket{\Lambda_{1}^+}+\ket{\Lambda_{2}^-}+\ket{\Lambda_{4}^+} \quad \ket{\Lambda_{1}^+}-\ket{\Lambda_{2}^-}+\ket{\Lambda_{4}^+} \quad \ket{\Lambda_{1}^+}+\ket{\Lambda_{2}^-}-\ket{\Lambda_{4}^+} \quad .
\end{equation}
The corresponding stabilizers are listed in \autoref{stab1+2-4+62so24sl2}. The stabilizing algebra is ($\mathfrak{so}(3)\ltimes \mathbb{R}^{3})\times
\mathbb{R}$ for the fist two bound states, and $(\mathfrak{so}(1,2)\ltimes \mathbb{R}^{3})\times
\mathbb{R}$ for the last bound state. In the notation of \cite{Borsten:2011ai}, the first two cases correspond to the rank-$3a^+$ and $3a^-$ orbits, where the first is 1/2-BPS and the second in non-supersymmetric, while the last corresponds to the rank-$3b$ non-supersymmetric orbit.

\begin{table}[t!]
\renewcommand{\arraystretch}{1.3}
\par
\begin{center}
\scalebox{0.9}{
\begin{tabular}{|c||c|c|c|c|}
\hline
Common & 2-conj. & $\ket{\Lambda_{1}^+}+\ket{\Lambda_{2}^-}+\ket{\Lambda_{4}^+}$ & $\ket{\Lambda_{1}^+}-\ket{\Lambda_{2}^-}+\ket{\Lambda_{4}^+}$ & $\ket{\Lambda_{1}^+}+\ket{\Lambda_{2}^-}-\ket{\Lambda_{4}^+}$\\ \hline\hline
$E_{\alpha_{1}}$ & $\Lambda_{1}^+ ,\Lambda_{2}^-$ & $E_\beta - E_{-\alpha_2}$ & $E_\beta + E_{-\alpha_2}$ & $E_\beta - E_{-\alpha_2}$\\
\cline{2-5}
$E_{\alpha_{3}}$ & & $E_{\alpha_1 + \alpha_2} - E_{-\alpha_2 -\alpha_3}$ & $E_{\alpha_1 + \alpha_2} - E_{-\alpha_2 -\alpha_3}$ & $E_{\alpha_1 + \alpha_2} + E_{-\alpha_2 -\alpha_3}$  \\
$H_{\alpha_1} -H_{\alpha_3}$ & $\Lambda_{1}^+ ,\Lambda_{4}^+$ & $E_{\alpha_2 + \alpha_3} - E_{-\alpha_1 -\alpha_2}$ & $E_{\alpha_2 + \alpha_3} - E_{-\alpha_1 -\alpha_2}$ & $E_{\alpha_2 + \alpha_3} + E_{-\alpha_1 -\alpha_2}$  \\
& & $E_{\alpha_1 + \alpha_2 + \alpha_3} + E_{-\alpha_2}$  & $E_{\alpha_1 + \alpha_2 + \alpha_3} + E_{-\alpha_2}$  & $E_{\alpha_1 + \alpha_2 + \alpha_3} - E_{-\alpha_2}$  \\ \cline{2-5}
& $\Lambda_{2}^- ,\Lambda_{4}^+$ & $E_{\alpha_1 + \alpha_2 + \alpha_3} + E_{\beta} $  & $E_{\alpha_1 + \alpha_2 + \alpha_3} - E_{\beta} $  & $E_{\alpha_1 + \alpha_2 + \alpha_3} - E_{\beta} $ \\
\hline
\end{tabular}
}
\end{center}
\caption{The stabilizers of the bound states  $\ket{\Lambda_{1}^+} +\ket{\Lambda_{2}^-} +\ket{\Lambda_{4}^+}$,  $\ket{\Lambda_{1}^+} -\ket{\Lambda_{2}^-} +\ket{\Lambda_{4}^+}$ and  $\ket{\Lambda_{1}^+} +\ket{\Lambda_{2}^-} -\ket{\Lambda_{4}^+}$. The generators in the third, fourth and fifth column are 2-conjunction stabilizers for the weight vectors whose weights are listed in the second column. The stabilizing algebra is $(\mathfrak{so}(3)\ltimes \mathbb{R}^{3})\times
\mathbb{R}$ in the fist two cases, and $(\mathfrak{so}(1,2)\ltimes \mathbb{R}^{3})\times
\mathbb{R}$ in the last case.}
\label{stab1+2-4+62so24sl2}
\end{table}

\begin{table}[b!]
\renewcommand{\arraystretch}{1.3}
\par
\begin{center}
\scalebox{0.63}{
\begin{tabular}{|c||c|c|c|c|c|}
\hline
Common & 2-conj. &$\ket{\Lambda_1^+}\! +\! \ket{\Lambda_2^-}\! +\! \ket{\Lambda_3^-}\! +\! \ket{\Lambda_4^+}$ & $\ket{\Lambda_1^+}\! -\! \ket{ \Lambda_2^-} \!+ \! \ket{ \Lambda_3^-}\! +\! \ket{\Lambda_4^+}$ & $\ket{\Lambda_1^+} \!-\!\ket{ \Lambda_2^- }\! -\!  \ket{\Lambda_3^- }\! +\! \ket{\Lambda_4^+}$ &  $\ket{\Lambda_1^+} \! -\! \ket{ \Lambda_2^- }\! + \! \ket{\Lambda_3^- } \! -\! \ket{\Lambda_4^+}$ \\
\hline
\hline
$H_{\alpha_1} - H_{\alpha_3}$ & ${\Lambda_1^+},{\Lambda_2^-}$ & $F_{\alpha_2}^- + F_{\beta}^-$ & $F_{\alpha_2}^+ + F_{\beta}^+$ & $F_{\alpha_2}^- - F_{\beta}^-$ & $F_{\alpha_2}^- - F_{\beta}^-$ \\
\cline{2-6}
& ${\Lambda_1^+},{\Lambda_3^-}$ & $F_{\alpha_1+ \alpha_2+\alpha_3}^- + F_{\beta}^-$  & $F_{\alpha_1+ \alpha_2+\alpha_3}^+ - F_{\beta}^+$  & $F_{\alpha_1+ \alpha_2+\alpha_3}^- - F_{\beta}^-$  & $F_{\alpha_1+ \alpha_2+\alpha_3}^- + F_{\beta}^-$\\
\cline{2-6}
 & &  $F_{\alpha_1+ \alpha_2+\alpha_3}^- - F_{\alpha_2}^-$  &  $F_{\alpha_1+ \alpha_2+\alpha_3}^+ + F_{\alpha_2}^+$  &  $F_{\alpha_1+ \alpha_2+\alpha_3}^- - F_{\alpha_2}^-$  &   $F_{\alpha_1+ \alpha_2+\alpha_3}^- + F_{\alpha_2}^-$ \\
&  ${\Lambda_1^+},{\Lambda_4^+}$ & $E_{\alpha_1+\alpha_2} - E_{-\alpha_2 -\alpha_3}$  & $E_{\alpha_1 +\alpha_2}  - E_{-\alpha_2 -\alpha_3}$  & $E_{\alpha_1 +\alpha_2} - E_{-\alpha_2 -\alpha_3}$  & $E_{\alpha_1 +\alpha_2} + E_{-\alpha_2 -\alpha_3}$ \\
 & & $E_{\alpha_2 +\alpha_3}-  E_{-\alpha_1 -\alpha_2}$ & $E_{\alpha_2 +\alpha_3}-  E_{-\alpha_1 -\alpha_2}$& $E_{\alpha_2 +\alpha_3}-  E_{-\alpha_1 -\alpha_2}$& $E_{\alpha_2 +\alpha_3}+  E_{-\alpha_1 -\alpha_2}$\\
 \cline{2-6}

 & &  $F_{\alpha_1+ \alpha_2+\alpha_3}^- - F_{\alpha_2}^-$  &  $F_{\alpha_1+ \alpha_2+\alpha_3}^+ + F_{\alpha_2}^+$  &  $F_{\alpha_1+ \alpha_2+\alpha_3}^- - F_{\alpha_2}^-$  &   $F_{\alpha_1+ \alpha_2+\alpha_3}^- + F_{\alpha_2}^-$ \\
&  ${\Lambda_2^-},{\Lambda_3^-}$ & $E_{\alpha_1} - E_{-\alpha_3}$  & $E_{\alpha_1} + E_{-\alpha_3}$  & $E_{\alpha_1} - E_{-\alpha_3}$  & $E_{\alpha_1} + E_{ -\alpha_3}$ \\
 & & $E_{\alpha_3}-  E_{-\alpha_1 }$ & $E_{\alpha_3}+ E_{-\alpha_1 }$& $E_{\alpha_3}-  E_{-\alpha_1 }$& $E_{\alpha_3}+  E_{-\alpha_1 }$\\
 \cline{2-6}
 & ${\Lambda_2^-},{\Lambda_4^+}$ & $F_{\alpha_1+ \alpha_2+\alpha_3}^- + F_{\beta}^-$  & $F_{\alpha_1+ \alpha_2+\alpha_3}^+ - F_{\beta}^+$  & $F_{\alpha_1+ \alpha_2+\alpha_3}^- - F_{\beta}^-$  & $F_{\alpha_1+ \alpha_2+\alpha_3}^- + F_{\beta}^-$\\
\cline{2-6}
 & ${\Lambda_3^-},{\Lambda_4^+}$ & $F_{\alpha_2}^- + F_{\beta}^-$ & $F_{\alpha_2}^+ + F_{\beta}^+$ & $F_{\alpha_2}^- - F_{\beta}^-$ & $F_{\alpha_2}^- - F_{\beta}^-$ \\
 \hline
\end{tabular}
}
\end{center}
\caption{The stabilizers of the 4-charge orbits of the ${\mathbf{(6,2)}}$ of $\mathfrak{so}(2,4)\oplus \mathfrak{sl}(2,\mathbb{R})$. In the second column we
list the pair of states for which the  corresponding operators are 2-conjunction stabilizers. In any column, there are only two independent
generators among those in the first  line of each row. The stabilizing
algebras are $\mathfrak{so}(2)\oplus \mathfrak{so}(4)$ in the first and third case, $\mathfrak{so}(1,1)\oplus \mathfrak{so}(1,3)$ in the second case, and $\mathfrak{so}(2)\oplus \mathfrak{so}(2,2)$   in the fourth case. }
\label{fourchargeorbits62so24sl2}
\end{table}

\subsubsection{4-charge orbits}

We finally consider the 4-charge orbits. There are four possible bound states, namely
\begin{eqnarray}
& &
\ket{\Lambda_1^+}\! +\! \ket{\Lambda_2^-}\! +\! \ket{\Lambda_3^-}\! +\! \ket{\Lambda_4^+} \qquad \ket{\Lambda_1^+}\! -\! \ket{ \Lambda_2^-} \!+ \! \ket{ \Lambda_3^-}\! +\! \ket{\Lambda_4^+} \nonumber \\
& & \ket{\Lambda_1^+} \!-\!\ket{ \Lambda_2^- }\! -\!  \ket{\Lambda_3^- }\! +\! \ket{\Lambda_4^+} \qquad \ket{\Lambda_1^+} \! -\! \ket{ \Lambda_2^- }\! + \! \ket{\Lambda_3^- } \! -\! \ket{\Lambda_4^+} \quad .
\end{eqnarray}
The procedure to determine the stabilizing algebra is analogous to the one used for the 4-charge orbits in the previous two sections: for each pair of weight vectors, we combine their 2-conjunction stabilizers in order to obtain a generator that is also a stabilizer for the other pair. On top of this, one has to consider the common stabilizer, which is the compact Cartan generator $H_{\alpha_1} - H_{\alpha_3}$. The final result is listed in \autoref{fourchargeorbits62so24sl2}. One can see from this table that the stabilizing algebra for the first and third bound states is the compact algebra  $\mathfrak{so}(2)\oplus \mathfrak{so}(4)$, while one gets $\mathfrak{so}(1,1)\oplus \mathfrak{so}(1,3)$ for the second bound state, and $\mathfrak{so}(2)\oplus \mathfrak{so}(2,2)$ for the fourth. In the notation of \cite{Borsten:2011ai}, the first and the third bound states correspond to the rank-$4a^+$ and $4a^-$ orbits (where the first is 1/2-BPS and the second is non-supersymmetric), the second bound state corresponds to the rank-$4c$ non-supersymmetric dyonic orbit and the fourth bound state corresponds to the rank-$4b$ non-supersymmetric orbit. As in the other 4-charge orbits discussed in the previous sections, the dyonic orbit is special because among the generators that become non-compact compared to the supersymmetric orbit, there are some that are conjunction stabilizers connecting real-weight vectors to real-weight vectors.

\vskip .7cm

This completes the analysis of the orbits of the $\mathcal{N}=2$, $D=4$ theory coupled to 4 vector multiplets and based on $\mathbb{R}\oplus \mathbf{\Gamma }_{1,3}$. The above procedure can be applied to any other case, reproducing the correct result also for the $\mathcal{N}=4$, $D=4$ sequence (\textit{cfr.} Tables VII and VIII of \cite{Borsten:2011ai}).

\section{\label{centralcharges}Central charges and defect branes}

In \cite{Bergshoeff:2011zk,Bergshoeff:2011se,Bergshoeff:2012pm} it was shown that in the maximal supergravity theories the ``non-standard'' branes (branes with two or less transverse directions), whose representations contain weights of different length \cite{Bergshoeff:2013sxa}, always have degenerate BPS conditions. The standard branes instead always belong to representations whose weights have all the same length, and there is a one-to-one relation between the charge of these branes and the corresponding central charge in the supersymmetry algebra. This in particular implies that in the maximal theory a black hole which is a bound state preserves less supersymmetry than its constituents \cite{Ferrara:1997ci,Ferrara:1997uz,Lu:1997bg}.

The group-theoretical properties of standard branes of  the symmetric ${\cal N}=2$ theories
considered in previous sections resemble those of the non-standard branes of the maximal theories. As we have discussed at length, these branes belong to representations that contain either short (in the case of $J_{3}^{\mathbb{R}}$) or complex weights (in the case of $J_{3}^{\mathbb{C}}$, $J_{3}^{\mathbb{H}}$ and $J_{3}^{\mathbb{O}}$), and the complex weights are projected onto the short weights of the restricted-root algebra, which is the algebra of the theory based on $J_{3}^{\mathbb{R}}$, by means of the Tits-Satake projection \cite{Bergshoeff:2014lxa}.  Moreover, the 1/2-BPS branes always
have a degenerate BPS condition \cite{Bergshoeff:2014lxa}, which implies that a bound state of degenerate 1/2-BPS branes can still be 1/2-BPS.
In particular, in five dimensions The R-symmetry is $USp(2) \equiv SU(2)$ and the scalar central charge is a singlet, which means that any supersymmetric bound state must preserve the same amount of supersymmetry, {\it i.e.} it has to be 1/2-BPS, as originally shown in \cite{Ferrara:1997uz}. A similar argument applies to the four-dimensional case, where the R-symmetry is $U(2)$ and there are two singlet central charges.

The BPS degeneracy of non-standard branes in the maximal case was used in  \cite{Bergshoeff:2013sxa} to compute the orbits of their bound states and to determine the supersymmetry that these bound states preserve in some examples. In this section we want to further comment in this direction, following the group-theory analogy between these branes and the standard branes of the ${\cal N}=2$  theories. We will focus in particular on defect branes, {\it i.e.} branes of codimension 2. In the ten-dimensional IIB theory there are two such branes, namely the D7-brane and its S-dual, and they preserve the same supersymmetry, so that one can construct a bound state which is still 1/2-BPS. It is important to recall that these solutions are not asymptotically flat, and their (quantized) charge is given by the monodromy, which determines how the scalars transform under $G(\mathbb{Z})$ (which is the
U-duality symmetry of the full quantum theory) when going around the brane in transverse space. We refer to \cite{deBoer:2012ma} for a careful discussion on these issues.

\begin{table}[t!]
\renewcommand{\arraystretch}{1.6}
\begin{center}
\begin{tabular}{|c|c|c|}
\hline
\multicolumn{2}{|c|}{Charge} & BPS \\ \hline\hline
\multirow{-1}{*}{\rotatebox{90}{\begin{tiny}1-state\end{tiny}}}
& $Q_1{}^2$ & $1/2$ \\
\hline
& $Q_1{}^2 + Q_2{}^1$ & $1/2$ \\
& $Q_1{}^2+ Q_2{}^3$ & $1/2$ \\
\multirow{-3}{*}{\rotatebox{90}{\begin{tiny}2-state\end{tiny}}}
& $Q_1{}^2+ Q_3{}^4$ & $1/4$ \\ \hline
& $Q_1{}^2 + Q_2{}^3 + Q_3{}^1$ & $1/2$ \\
& $Q_1{}^2 + Q_2{}^3 + Q_3{}^4$ & $1/4$ \\
& $Q_1{}^2 +Q_2{}^3 + Q_4{}^5$ & $1/4$ \\
\multirow{-4}{*}{\rotatebox{90}{\begin{tiny}3-state\end{tiny}}}
& $Q_1{}^2 + Q_2{}^1 + Q_3{}^4$ & $1/4$ \\ \hline
& $Q_1{}^2 + Q_2{}^3 + Q_3{}^4 + Q_4{}^5$ & $1/4$ \\
& $Q_1{}^2 + Q_2{}^3 + Q_3{}^4+ Q_4{}^1$ & $1/4$ \\
& $Q_1{}^2 +Q_2{}^3 + Q_3{}^1+ Q_4{}^5$ & $1/4$ \\
& $Q_1{}^2+ Q_2{}^1 + Q_3{}^4 + Q_4{}^3$ & $1/4$\\
\multirow{-5}{*}{\rotatebox{90}{\begin{tiny}4-state\end{tiny}}}
& $Q_1{}^2 + Q_2{}^1 + Q_3{}^4 + Q_4{}^5$ & $1/4$ \\ \hline
& $Q_1{}^2 +Q_2{}^3 + Q_3{}^4 + Q_4{}^5 + Q_5{}^1$ & $1/4$ \\
\multirow{-2}{*}{\rotatebox{90}{\begin{tiny}5-state\end{tiny}}}
& $Q_1{}^2 + Q_2{}^1 + Q_3{}^4 + Q_4{}^5 + Q_5{}^3$ & $1/4$ \\ \hline
\end{tabular}
\end{center}
\caption{The different bound states of 4-branes in maximal supergravity in $D=7$ and the amount of supersymmetry that in each case we conjecture to be preserved. }
\label{conjectureddefectboundstates}
\end{table}

The single 1/2-BPS defect branes of the maximal theories were classified in \cite{Bergshoeff:2011se}. The charges belong to the adjoint of the duality symmetry group, and the single 1/2-BPS branes correspond to the root vectors (the roots are obviously the long weights of the adjoint representation). The BPS condition of these branes always has degeneracy two \cite{Bergshoeff:2011se}. We can consider as an example the 4-branes of the seven-dimensional maximal theory, whose global symmetry group is $SL(5,\mathbb{R})$. We denote the 4-brane charges in the ${\bf 24}$ as $Q_M{}^N$, where $M$ and $N$ are indices of the fundamental and antifundamental of  $SL(5,\mathbb{R})$ respectively. There are twenty 1/2-BPS branes, corresponding to the components  with $M\neq N$. The R-symmetry is $SO(5)$, and the 4-brane central charge belongs to the ${\bf 10}$, which implies a degeneracy two for each central charge as already anticipated. In particular, the charges $Q_M{}^N$ and $Q_N{}^M$ correspond to branes preserving the same supersymmetry.

We can identify the different bound states in terms of their charge components precisely as we did in the previous sections. Identifying with $Q_1{}^2$ the charge corresponding to the highest weight, the charges that are disconnected from it are $Q_M{}^N$ with $M \neq 1$ and $N \neq 2$. One finds that there are three different types of rank-2 orbits, and we can choose their representatives to be $Q_1{}^2 + Q_2{}^1$, $Q_1{}^2 + Q_2{}^3$ and $Q_1{}^2 + Q_3{}^4$ respectively. In terms of the R-symmetry $SO(5)$, we conjecture that only the rank-2 bound states with indices all different are 1/4-BPS, while the other bound states preserve the same supersymmetry as the rank-1 state. This is motivated by the fact that only if the indices are all different one obtains a quantity $\epsilon^{ABCDE} Z_{AB} Z_{CD}$ which is non-vanishing, where $Z_{AB}$ is the central charge and the indices $A,B,...$  are vector indices of the R-symmetry $SO(5)$.

We list in \autoref{conjectureddefectboundstates} all the a-priori different bound states that one obtains, together with the conjectured supersymmetry that they preserve according to our criterion.
A test of the validity of these conjectures, as well as
a careful analysis of the orbits corresponding to the different bound states in \autoref{conjectureddefectboundstates}, is beyond the scope of this paper. Recently,
in \cite{deBoer:2014iba} a detailed analysis of the supersymetric solutions of maximal supergravity in three dimensions has been performed. These solutions correspond to 0-branes, which are defect branes in three dimensions. In that analysis, the classification of the solutions in terms of nilpotent orbits plays a crucial role. Although our results are too preliminary to allow any detailed comparison, it would be interesting to analyze the representatives of the supersymmetric orbits in Table 2 of  \cite{deBoer:2014iba} in relation with the bound states of root vectors of $E_{8(8)}$ that one can construct using our method.
We hope to report in this direction in the near future.

Finally, we can comment of the possibility of extending this analysis to the defect branes of the symmetric ${\cal N}=2$ theories discussed in previous sections. In five dimensions, these theories do not contain any single 1/2-BPS 3-brane, and it would thus be crucial to analyze whether supersymmetric 3-brane solutions can be constructed at all, in order to test the validity of our conjectures. In four dimensions the situation is more complicated because in general the adjoint representation contains long (or real) roots, short (or complex) roots and Cartan generators. This means that the representation contains weights of three different lengths, and therefore a more careful analysis is required.

\section{\label{conclusions}Conclusions}

In this paper we have shown that the analysis performed in \cite{Lu:1997bg} to compute the duality orbits of BHs in maximal supergravities can be extended \textit{at least} to all the  ${\cal N}=2$ Maxwell-Einstein supergravity theories with symmetric scalar manifolds in $D=4$ and $D=5$, and based on rank-3 simple and semisimple Jordan algebras.
A crucial ingredient was the conjecture made in \cite{Bergshoeff:2014lxa} that the charge of any single 1/2-BPS $p$-brane in such theories corresponds to a  real-long-weight vector, where the reality properties of the weights are determined by the Tits-Satake diagram that identifies the real form of the Lie algebra of the duality symmetry. Focusing on asymptotically-flat branes in $D=5$ and $D=4$, we have first considered in detail the magic supergravity theories based $J_{3}^{\mathbb{R}}$ and $J_{3}^{\mathbb{C}}$. In the $J_{3}^{\mathbb{R}}$ case the symmetry algebras are split, and therefore all the weights are real. The relevant representations contain weights of two  different lengths, and we have shown that all the orbits can be computed as bound states of long-weight vectors.
In the $J_{3}^{\mathbb{C}}$ case, instead, all the weights have the same length, but the algebras are not split and hence some of the weights are not real. We have shown that the various BH orbits correspond to different bound states of real-weight vectors. We have then shown how it is straightforward to extend the analysis
to the ${\cal N}=4$ and ${\cal N}=2$ theories based on infinite sequences of rank-3 semisimple Jordan algebras. The fact that all our computations correctly reproduce the results known in the literature can be considered as a proof that the conjecture made in
\cite{Bergshoeff:2014lxa} is correct.

In general, a crucial feature of the  theories considered in this paper is the fact that all the BH solutions whose relevant orbit representative has more than one charge (\textit{i.e.}, it is of rank $>1$ \cite{Ferrar, Krutelevich}) split into different duality orbits, preserving different amount of supersymmetry.   We have shown that this stratification can essentially be traced back to the aforementioned structure of the weights  of the representations. In particular, in the $J_{3}^{\mathbb{R}}$  case, this is due to the presence of short weights, while  in the $J_{3}^{\mathbb{C}}$ case it is due to the presence of complex weights. Indeed, it is the compactness of the so-called conjunction stabilizers on such weights that flips by changing the relative sign of the charges in the bound state.
The theories based on $J_{3}^{\mathbb{C}}$, $J_{3}^{\mathbb{H}}$ and $J_{3}^{\mathbb{O}}$ all share the feature that by projecting the roots of the corresponding Lie algebras on their real part by means of the Cartan involution one always obtains the roots of the Lie algebras of the theories based on $J_{3}^{\mathbb{R}}$. Moreover, under the same projection the representations of all the brane charges are such that the real weights are fixed and become the long weights of the representations of the $J_{3}^{\mathbb{R}}$ theories, while the complex weights are mapped to the short weights \cite{Bergshoeff:2014lxa}. This explains why the stratification of the orbits is exactly the same in all the magic theories. In the maximal theories, the duality symmetries are always split, and the BH charges belong to representations whose weights all have the same length, which is the reason why the same stratification does not occur in that case.

A special attention should be paid to the 4-charge orbits in four dimensions. In the ${\cal N}=2$ magic theories, the rank-4 solutions stratify into the 1/2-BPS time-like orbit, the non-BPS space-like dyonic orbit and the non-BPS time-like orbit. We have shown that all these orbits correspond to bound states of the same four long-weight or real-weight vectors. The stabilizers that become non-compact in the non-BPS time-like orbit with respect to the 1/2-BPS case are conjunction stabilizers on short-weight or complex-weight vectors, and again this explains why this orbit does not exist in the maximal case. On the other hand, the stabilizers that become non-compact in the non-BPS space-like dyonic orbit with respect to the 1/2-BPS case are conjunction stabilizers on both short-weight and long-weight vectors (for the theory based on $J_{3}^{\mathbb{R}}$), or both complex-weight and real-weight vectors (for the other theories). This is the reason why this orbit exists also in the maximal theory.

The charges of the defect branes of the maximal supergravity theories share with the charges of the black holes of the theory based on $J_{3}^{\mathbb{R}}$  the property that their representation contain weights of different length. This led us also to conjecture that our method should be applied to classify the different bound states of such defect branes and the amount of supersymmetry that they preserve. It would be interesting to compare our results to those of
\cite{deBoer:2014iba} and also to generalize them to the case of theories with less supersymmetry, and in particular to the ${\cal N}=2$ theories considered in this paper. We leave this as an open problem.

The powerful method used to analyze and derive the structure of the
stratification of a representation space of a given Lie algebra, used in
\cite{Lu:1997bg} and exploited in the present paper, can be applied to a
variety of frameworks.
For instance, our analysis could be extended also to those Maxwell-Einstein
supergravity theories which have symmetric scalar manifolds but are not
related to rank-3 Jordan algebras, such as $\mathcal{N}=2$ minimally
coupled, $\mathcal{N}=3$ and $\mathcal{N}=5$ theories in $D=4$, and the
so-called non-Jordan symmetric sequence in $D=5$. Furthermore, it would be
straightforward to apply the above method to analyze the orbits of dyonic
black strings (as well as of electric black holes and magnetic black
2-branes) in $D=6$, both chiral and non-chiral,\footnote{Various results on duality orbits in $D=6$ were derived in \cite{Ferrara:1997ci,Lu:1997bg,Ferrara:2006xx,Andrianopoli:2007kz,Borsten:2010aa}.} theories.
Moreover, it would be interesting to consider non-supersymmetric
Maxwell-Einstein theories in $D=4,5,6$ dimensions, based on the rank-3
simple Jordan algebras $J_{3}^{\mathbb{C}_{s}}$ and $J_{3}^{\mathbb{H}_{s}}$
over the split complex numbers $\mathbb{C}_{s}$ and the split quaternions $\mathbb{H}_{s}$. These theories have a split U-duality algebra \cite{Breitenlohner:1987dg, Gunaydin:2000xr, Gunaydin:2009zza}, and they would
share the same structure of stratification as maximal supergravity (which is
based on the rank-3 simple Jordan algebra $J_{3}^{\mathbb{O}_{s}}$ over the
split complex numbers $\mathbb{O}_{s}$).

The analysis presented in this work may also turn to be useful to analyze
and gain insights in the structure of duality orbits of multi-centered
extremal black $p$-branes in supergravity, for example of two-centered
extremal black holes in $D=4$ Maxwell-Einstein theories. Some results on the
stabilizing subalgebras of such orbits have been derived in literature \cite{Ferrara:2010ug, Andrianopoli:2011gy, Ceresole:2011xd, Ferrara:2011di,
Ferrara:2012yp, Cacciatori:2012tj} (see also \cite{Bossard:2013nwa}), but
they only concern ``generic'' orbits.

Another framework in which the approach of this paper could be exploited is
provided by flux compactifications. For instance, it may be applied to the
determination of the stratification of the representation space of
non-geometric fluxes, along the lines and motivations \textit{e.g.} of \cite{Dibitetto:2012rk}.

Finally, it would be interesting to study the orbit structure pertaining to
the recently introduced ``Born-Infeld attractors'' in the
context of new types of $U(1)^{n}$ Born-Infeld actions based on rigid $\mathcal{N}=2$ special geometry in four dimensions \cite{Ferrara:2014oka,
Ferrara:2014nwa}.

\vskip .7cm

\section*{Acknowledgments}
F.R. would like to thank G. Pradisi for discussions.
The work of A.M. has been supported by a Senior Grant of the ``Enrico Fermi'' Center, Rome, in association
with the Department of Physics and Astronomy ``Galileo Galilei'' of the University of Padova.

\vskip .7cm

\appendix

\section{\label{appendixshortweights}Short weights and $F_{\alpha}^{\pm}$ stabilizers}

The analysis developed in \autoref{magicn2sugraR} for split Lie algebras shows
a variety of brane bound states richer than the ones found in the
maximal theory \cite{Lu:1997bg}. This variety, that yields a
stratification of the corresponding representation space into orbits, characterized by different compactness of
the semisimple part of their stabilizing subalgebra, can ultimately be tracked back to algebraic properties of the
representations we have analyzed (namely, the $\mathbf{6}$ of $\mathfrak{sl}(3,\mathbb{R})$ in $D=5$ and the $\mathbf{14}^{\prime }$ of $\mathfrak{sp}(6,\mathbb{R})$ in $D=4$). As we have outlined in \autoref{magicn2sugraR}, the only generators
able to modify the compactness of the stabilizing algebra are the $F_{\alpha }^{\pm }$'s. On the other hand, the aforementioned U-duality representations of
BHs in the magic $\mathcal{N}=2$ supergravity based on $J_{3}^{\mathbb{R}}$ differ from the maximal theory ones (namely, $\mathbf{27}$ of $\mathfrak{e}_{6(6)}$ in $D=5$ and $\mathbf{56}$ of $\mathfrak{e}_{7(7)}$ in $D=4$) for the presence of
weights of different lengths.

In this appendix, we aim at making the link
between the presence of short weights and orbit stratification explicit.
This link essentially resides in the relation between short weights and $F_{\alpha
}^{\pm }$ stabilizers, realizing the statement that if the bound state of two long-weight vectors of a representation, say $\ket{\Lambda _{1}},\ket{\Lambda _{2}}$, is stabilized by
an operator like $F_{\alpha }^{\pm }$, assuming that
\begin{flalign}
     &E_{\alpha}\ket{\Lambda_{1}}=\ket{\Sigma} \nonumber \\
   &E_{-\alpha}\ket{\Lambda_{1}}=0\nonumber \\
   &E_{\alpha}\ket{\Lambda_{2}}=0\nonumber \\
   &E_{-\alpha}\ket{\Lambda_{2}}=\ket{\Sigma} \ \ ,\label{equationdefiningpandq}
      \end{flalign}
then the
weight $\Sigma=\Lambda _{1}+\alpha =\Lambda _{2}-\alpha $ must be a short weight,
namely
\begin{flalign}
 & (\Sigma , \Sigma) = (\Lambda_{1}+\alpha,\Lambda_{1}+\alpha ) = (\Lambda_{2}-\alpha,\Lambda_{2}-\alpha ) <  ( \Lambda_{1},\Lambda_{1} )= (\Lambda_{2},\Lambda_{2} ) \ .\label{sigmaisshort}
\end{flalign}
To prove this, we show that if this inequality did not hold, this would lead to an inconsistency.
In particular, let us suppose that  $(\Lambda_{1}+\alpha,\Lambda_{1}+\alpha ) \geq  ( \Lambda_{1},\Lambda_{1} )$ and $(\Lambda_{2}-\alpha,\Lambda_{2}-\alpha ) \geq  ( \Lambda_{2},\Lambda_{2} )$. This would imply that
\begin{flalign}
 & \frac{2( \alpha,\Lambda_{1})}{( \alpha,\alpha )} \geq -1 \nonumber \\
 - & \frac{2(\alpha,\Lambda_{2})}{(\alpha,\alpha )} \geq -1 \ \ ,
\end{flalign}
or equivalently
\begin{flalign}
 &(p_{\Lambda_{1}}-q_{\Lambda_{1}})_{\alpha} \geq -1 \nonumber \\
 &(q_{\Lambda_{2}}-p_{\Lambda_{2}})_{\alpha}\geq -1 \ \ ,
\end{flalign}
where we denote with $(p_{\Lambda })_{\alpha }$ and $%
(q_{\Lambda })_{\alpha }$  the steps down and up respectively that one can make
starting from $\Lambda $ along the direction $\alpha $ and reaching another
weight vector.
This relation is clearly inconsistent because \autoref{equationdefiningpandq} implies
\begin{flalign}
&(p_{\Lambda_{1}})_{\alpha}=0 \nonumber \\
&(q_{\Lambda_{1}})_{\alpha}=2 \nonumber \\
&(q_{\Lambda_{2}})_{\alpha}=0\nonumber \\
&(p_{\Lambda_{2}})_{\alpha}=2 \ \ .
\end{flalign}
Hence, this proves \autoref{sigmaisshort}, and therefore $\Sigma$ must be a short weight.

To conclude,  the presence of the $%
F_{\alpha }^{\pm }$ as stabilizers is directly related to the presence of weights of different
length  in the representation. As shown in the paper, the stratification of the 2-charge and 3-charge orbits is only  due to the presence of such generators in the stabilizing algebra.
This explains why in the
maximal supergravity theories, whose BH charge representations have weights all with the same length, such stratification does not occur.

\section{\label{appendixextraspecial}Cartan involution and extraspecial pairs}

In the paper we have derived the duality orbits of extremal BH solutions using the following procedure:
\begin{itemize}

\item

the identification of the stabilizing algebra, in
particular of its semisimple part, by studying the commutation
relations of the stabilizers from the commutation relations of
the U-duality Lie algebra itself. This can be done reconstructing the Chevalley basis
and sometimes requires a complexification, since not all real forms
admit Chevalley basis;

\item

the identification of the real form of the stabilizing algebra from the action of the Cartan involution on its generators.

\end{itemize}
In these two steps a relevant role is played by the structure constants of
the U-duality algebra. In particular, while the identification of an algebra
starting from a set of generators essentially reduces to identifying the structure constants,
the connection between Cartan involution and structure constants is more
subtle.

In order to determine the action of the Cartan involution on the stabilizing generators, we have derived the dual action on the corresponding roots from the Tits-Satake diagram of the real form of the U-duality algebra through the scheme elucidated in \autoref{secmagicalCHO}.
In particular, a black painted node, corresponding to an imaginary simple root, which is fixed under $\theta$, gives a generator which is also fixed under $\theta$, and hence compact. On the other hand, for an unpainted node, either isolated or linked by an arrow to another unpainted node,  the action of the Cartan involution on the
corresponding root vector is fixed only up to a sign:
\begin{flalign}
 &\theta E_{\alpha}=\rho_{\alpha}E_{\theta\alpha} \quad ,\label{signambiguityrho}
\end{flalign}
where $\rho_{\alpha}$ can be chosen to be $\pm 1$. Once the action of the Cartan
involution on the simple-root generators is defined, its action on the other generators can be deduced by using the commutation rules
\begin{flalign}
 &[E_{\alpha},E_{\beta}]=\left\{\begin{array}{ll}
 N_{\alpha,\beta}E_{\alpha+\beta}& \mbox{if } \alpha+\beta \ \mbox{root}\\
 H_{\alpha}  & \mbox{if } \alpha+\beta=0 \\
 0&\mbox{if }\alpha+\beta \ \mbox{not \ root} \label{definitionofNalphabeta}
 \end{array},
 \right.
\end{flalign}
by means of
\begin{flalign}
 &\theta [E_{\alpha},E_{\beta}]=N_{\alpha,\beta} \theta E_{\alpha+\beta}= [\theta E_{\alpha},\theta E_{\beta}]= \rho_{\alpha} \rho_{\beta} [ E_{\theta\alpha}, E_{\theta\beta}]=\rho_{\alpha} \rho_{\beta}N_{\theta\alpha,\theta\beta} E_{\theta\alpha+\theta\beta} \quad ,
\end{flalign}
yielding
\begin{flalign}
 & \theta E_{\alpha+\beta}=\rho_{\alpha} \rho_{\beta}\frac{N_{\theta\alpha,\theta\beta}}{N_{\alpha,\beta}} E_{\theta\alpha+\theta\beta} \quad .\label{cartaninvstruct}
\end{flalign}
This relation shows that the action of the Cartan involution
on any generator can be deduced from its action on the simple roots, by
knowing the structure constants of the underlying Lie algebra.

Stating from the commutation relations in \autoref{definitionofNalphabeta}, by
using the
antisymmetry of the commutator and the Jacobi identity one obtains that the
structure constants must satisfy the following identities (for further details see {\it e.g.} \cite{Carter}):
\begin{flalign}
 &N_{\alpha,\beta}=-N_{\beta,\alpha}\nonumber \\
 &\frac{N_{\alpha_{1},\alpha_{2}}}{\langle\alpha_{3}, \alpha_{3}\rangle}=\frac{N_{\alpha_{3},\alpha_{1}}}{\langle\alpha_{2}, \alpha_{2}\rangle}=\frac{N_{\alpha_{2},\alpha_{3}}}{\langle\alpha_{1}, \alpha_{1}\rangle}\nonumber \\
 &N_{\alpha,\beta}N_{-\alpha,-\beta}=-(p+1)^{2}\nonumber \\
 &\frac{N_{\alpha_{1},\alpha_{2}}N_{\alpha_{3},\alpha_{4}}}{\langle \alpha_{1}+\alpha_{2},\alpha_{1}+\alpha_{2}\rangle}+\frac{N_{\alpha_{2},\alpha_{3}}N_{\alpha_{1},\alpha_{4}}}{\langle \alpha_{2}+\alpha_{3},\alpha_{2}+\alpha_{3}\rangle}
+\frac{N_{\alpha_{3},\alpha_{1}}N_{\alpha_{2},\alpha_{4}}}{\langle \alpha_{3}+\alpha_{1},\alpha_{3}+\alpha_{1}\rangle}=0 \quad .\label{allreln14}
\end{flalign}
where $p$ is defined as the largest integer such that $\beta -p\alpha$ is a root.
It is possible to satisfy these equations with the
Chevalley basis elements defined in such a way that \cite{Carter}
\begin{equation}
N_{\alpha,\beta}=\pm(p+1) \quad .\label{signambiguity}
\end{equation}
The sign ambiguity in \autoref{signambiguity} can only partially be fixed by the relations in \autoref{allreln14}, and the pairs of roots $\alpha$ and $\beta$ such that  the sign of $N_{\alpha\beta}$ is undetermined are called \textit{extraspecial pairs}. Once one makes a choice for these signs, all the other structure
constants signs are fixed.
In general, the roots $\alpha$ and $\beta$ form a  {\textit{special pair}} if $%
0<\alpha<\beta$ and $\alpha+\beta$ is a root. Such a pair in an extraspecial pair if, for all the other
special pairs $(\alpha^{\prime },\beta^{\prime })$ enjoying $\alpha+\beta=%
\alpha^{\prime }+\beta^{\prime }$, it holds that $\alpha < \alpha^{\prime }$. For the particular case of the Lie algebra $\mathfrak{sp}(6,\mathbb{R})$ that was discussed in \autoref{magicn2sugraR}, we list in \autoref{structureconstantsofsp6} the structure constants $N_{\alpha\beta}$, that are all uniquely determined once the choice for the signs of the structure constants associated to the extraspecial pairs is fixed as done in the table.

\begin{table}[h!]
\renewcommand{\arraystretch}{1.8}
\par
\begin{center}
\resizebox{\textwidth}{!}{
  \begin{tabular}{|c|c|c|c|c|c|c|c|c|c|c|c|c|c|c|c|c|c|c|c|} \hline
\centering\textcolor{black}{$N_{\alpha,\beta}$}&\multicolumn{19}{|c|}{\textcolor{black}{$\alpha$}}\\ \cline{1-20}
& &\rotatebox{90}{( 1 0 0 )} &\rotatebox{90}{( 0 1 0 )} &\rotatebox{90}{( 0 0 1 )} &\rotatebox{90}{( 1 1 0 )} &\rotatebox{90}{( 0 1 1 )} &\rotatebox{90}{( 1 1 1 )} &\rotatebox{90}{( 0 2 1 )} &\rotatebox{90}{( 1 2 1 )} &\rotatebox{90}{( 2 2 1 )} &\rotatebox{90}{( -1 0 0 )} &\rotatebox{90}{( 0 -1 0 )} &\rotatebox{90}{( 0 0 -1 )} &\rotatebox{90}{( -1 -1 0 )} &\rotatebox{90}{( 0 -1 -1 )} &\rotatebox{90}{( -1 -1 -1 )} &\rotatebox{90}{( 0 -2 -1 )} &\rotatebox{90}{( -1 -2 -1 )} &\rotatebox{90}{( -2 -2 -1 )}\tabularnewline  \cline{2-20}
&( 1 0 0 )&\cellcolor{black}&\cellcolor{gray!40}1 &0 &0 &1 &0 &1 &-2 &0 &\cellcolor{black}&0 &0 &-1 &0 &-1 &0 &-2 &1 \tabularnewline \cline{2-20}

&( 0 1 0 )&-1 &\cellcolor{black}&\cellcolor{gray!40}1 &0 &-2 &-1 &0 &0 &0 &0 &\cellcolor{black}&0 &1 &-2 &0 &1 &1 &0 \tabularnewline \cline{2-20}

&( 0 0 1 )&0 &-1 &\cellcolor{black}&-1 &0 &0 &0 &0 &0 &0 &0 &\cellcolor{black}&0 &1 &1 &0 &0 &0 \tabularnewline \cline{2-20}

&( 1 1 0 )&0 &0 &\cellcolor{gray!40}1 &\cellcolor{black}&-1 &2 &0 &0 &0 &-1 &1 &0 &\cellcolor{black}&0 &-2 &0 &1 &-1 \tabularnewline \cline{2-20}

&( 0 1 1 )&-1 &\cellcolor{gray!40}2 &0 &1 &\cellcolor{black}&0 &0 &0 &0 &0 &-2 &1 &0 &\cellcolor{black}&1 &-1 &-1 &0 \tabularnewline \cline{2-20}

&( 1 1 1 )&0 &\cellcolor{gray!40}1 &0 &-2 &0 &\cellcolor{black}&0 &0 &0 &-1 &0 &1 &-2 &1 &\cellcolor{black}&0 &-1 &1 \tabularnewline \cline{2-20}

&( 0 2 1 )&-1 &0 &0 &0 &0 &0 &\cellcolor{black}&0 &0 &0 &1 &0 &0 &-1 &0 &\cellcolor{black}&1 &0 \tabularnewline \cline{2-20}

&( 1 2 1 )&\cellcolor{gray!40}2 &0 &0 &0 &0 &0 &0 &\cellcolor{black}&0 &-2 &1 &0 &1 &-1 &-1 &1 &\cellcolor{black}&-1 \tabularnewline \cline{2-20}

\multirow{-10}{*}{\rotatebox{90}{\textcolor{black}{$\beta$}}}&( 2 2 1 )&0 &0 &0 &0 &0 &0 &0 &0 &\cellcolor{black}&1 &0 &0 &-1 &0 &1 &0 &-1 &\cellcolor{black}\tabularnewline

\hline\end{tabular}}
\end{center}
\caption{The structure constants of $\mathfrak{sp}(6,\mathbb{R})$. Given the simple roots $\alpha_1$, $\alpha_2$ and $\alpha_3$, each root $a\alpha_1 + b  \alpha_2 + c \alpha_3$ is identified in terms of the three integers $a,b,c$. The grey entries are those associated to the extraspecial pairs of roots.
We only list $N_{\alpha\beta}$ for $\beta$ a positive root. The other cases can be deduced using the relations in \autoref{allreln14}.} \label{structureconstantsofsp6}
\end{table}

It should be stressed that in general
the action of the Cartan involution depends on how the signs of the
structure constants are chosen, as can be seen from \autoref{cartaninvstruct}. An exception to this general rule occurs for split real forms, in which case $\theta$ acts on all simple roots as $\theta\alpha=-\alpha$, and using the third of \autoref{allreln14} and \autoref{signambiguity} one can show that \autoref{cartaninvstruct} can be recast as
\begin{flalign}
 & \theta E_{\alpha+\beta}=\rho_{\alpha} \rho_{\beta}\frac{N_{-\alpha,-\beta}}{N_{\alpha,\beta}} E_{-\alpha-\beta}=-\rho_{\alpha} \rho_{\beta} E_{-\alpha-\beta} \quad ,\label{thetaEalphabetafromsimplerootsandrho}
\end{flalign}
where $\rho_\alpha$ is the arbitrary sign in the Cartan involution for the simple roots associated to the unpainted nodes defined in \autoref{signambiguityrho}.
The relation in \autoref{thetaEalphabetafromsimplerootsandrho}
shows that the action of the Cartan involution for split real forms  does not depend on the structure constants, but only
on the choice of the $\rho_{\alpha}$'s. In particular, in this paper we made the choice $\rho_{\alpha_i} =-1$ for all simple roots, so that $\theta E_\alpha = - E_{-\alpha}$ for any root.

\section{\label{appendixpictures}Pictorial derivation of the conjunction stabilizers}

\begin{figure}[t!]
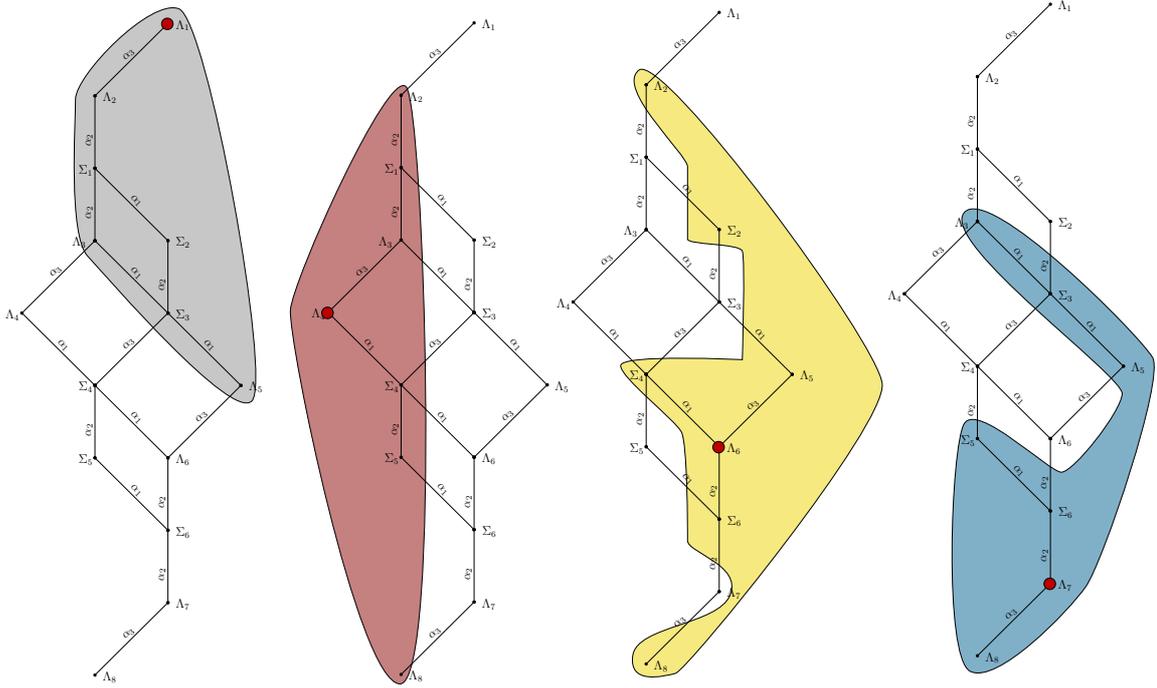

\centering


\scalebox{0.4} 
{

}
\caption{Figure showing in the colored sets the weights that are connected to the long weights of the ${\bf 14^\prime}$  of $\mathfrak{sp}(6,\mathbb{R})$ considered in the paper, that are painted in red.} \label{drawingoftheorbitsofthe14}
\end{figure}

\begin{figure}[t!]
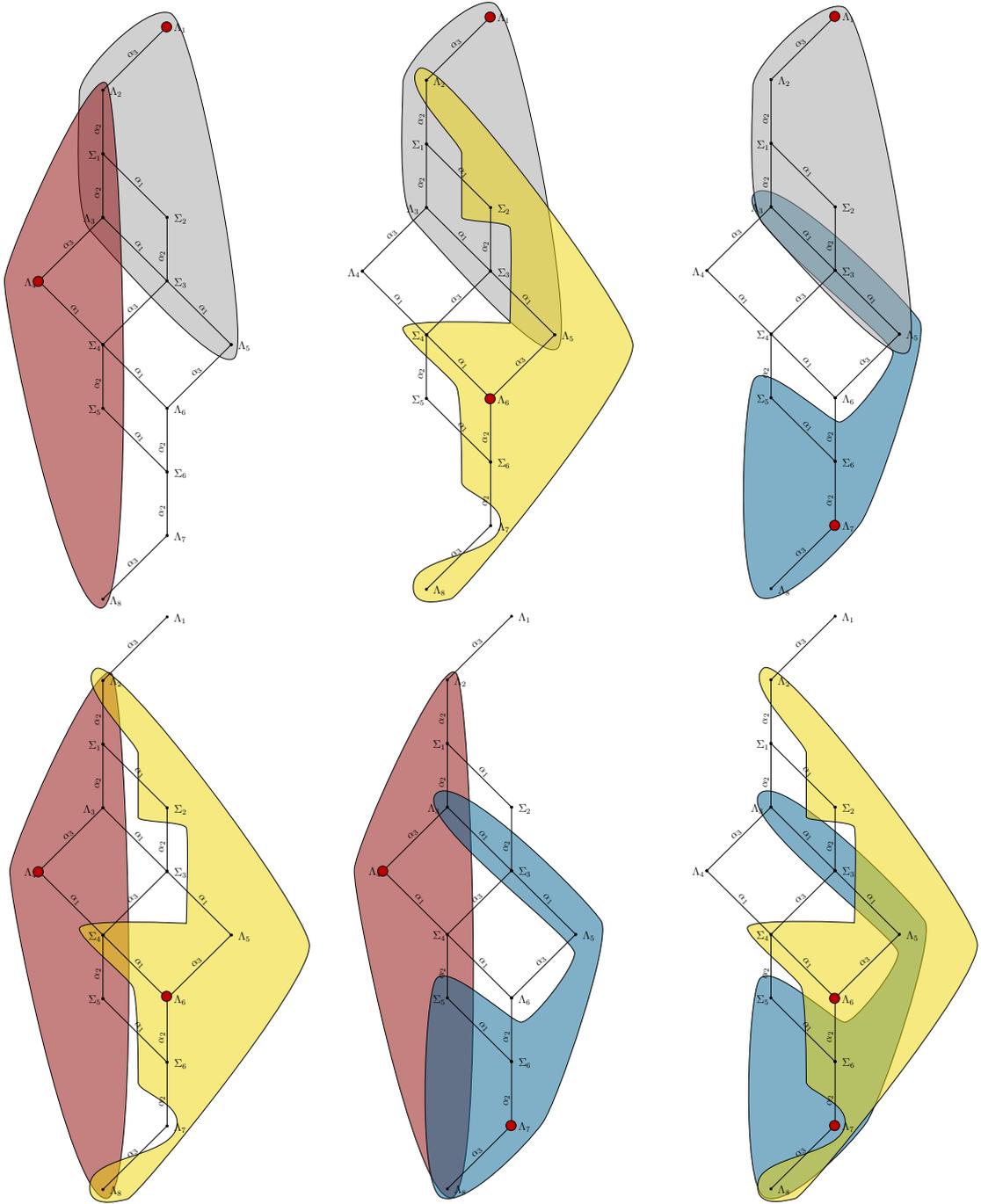

\scalebox{0.4} 
{
%
}

\caption{Figure showing the intersections between any pair of colored sets of \autoref{drawingoftheorbitsofthe14}.  }\label{intersectionsofcolors}
\end{figure}

In this paper we have derived the orbits of various multi-charge BH solutions from the stabilizers of bound states of suitably chosen weight vectors. The condition that these weight vectors have to satisfy is that they correspond to long weights (for split real forms) or real weights (for other real forms) and that they are independent, {\it i.e.} not connected by transformations of the algebra. The aim of this appendix  is to make it clear that the results obtained  do not depend on the particular choice of weights.

We consider the particular case of the 2-charge orbits of the  ${\bf 14^\prime}$ of $\mathfrak{sp}(6,\mathbb{R})$, whose weights are shown in \autoref{14ofsp6}. In the paper we have considered the weights $\Lambda_1$, $\Lambda_4$, $\Lambda_6$ and $\Lambda_7$. We draw in \autoref{drawingoftheorbitsofthe14} all the weights that are connected to each of these four weights by transformations of $\mathfrak{sp}(6,\mathbb{R})$.  In each case, all such weights are those inside the colored set. From the figure it is clear that the stabilizing algebra for each of these weights is the same because the number of long weights and short weights that are connected to a given long weight is always the same.

In \autoref{intersectionsofcolors} we draw the intersection of colored sets for any pair of diagrams in \autoref{drawingoftheorbitsofthe14}. In particular, the first diagram corresponds to the intersection between the set of weights connected to $\Lambda_1$ and those connected to $\Lambda_4$. We can see from the diagram that the only weights belonging to the intersection are the short weight $\Sigma_1$ and the long weights $\Lambda_2$ and $\Lambda_3$. By comparing with \autoref{14stab2states}, we see that these three weights correspond to the three conjunction stabilizers. If we now consider all the other remaining five pairs, we see that in all cases the intersection set is made of one short and two long weights, which  implies that the stabilizing algebra for any pair of two long weights is the same.

\vskip .7cm

\bibliographystyle{plain}
\bibliography{orbitsbibliography}
{}

\end{document}